\documentclass[10pt]{article}
\usepackage[top=2cm,left=2cm,right=2cm]{geometry}
\usepackage[numbers,square]{natbib}

\usepackage[english]{babel}
\usepackage{epsfig}
\usepackage{hyperref}
\usepackage{amssymb}
\usepackage{amsfonts}
\usepackage{epsf}
\usepackage{rotating}
\usepackage{graphicx}
\usepackage{amsmath}
\usepackage{fancyhdr}
\usepackage{lineno}
\usepackage{subfigure}
\usepackage{babel}
\usepackage{graphics}
\usepackage{geometry}
\usepackage{pstricks}
\usepackage{color}

\makeatletter
\def\cleardoublepage{\clearpage\if@twoside
\ifodd\c@page
\else\hbox{}\thispagestyle{empty}\newpage
\if@twocolumn\hbox{}\newpage\fi\fi\fi}
\makeatother

\let\a=\alpha   \let\b=\beta   \let\g=\gamma   \let\d=\delta
         
    \let\k=\kappa    
                 \let\r=\rho
\let\s=\sigma

\def\a{\alpha}
\def\b{\beta}

\def\g{\gamma}

\def\r{\rho}

\def\ds#1{#1\kern-1ex\hbox{/}}
\def\dsh{h\kern-1.2ex /}

\newcommand{\bea}{\begin{eqnarray}}
\newcommand{\eea}{\end{eqnarray}}

\def\nn{\nonumber}

\def\beq{\begin{equation}}
\def\eeq{\end{equation}}

\def\ba{\begin{eqnarray}}
\def\ea{\end{eqnarray}}

\def\slash#1{#1\hskip-6pt/\hskip6pt}

\newcommand{\beqa}{\begin{eqnarray}}
\newcommand{\eeqa}{\end{eqnarray}}

\newcommand{\si}{\sigma}

\newcommand{\pd}{\partial}

\newcommand{\gsim}{\lower.7ex\hbox{$\;\stackrel{\textstyle>}{\sim}\;$}}
\newcommand{\lsim}{\lower.7ex\hbox{$\;\stackrel{\textstyle<}{\sim}\;$}}

\begin{document}
\begin{center}
\vspace{4.cm}
{\bf \large  Graviton Vertices and the Mapping of Anomalous Correlators  \\ to Momentum Space for a General
Conformal Field Theory}

\vspace{1cm}

{\bf $^a$Claudio Corian\`{o}, $^a$Luigi Delle Rose, $^b$Emil Mottola and $^a$Mirko Serino}

\vspace{1cm}

{\it$^a$ Dipartimento di Matematica e Fisica \\
Universit\`a del Salento\\ and \\
INFN
Lecce, Via Arnesano 73100 Lecce, Italy\\}
\vspace{0.5cm}
{\it$^b$ Theoretical Division, Los Alamos National Laboratory \\Los Alamos, NM, 87545, USA\footnote{claudio.coriano@unisalento.it, luigi.dellerose@le.infn.it, emil@lanl.gov, mirko.serino@le.infn.it}}

\begin{abstract}

We investigate the mapping of conformal correlators and of their anomalies from configuration to momentum space for 
general dimensions, focusing on the anomalous correlators $TOO$, $TVV$ - involving the energy-momentum tensor $(T)$ 
with a vector $(V)$ or a scalar operator ($O$) - and the 3-graviton vertex $TTT$. We compute the $TOO$, $TVV$ and $TTT$ 
one-loop vertex functions in dimensional regularization for free field theories involving conformal scalar, fermion and vector fields. 
Since there are only one or two independent tensor structures solving all the conformal Ward identities 
for the $TOO$ or $TVV$ vertex functions respectively, and three independent tensor structures for the $TTT$ vertex,
and the coefficients of these tensors are known for free fields, it is possible to identify the corresponding tensors 
in momentum space from the computation of the correlators for free fields. This works in general $d$ dimensions 
for $TOO$ and $TVV$ correlators, but only in $4$ dimensions for $TTT$, since vector fields are conformal only in $d=4$. 
In this way the general solution of the Ward identities including anomalous ones for these correlators in (Euclidean) 
position space, found by Osborn and Petkou is mapped to the ordinary diagrammatic one in momentum space.
We give simplified expressions of all these correlators in configuration space which are explicitly Fourier integrable 
and provide a diagrammatic interpretation of all the contact terms arising when two or more of the points coincide.
We discuss how the anomalies arise in each approach. We then outline a general algorithm for mapping correlators 
from position to momentum space, and illustrate its application in the case of the $VVV$ and $TOO$ vertices.  
The method implements an intermediate regularization - similar to differential regularization - for the identification 
of the integrands in momentum space, and one extra regulator. The relation between the ordinary Feynman expansion 
and the logarithmic one generated by this approach are briefly discussed.

\end{abstract}

\end{center}
\newpage

 \section{Introduction}
The analysis of correlation functions in $d$-dimensional quantum field theory possessing conformal invariance has found widespread interest over the years (see \cite{Fradkin:1997df} for an overview). Given the infinite dimensional character of the conformal algebra in 2-dimensions, conformal field theories (CFT's) in 2-dimensions have received the most attention, although 4-dimensional conformal theories have also been studied (see for instance \cite{Stanev:1987dh, Stanev:1988ft}). In $d$ dimensional CFT's the structure of generic conformal correlators is not entirely fixed just by conformal symmetry, but for 2- and 3-point functions the situation is rather special and these can be significantly constrained, up to a small number of constants. 

In several recent works \cite{Giannotti:2008cv} \cite{Armillis:2009pq, Armillis:2010qk} certain correlation functions describing the interaction between a gauge theory and gravity with massless fields in the internal loop and related therefore to the axial and trace anomalies in these theories have been analyzed. The interesting property that such anomalous amplitudes contain massless poles in 2-particle intermediate states has been exposed in these investigations. In particular this has been demonstrated in the TVV amplitude in massless QED and QCD, characterized by the insertion of the energy momentum tensor (T) on 2-point functions of vector gauge currents (V). This amplitude gives the leading order contribution to the interaction between a gauge theory and gravity, mediated by the trace anomaly.

The complete evaluation of this amplitude in the Standard Model \cite{Coriano:2011zk} confirms the conclusion of \cite{Giannotti:2008cv}, namely the presence of an effective massless scalar ``dilaton-like" degree of freedom in intermediate 2-particle states intimately connected with the trace anomaly, in the sense that the non-zero residue of the pole is necessarily proportional to the coefficient of the anomaly. The perturbative results of \cite{Giannotti:2008cv} \cite{Armillis:2009pq, Armillis:2010qk} are also in agreement with the anomaly-induced gravitational effective action in 4 dimensions whose non-local form was found in \cite{Riegert:1984kt}, and whose local covariant form necessarily implies effective massless scalar degree(s) of freedom 
\cite{MazurMottola:2001,Mottola:2006ew,Mottola:2010}. This is the 4-dimesnional analog of the anomaly-induced action in 2-dimensional CFT's coupled to a background metric generated by the 2-dimensional trace anomaly and related to the central term in the infinite dimensional Virasoro algebra \cite{Polyakov:1981}. The anomaly-induced scalar in the 2-dimensional case is the Liouville mode of non-critical string theory on the 2-dimensional world sheet of the string.

In even dimensions greater than 2 it is important to recognize that the anomaly-induced effective action discussed in  
\cite{Riegert:1984kt,MazurMottola:2001,Mottola:2006ew, Mottola:2010} is determined only up to Weyl invariant terms. The full quantum effective action is not determined by the trace anomaly alone, and hence only when certain anomalous contributions to the TJJ or other amplitudes are isolated from their non-anomalous parts should any comparison with the anomaly-induced effective action be made. The non-anomalous components are dependent upon additional Weyl invariant terms in the quantum effective action and thus even in the CFT limit need not agree with the anomaly-induced action, without contradicting its validity for determining the anomalous terms \cite{MazurMottola:2001}. On the other hand these additional Weyl invariant terms for simple amplitudes such as TVV can be determined in principle by the Ward identities of $SO(d,2)$ conformal invariance, together with those of gauge invariance for the vector currents. Other triangle amplitudes in 4 dimensions such as the graviton-fermion-antifermion vertex function, for which similar considerations should apply have been investigated primarily for phenomenological reasons \cite{Degrassi:2008mw}, although this amplitude is anomaly-free. 

From the CFT side some important information is available \cite{Osborn:1993cr,Erdmenger:1996yc,Cappelli:2001pz}. These results  concern 
the $TOO$ - with $O$ denoting a generic scalar - $TVV$ and $TTT$ vertices, which are determined by applying the conformal Ward 
identities in Euclidean position space. Some of the vertices,  such as the $TTT$, for $d=4$ are shown in the analysis of 
\cite{Osborn:1993cr, Erdmenger:1996yc} to be expressible in terms of three linearly independent tensor structures. Imposing the 
conformal Ward identities and identifying these tensor structures directly in momentum space turns out the be technically quite 
involved. The main goal of the present work is to initiate a systematic study enabling comparison of general results of 4-dimensional 
CFT's based on position space analysis such as \cite{Osborn:1993cr, Erdmenger:1996yc} with explicit realizations of anomalous 3-point 
vertices in free field theory, most commonly expressed in momentum space. Recent results of studies of three- and four- point functions in  $d=3$ in the context of the $ADS_4/CFT_3$ correspondence are contained in 
\cite{Bzowski:2011ab,Maldacena:2011nz,Raju:2012zs}.

For general $d$ dimensions and, specifically, in $d=4$, rather than trying to identify these tensor structures 
directly in momentum space, which is quite cumbersome, it is much simpler to calculate explicitly the $TTT$ correlator for specific 
free-field theories of scalars, spinors, and vectors in one-loop Lagrangian perturbation theory, thereby identifying the three 
linearly independent tensor structures {\it a posteriori} with the general CFT analysis of \cite{Osborn:1993cr, Erdmenger:1996yc}. A 
similar method works for the $TVV$, $VVV$ and $TOO$ vertices for any dimension, while in the $TTT$ case the contribution coming from 
the exchange of a spin 1 field in the loop diagrams is  conformally invariant only in $d=4$.

While the imposition of the conformal Ward identities is technically simpler in position space, the appearance of massless poles 
associated with anomalies is very much obscured. Indeed conformal anomalies necessarily arise quite differently in momentum space and 
in Euclidean position space, where the only possibility for anomalous terms lies in appearance of ultralocal divergences proportional 
to delta functions or derivatives thereof at cooincident spatial points. Thus a very careful regularization procedure is required to 
determine these anomalous ultralocal contributions which are absent for any finite point separation. The special strategy followed in 
determining these anomalous ultralocal contributions in position space, developed in \cite{Osborn:1993cr}, merits some comments for 
its peculiarity.
In \cite{Osborn:1993cr, Erdmenger:1996yc} the Ward identities are solved in each case by combining a homogeneous solution - obtained 
for separate (non-coincident) points of the correlator - with inhomogenous terms, identified via a regularization of the same 
correlator in the coincidence limit and with the inclusion of contact terms. The contact terms proportional to delta functions and 
derivatives thereof determine the anomalies. Such a separation, based on homogeneous and inhomogeneous terms in the Ward identities 
cannot be easily carried out in momentum space. Moreover in the approach of \cite{Giannotti:2008cv} the origin of the conformal 
anomaly as an {\it infrared} effect (rather than a result of any UV regularization procedure) following from the imposition of all 
non-anomalous Ward identities and the spectral representation of the amplitude was emphasized. In this approach massless anomaly 
poles at $k^2=0$ play an essential role. At first glance this appears to be quite different than the ultralocal delta function terms 
obtained in the position space approach of \cite{Osborn:1993cr, Erdmenger:1996yc}. Thus the relationship of the several approaches 
requires some clarification, and this is a principal motivation for the present work. The eventual agreement of the two approaches 
may seem less surprising if it is remembered that cooincident point singularities in Euclidean position space become light cone 
singularities in Minkowski spacetime, and these lightcone singularities are associated with the propagation of massless fields, which 
generally have long range infrared effects.

Our work is composed of two main parts. In the first part, building on the results of \cite{Osborn:1993cr, Erdmenger:1996yc}, we 
compute the complete structure of the 3-point correlators in configuration and in momentum space for a general CFT. In particular we 
generalize our previous studies of the $TVV$ correlator, formally studied by us in 4 dimensions \cite{Giannotti:2008cv} 
\cite{Armillis:2009pq, Armillis:2010qk} in QED and QCD, to $d$ dimensions and for any CFT. We also study the $TTT$ vertex and perform 
a complete investigation of this correlator by the same approach. The analysis is performed in perturbation theory and the result is 
secured by a successful test of all the Ward identities satisfied by this vertex, outlining their derivation and their perturbative 
implementation, and using a symbolic manipulation program written by us. Both for the $TVV$ and $TTT$ cases our computations have 
been performed under the most general (off-shell) conditions, but the remarkable complexity of the general result allows us to 
present here, in a compact form, only the expression for the 2-particle on-shell case. We give particular emphasis to the discussion 
of the connection between the general approach of \cite{Osborn:1993cr} and the perturbative picture. In particular, we give a 
diagrammatic interpretation of the various contact terms introduced by Osborn and Petkou in order to solve the Ward identities for 
generic positions of the points of the correlators. This allows to close a gap between the bootstrap method of \cite{Osborn:1993cr}, 
our previous investigations of the TVV \cite{Giannotti:2008cv} \cite{Armillis:2009pq, Armillis:2010qk}, and the current study of the 
$TTT$ vertex. We show that the perturbative analysis in momentum space in dimensional regularization is in complete agreement with 
their results.

It should be remarked that, in general, the momentum space formulation of the correlators of a CFT remains largely unexplored, since 
in many cases there is no Lagrangian description which may justify such an effort, and the spacetime formulation remains the only 
significant one. The use of symmetry principles to infer the general solution to conformal Ward identities from some specific 
correlation functions computed in momentum space perturbation theory, allows to collect information about a conformal theory even 
when a Lagrangian formulation of the same correlators is not readily found or may not exist at all. 

This brings us to the second part of our work, contained in section 8, where we discuss a general and very efficient procedure to 
map to momentum space any massless correlator, not necessarily related to a Lagrangian description. This part is motivated by the 
attempt of transforming to momentum space any massless correlator given in position space, independently from whether this is Fourier 
integrable or not. 

The investigation of these correlators in momentum space reveals, in general, some specific facts, such as the presence of single and 
multi-logarithmic integrands which, in general, can't be re-expressed in terms of ordinary master integrals, typical of the Feynman 
expansion. To address these points, one has to formulate an alternative and general approach to perform the transforms, {\em not 
directly linked to the free-field realization}, since in this case such representation, as we have just mentioned, may not exist. 

The method that we propose combines a $d$-dimensional version of differential regularization, similar to the approach suggested in 
\cite{Osborn:1993cr, Erdmenger:1996yc}. In our case we use the standard technique of "pulling out" derivatives (via partial 
integration) in very singular correlators in such a way to make them Fourier integrable, {\it i.e.} expressible as integrals in 
momentum space which are well-defined for non-coincident points. This is combined with {\em the method of uniqueness} 
\cite{Kazakov:1986mu},  here generalized to tensor structures, in order to formulate a complete and self-consistent procedure. As in 
\cite{Osborn:1993cr, Erdmenger:1996yc} we need an extra regulator ($\omega$), unrelated to the dimensional regularization parameter 
($\epsilon$). Our approach is defined as a generic algorithm which can handle rather straightforwardly any massless correlator 
written in configuration space. The algorithm has been implemented in a symbolic manipulation program and can handle correlators of 
any rank.

The aim of the method is to test the Fourier integrability of a given correlator, by checking the cancellation of the singularities in the extra regulator directly in momentum space, and to provide us with the direct expression of the transform. After a few non trivial examples,
we will show how to reproduce, by this method, some of the results of the conformal correlators discussed in the first part, the $VVV$ and the $TOO$ being two examples.

Given the large space and scope of this analysis, which is technically quite involved, we will not attempt in this work to address the issue of the presence of anomaly poles in the $TTT$ correlator in analogy to what discussed in \cite{Giannotti:2008cv} \cite{Armillis:2009pq, Armillis:2010qk} for the TVV case.  Although this is an important motivation for initiating this study, demonstrating the existence of the pole(s) requires additional analysis which we do not attempt in this paper. We expect to address this final point in a related work making use of the technical framework and building upon the results of the present study.

\section{Conformal Correlators and the Trace Anomaly}

\subsection{Conventions and the trace anomaly equation}

Before coming to a discussion of the main correlators investigated in our work we introduce here our definitions
and conventions which will be used throughout.

The basic trace anomaly equation for a conformal theory in $d=4$ is \cite{Duff:1977ay}\cite{Duff:1993wm}

\bea \label{TraceAnomaly}
g_{\mu\nu}(z)\langle T^{\mu\nu}(z) \rangle
&=&
\sum_{I=f,s,V}n_I \bigg[\beta_a(I)\, F(z) + \beta_b(I)\, G(z) + \beta_c(I)\,\Box R(z) + \beta_d(I)\, R^2(z) \bigg]
+ \frac{\kappa}{4}n_V F^{a\,\mu\nu}\,F^a_{\mu\nu} (z) \nn \\
&\equiv& 
\mathcal{A}(z,g) \, ,
\eea
whose coefficients $\beta_a, \, \beta_b, \, \beta_c$ and $\beta_d$ depend on the field content of the Lagrangian
(fermion, scalar, vector) and we have a multiplicity factor $n_I$ for each particle species \footnote{Equivalent and more popular 
notations are $c\equiv 16 \pi^2 \beta_a$ and $a\equiv -16 \pi^2 \beta_b$}. Actually 
the coefficient of $R^2$ must vanish identically
\beq
\beta_d \equiv 0
\eeq
since a non-zero $R^2$ in this basis cannot be obtained from any effective action (local or not) 
\cite{Bonora:83,AntMazMott:1992,MazurMottola:2001}. In addition, the value of $\beta_c$ is regularization dependent, corresponding to 
the fact that it can be changed by the addition of an arbitrary local $R^2$ term in the effective action. Thus only $\beta_a$,
$\beta_b$ and $\kappa$ correspond to true anomalies in trace of the stress tensor. In dimensional regularization one finds
\beq\label{constraints}
\beta_c = -\frac{2}{3}\,\beta_a\,.
\eeq
In table \ref{AnomalyCoeff} we list the values of the coefficients for the three theories of spin $0, \frac{1}{2}, 1$
mentioned, that we are going to consider extensively throughout the paper.
\begin{table}
$$
\begin{array}{|c|c|c|c|}\hline
I & \beta_a(I)\times 2880\,\pi^2 & \beta_b(I)\times 2880\,\pi^2 & \beta_c(I)\times 2880\,\pi^2
\\ \hline\hline
S & \frac{3}{2} & -\frac{1}{2} & -1
\\ \hline
F & 9 & -\frac{11}{2} & -6
\\ \hline
V & 18 & -31 & -12
\\
\hline
\end{array}
$$
\caption{Anomaly coefficients for a conformally coupled scalar, a Dirac Fermion and a vector boson}
\label{AnomalyCoeff}
\end{table}
$\mathcal{A}(z,g)$ contains the field-strength of the background gauge field, $F^a_{\mu\nu}$,
and the invariants built out of the Riemann tensor,
${R^{\alpha}}_{\beta\gamma\delta}$, as well as the Ricci tensor $R_{\alpha\beta}$ and the scalar curvature $R$.
G and $F$ in Eq. (\ref{TraceAnomaly}) are the Euler density and the square of the Weyl tensor respectively.\\ 
All our conventions are listend in appendix \ref{ComputeTTT}.\\
Eq. (\ref{TraceAnomaly}) plays the role of a generating functional for the anomalous Ward identities of any underlying 
Lagrangian field theory. These conditions are not necessarily linked to any
Lagrangian, since the solution of these and of the other (non anomalous) Ward identities - which typically define 
a certain correlator - are based on generic requirements of conformal invariance.
For our purposes, all these identities can be extracted from an ordinary generating functional, defined in terms 
of a generic Lagrangian $\mathcal{L}$ which offers a convenient device to identify such relations.
For this reason we introduce the ordinary definition of the energy-momentum tensor
\beq \label{EMT}
T^{\mu\nu}(z) = -\frac{2}{\sqrt{-g_z}}\frac{\delta\,\mathcal{S}}{\delta g_{\mu\nu}(z)} 
= g^{\mu\alpha}(z)\,g^{\nu\beta}(z)\,\frac{2}{\sqrt{-g_z}}\,\frac{\delta\mathcal S}{\delta g^{\alpha\beta}(z)} \, ,
\eeq
in terms of the quantum action $\mathcal{ S}$, so that its quantum average is
\beq \label{VEVEMT}
\langle T^{\mu\nu}(z) \rangle = \frac{2}{\sqrt{-g_z}}\frac{\delta\, \mathcal W}{\delta\, g_{\mu\nu}(z)}\,,
\eeq
(with $\textrm{det}\, g_{\mu\nu}(z)\equiv g_z$) where $\mathcal W$ is the Euclidean generating functional of the theory
\footnote{$\mathcal W$ depends, in general, from the background metric $g_{\mu\nu}(x)$, 
the gauge fields $A^a(x)$ and scalar sources $J(x)$
In the equations below, only those dependences which are relevant for the case at hand will be explicitly indicated.}
\beq\label{Generating}
\mathcal W = \frac{1}{\mathcal{N}} \, \int \, \mathcal D\Phi \, e^{-\, \mathcal{S}} \, ,
\eeq
where $\mathcal{N}$ a normalization factor and $\Phi$ denotes all the quantum fields of the theory.\\
Inserting these definitions in (\ref{TraceAnomaly}) and multiplying both sides by $\sqrt{-g_z}$  we obtain
\beq
\label{TraceAnomalySymm}
2 \, g_{\mu\nu}(z)\frac{\delta\, \mathcal{W}}{\delta \, g_{\mu\nu}(z)}=
\sqrt{-g_z} \, \mathcal A(z,g) \, .
\eeq
From (\ref{TraceAnomaly}) and (\ref{TraceAnomalySymm}) we can extract an identity for the anomaly for correlators involving $n$ 
insertions of energy momentum tensors, by taking $n$ functional derivatives with respect to the metric of both sides of 
(\ref{TraceAnomalySymm}) and setting $g_{\mu\nu}=\delta_{\mu\nu}$ at the end.
In the same way, the anomalous Ward identity for the $TVV$ can be obtained by
functional differentiation of the same equation respect to the background gauge fields.
In perturbation theory, however, imposing the conservation Ward identity for the energy-momentum
tensor and of the Ward identity for the vector currents - whenever these are present - is sufficient to obtain
the corresponding anomalous Ward identity. In the case of the $TVV$,
for instance, this is a common practice, since only one term ($F^{a\,\mu\nu}(z)\,F^a_{\mu\nu} (z)$)
can appear in the anomaly. Therefore the anomaly condition comes as a necessary consequence of the other Ward
identities and can be checked at the end of the computation to correspond to the one derived from Eq. (\ref{TraceAnomaly}). Things 
are far more involved for vertices with multiple insertions of gravitons, such as the $TTT$ vertex,
and a successful test of the anomalous Ward identity is crucial in order to secure the correctness of the result of
the computation.

\subsection{Definition of the correlators and Ward identities for the $TVV$ and $TOO$ vertices}

We provide the basic definition of the correlators that we are going to investigate, in analogy to \cite{Osborn:1993cr}.
We start from the $TVV$ vertex and use the Euclidean convention.
We recall that in this case the functional average of the gauge current $V$ is obtained by functional differentiation
of the generating functional with respect to the background gauge field $A_\mu^a$
\bea
\langle V^{a \, \mu}(x) \rangle = -\frac{1}{\sqrt{-g_x}} \frac{\delta \mathcal{W}}{\delta A^a_\mu(x)} \bigg |_{g=\delta, A=0}  \,.
\eea
To construct the $TVV$ correlator we can first perform a functional derivative with respect to the metric followed by the flat 
space-time limit ($g_{\mu\nu}=\delta_{\mu\nu}$) and then insert the vector currents by taking derivatives with respect to the 
gauge field source $A$
\bea
\langle T^{\mu\nu}(x_1) V^{a \, \alpha} (x_2) V^{b \, \beta} (x_3) \rangle 
&=&  
\bigg\{\frac{\delta^2}{\delta A^a_\alpha(x_2) \delta A^b_\beta(x_3)} 
\bigg[\frac{2}{\sqrt{-g_{x_1}}} \frac{\delta \mathcal{W}}{\delta g_{\mu\nu}(x_1)} \bigg]_{g=\delta} \bigg\}_{A=0} \nn \\
&=&
\langle T^{\mu\nu}[A](x_1) V^{a\,\alpha}(x_2) V^{b\,\beta}(x_3) \rangle_{A=0}  + 
\langle \frac{\delta T^{\mu\nu}[A](x_1)}{\delta A^a_\alpha(x_2)} V^{b\,\beta}(x_3) \rangle_{A= 0} \nn \\
&+& 
\langle \frac{\delta T^{\mu\nu}[A](x_1)}{\delta A^b_\beta(x_3)}  V^{a\,\alpha}(x_2) \rangle_{A= 0} \nn \\
\eea
where $T_{\mu\nu}[A]$ is the energy-momentum tensor calculated in the presence of the background source $A_\mu^a$. The first term in 
the previous expression represents the insertion of the three operators, while the last two are contact terms, with the topology of 
2-point functions, exploiting the linear dependence of the energy-momentum tensor from the source field $A$. \\

The construction of the TOO correlator is analogous. If the scalar operator $O$ is coupled to the source $J$ we define
\bea
\langle O(x) \rangle = -\frac{1}{\sqrt{-g_x}} \frac{\delta \mathcal{W}}{\delta J(x)} \bigg |_{g=\delta, J=0}  \,
\eea
and then the three point function is generated as
\bea
\langle T^{\mu\nu}(x_1) O (x_2) O(x_3) \rangle 
&=&  
\bigg\{ \frac{\delta^2}{\delta J(x_2)  \delta J(x_3)} \bigg[  
\frac{2}{\sqrt{-g_{x_1}}} \frac{\delta \mathcal{W}}{\delta g_{\mu\nu}(x_1)} \bigg]_{g=\delta}\bigg\}_{J=0} \nn \\
&=& \langle T^{\mu\nu}[J](x_1) O(x_2) O(x_3) \rangle_{J=0}  + \langle \frac{\delta T^{\mu\nu}[J](x_1)}{\delta J(x_2)}  O(x_3) 
\rangle_{J= 0}  + \langle \frac{\delta T^{\mu\nu}[J](x_1)}{\delta J(x_3)}  O(x_2) \rangle_{J= 0}. \nn \\
\eea

The third correlator that we will analyze will be the $VVV$ vertex, which is defined by the third functional derivative of the 
generating functional with respect to the source gauge field $A^a_{\mu}(x)$
\bea
\langle V^{a\,\mu}(x_1) V^{b\,\nu}(x_2) V^{c\,\rho}(x_3) \rangle = 
- \frac{\delta^3 \mathcal{W}|_{g=\delta}}{\delta A^a_\mu(x_1) 
\delta A^b_\nu(x_2) \delta A^c_\rho(x_3)} \bigg |_{A=0}.
\eea
The $VVV$ is anomaly free, as is the $TVV$ for general ($d\neq 4$) dimensions. To derive the non-anomalous Ward identities for 
general dimensions we assume that the generating functional $W[g,A]$ is invariant under diffeomorphisms
\bea
\mathcal{W}[g, A] = \mathcal{W}[g',A'] \,,
\eea
where $g'$ and $A'$ are transformed metric and gauge field under the general infinitesimal coordinate transformation $x^\mu 
\rightarrow {x'}^\mu = x^\mu + \epsilon^\mu$
\bea
\delta g_{\mu\nu} = \nabla_{\mu} \epsilon_{\nu}  +  \nabla_{\nu} \epsilon_{\mu} \,, \qquad
\delta A^{a}_{\mu} = \epsilon^{\lambda} \nabla_{\lambda} A^a_{\mu} + A^{a \, \lambda} \nabla_{\mu} \epsilon_{\lambda} \,,
\eea
Diffeomerphism invariance and gauge invariance give the relation
\bea \label{BasicWard}
&&\qquad 
\nabla_{\mu} \langle T^{\mu\nu} \rangle + \nabla^{\nu} A^a_\mu \langle V^{a \, \mu} \rangle + \nabla_{\mu} \left( A^{a\,\nu} \langle V^{a \, \mu}\rangle \right) = 0 \,, \label{DiffWI} \\
&&  \qquad \nabla_{\mu} \langle V^{a \, \mu} \rangle + f^{abc} A^b_{\mu} \langle V^{c \mu}\rangle = 0 \,, \label{GaugeWI}
\eea
while naive scale invariance gives the traceless condition
\bea \label{NaiveScaleWI}
&&  \qquad g_{\mu\nu} \langle T^{\mu\nu} \rangle = 0.
\eea
This last Ward identity is naive, due to the appearance of an anomaly at quantum level, 
after renormalization of the correlator for $d=4$.
It is however the correct identity in the $TVV, TOO$ and $TTT$ cases away from $d=4$. In this respect, the functional differentiation 
of (\ref{BasicWard}) and (\ref{NaiveScaleWI}) allows to derive ordinary Ward identities for the various correlators.
In the $TVV$ case we obtain the conservation equation
\bea
\partial_{\mu}^{x_1} \langle T^{\mu\nu}(x_1) V^{a\,\alpha}(x_2) V^{b\,\beta}(x_3) \rangle  
&=&
\partial^{\nu}_{x_1} \delta^d(x_{12}) \langle V^{a\,\alpha}(x_1) V^{b\,\beta}(x_3) \rangle 
+ \partial^{\nu}_{x_1} \delta^d(x_{31}) \langle V^{a\,\alpha}(x_2) V^{b\,\beta}(x_1) \rangle \nn \\
&-& 
\delta^{\nu\alpha} \partial^{x_1\,}_{\mu} \left( \delta^d(x_{12})  \langle V^{a\,\mu}(x_1) V^{b\,\beta}(x_3) \rangle \right) - 
\delta^{\nu\beta} \partial^{x_1\,}_{\mu} \left( \delta^d(x_{31})  \langle V^{a\,\alpha}(x_2) V^{b\,\mu}(x_1) \rangle \right) \nn \\
\eea
and vector current Ward identities
\bea
\partial_{\alpha}^{x_2} \langle T^{\mu\nu}(x_1) V^{a\,\alpha}(x_2) V^{b\,\beta}(x_3) \rangle  &=& 0 \,, \qquad
\partial_{\beta}^{x_3} \langle T^{\mu\nu}(x_1) V^{a\,\alpha}(x_2) V^{b\,\beta}(x_3) \rangle  = 0 \,,
\eea
while the naive identity (\ref{NaiveScaleWI}) gives the non-anomalous condition
\bea
\delta_{\mu\nu} \, \langle T^{\mu\nu}(x_1) V^{a\,\alpha}(x_2) V^{b\,\beta}(x_3) \rangle  &=& 0 \,
\eea
for $d\neq 4$.

\subsection{Definitions for the $TTT$ Amplitude}
\label{DiagTTT}

For the multi-graviton vertices, it is convenient to define the corresponding correlation function as the n-th
functional variation with respect to the metric of the generating functional $\mathcal W$ evaluated
in the flat-space limit
\bea \label{NPF}
\langle T^{\mu_1\nu_1}(x_1)...T^{\mu_n\nu_n}(x_n) \rangle 
&=&
\bigg[\frac{2}{\sqrt{-g_{x_1}}}...\frac{2}{\sqrt{-g_{x_n}}} \,
\frac{\delta^n \mathcal{W}}{\delta g_{\mu_1\nu_1}(x_1)...\delta g_{\mu_n\nu_n}(x_n)}\bigg]
\bigg|_{g_{\mu\nu} = \delta_{\mu\nu}} \nonumber \\ 
&=&  
2^n\, \frac{\delta^n \mathcal{W}}{\delta g_{\mu_1\nu_1}(x_1)...\delta g_{\mu_n\nu_n}(x_n)}\bigg|_{g_{\mu\nu} = 
\delta_{\mu\nu}} \, ,
\eea
so that it is explicitly symmetric with respect to the exchange of the metric tensors.
As we are going to deal with correlation functions evaluated in the flat-space limit all through the paper
we will omit to specify it from now on, so as to keep our notation easy.
The 3-point function we are interested in studying is found by evaluating (\ref{NPF}) for $n=3$,
\bea\label{3PF}
\langle T^{\mu\nu}(x_1)T^{\rho\sigma}(x_2)T^{\alpha\beta}(x_3)\rangle
&=&
8 \, \bigg[ - \, \big\langle \frac{\delta \mathcal S}{\delta g_{\mu\nu}(x_1)}\frac{\delta \mathcal S}{\delta g_{\rho\sigma}(x_2)}
\frac{\delta \mathcal S}{\delta g_{\alpha\beta}(x_3)}\rangle \nonumber \\
&& \hspace{-15mm}
+ \, \langle \frac{\delta^2 \mathcal S }{\delta g_{\alpha\beta}(x_3)\delta g_{\mu\nu}(x_1)}
\frac{\delta \mathcal S }{\delta g_{\rho\sigma}(x_2)} \big\rangle
 + \big\langle \frac{\delta^2 \mathcal S }{\delta g_{\rho\sigma}(x_2)\delta g_{\mu\nu}(x_1)}
\frac{\delta \mathcal S }{\delta g_{\alpha\beta}(x_3)} \rangle
\nonumber \\
&&\hspace{-15mm}
+ \, \langle \frac{\delta^2 \mathcal S}{\delta g_{\rho\sigma}(x_2)\delta g_{\alpha\beta}(x_3)}
\frac{\delta \mathcal S }{\delta g_{\mu\nu}(x_1)}\big\rangle
- \,\big\langle \frac{\delta^3 \mathcal S}{\delta g_{\rho\sigma}(x_2)\delta g_{\alpha\beta}(x_3)
\delta g_{\mu\nu}(x_1)}\big\rangle \bigg] \, ,
\nonumber \\
\eea
where the angle brackets denote the vacuum expectation value.
Notice that the last term is identically zero in dimensional regularization, being proportional to a massless tadpole.
The correlator
\beq
\big\langle \frac{\delta\mathcal S}{\delta g_{\mu\nu}(x_1)}
            \frac{\delta\mathcal S}{\delta g_{\rho\sigma}(x_2)}
            \frac{\delta\mathcal S}{\delta g_{\alpha\beta}(x_3)}
            \big\rangle \, , \label{Triangle}\\
\eeq
has the diagrammatic representation of a triangle topology, while the contributions
\bea \label{bubbles}
\big\langle \frac{\d^2 \mathcal S}{\delta g_{\rho\sigma}(x_2)\delta g_{\alpha\beta}(x_3)}
            \frac{\delta\mathcal S}{\delta g_{\mu\nu}(x_1)} \big\rangle \, ,
\quad
\big\langle \frac{\d^2 \mathcal S}{\delta g_{\alpha\beta}(x_3)\delta g_{\mu\nu}(x_1)}
            \frac{\delta\mathcal S}{\delta g_{\rho\sigma}(x_2)} \big\rangle \, ,
\quad
\big\langle \frac{\d^2 \mathcal S}{\delta g_{\rho\sigma}(x_2)\delta g_{\mu\nu}(x_1)}
            \frac{\delta\mathcal S}{\delta g_{\alpha\beta}(x_3)} \big\rangle
\eea
are interpreted in the perturbative analysis as the "k", "q" and "p" bubble respectively, also termed "T-bubbles" in
\cite{Armillis:2009pq}.\\
In the perturbative realization of these expressions we will also establish a connection between these contributions and the
extra terms generated at the 2-point coincidence limit of the general 3-point vertices discussed in \cite{Osborn:1993cr}. 
For a 3-point vertex the dependence in configuration
space is labelled as $(x_1,x_2,x_3)$ with an incoming momentum $(k)$ at $x_1$ 
and two outgoing momenta $q, p$ at $x_2$ and $x_3$ respectively. 
These conventions are summarized by the transforms
\bea\label{3PFMom}
&&
\int \, d^4x_1\,d^4x_2\,d^4x_3\, 
\langle T^{\mu\nu}(x_1)T^{\rho\sigma}(x_2)T^{\alpha\beta}(x_3)\rangle \,
e^{-i(k\cdot x_1 - q\cdot x_2 - p\cdot x_3)} =
(2\pi)^4\,\delta^{(4)}(k-p-q)\, \langle T^{\mu\nu}T^{\rho\sigma}T^{\alpha\beta}\rangle(p,q)\, , \nn \\
\eea
and
\beq\label{2PFMom}
\int \, d^4x_2 \, d^4x_3 \, \langle T^{\rho\sigma}(x_2)T^{\alpha\beta}(x_3) \rangle\, e^{-i(q\cdot x_2 - p\cdot x_3)} =
(2\pi)^4\,\delta^{(4)}(p-q)\, \langle T^{\rho\sigma}T^{\alpha\beta} \rangle(p) \, ,
\eeq
for 3- and 2-point functions respectively.\\

\subsection{General covariance Ward identities for the $TTT$}
\label{DiagTTTWardCov}

The requirement of general covariance for the generating functional $\mathcal W$
immediately leads to the master Ward identity for the conservation of the energy momentum tensor
given in (\ref{BasicWard}) (of course we disregard background gauge fields here),
\beq\label{masterWI0}
\nabla_\nu \langle T^{\mu\nu}(x_1) \rangle
= \nabla_\nu \bigg(\frac{2}{\sqrt{-g_{x_1}}}\frac{\delta\mathcal W}{\delta g_{\mu\nu}(x_1)}\bigg)
= 0\, ,
\eeq
and expanding the covariant derivative we can write it as
\beqa
&&
\frac{2}{\sqrt{-g_{x_1}}}\bigg(\pd_{\nu}\frac{\delta\mathcal W}{\delta g_{\mu\nu}(x_1)}
- \Gamma^\lambda_{\lambda\mu}(x_1)\frac{\delta\mathcal W}{\delta g_{\mu\nu}(x_1)}
+ \Gamma^\mu_{\kappa\nu}(x_1)\frac{\delta\mathcal W}{\delta g_{\kappa\nu}(x_1)}
+ \Gamma^\nu_{\kappa\nu}(x_1)\frac{\delta\mathcal W}{\delta g_{\mu\kappa}(x_1)}\bigg) = 0, \nn\\
\eeqa
where the first of the three Christoffel symbols is generated by differentiation of ${1}/{\sqrt{-g_{x_1}}}$ in the definition of
$T_{\mu\nu}$ together with
\beq
\Gamma^\alpha_{\alpha\beta}(x_1)= \frac{1}{2} \, g^{\alpha\gamma}(x_1) \, \pd_\beta \, g_{\alpha\gamma}(x_1)\,
\eeq
or, equivalently, as
\beqa \label{masterWI}
&& 
2 \, \bigg(\pd_{\nu}\frac{\delta\mathcal W}{\delta g_{\mu\nu}(x_1)}
+ \Gamma^\mu_{\kappa\nu}(x_1)\frac{\delta\mathcal W}{\delta g_{\kappa\nu}(x_1)}\bigg)
= 0\, .
\eeqa
By taking one and two functional derivatives of (\ref{masterWI}) with respect to $g_{\rho\sigma}(x_2)$ and 
$g_{\rho\sigma}(x_2)$ and $g_{\alpha\beta}(x_3)$ respectively, one gets, in curved space-time,
\beqa
&&
4 \, \bigg[ \pd_\nu \frac{\delta^2\mathcal W}{\delta g_{\rho\sigma}(x_2)\delta g_{\mu\nu}(x_1)}
+ \frac{\delta \Gamma^\mu_{\kappa\nu}(x_1)}{\delta g_{\rho\sigma}(x_2)}\frac{\delta\mathcal W}{\delta g_{\kappa\nu}(x_1)}
+ \Gamma^\mu_{\kappa\nu}(x_1) \frac{\delta^2 \mathcal W}{\delta g_{\mu\nu}(x_1)\delta g_{\rho\sigma}(x_2)}\bigg]
= 0 \, \label{WI2PFCoordinate} \\
&&
8 \bigg[\pd_\nu\frac{\delta^3\mathcal W}{\delta g_{\alpha\beta}(x_3)\delta g_{\rho\sigma}(x_2)\delta g_{\mu\nu}(x_1)}
+ \frac{\delta \Gamma^\mu_{\kappa\nu}(x_1)}{\delta g_{\rho\sigma}(x_2)}
  \frac{\delta^2 \mathcal W}{\delta g_{\alpha\beta}(x_3)\delta g_{\kappa\nu}(x_1)}
+ \frac{\delta \Gamma^\mu_{\kappa\nu}(x_1)}{\delta g_{\alpha\beta}(x_3)}
  \frac{\delta^2 \mathcal W}{\delta g_{\rho\sigma}(x_2)\delta g_{\kappa\nu}(x_3)} 
\nonumber \\
&&
+ \frac{\delta^2\Gamma^\mu_{\kappa\nu}(x_1)}{\delta g_{\rho\sigma}(x_2)\delta g_{\alpha\beta}(x_3)}
  \frac{\delta\mathcal W}{\delta g_{\mu\nu}(x_1)}
+ \Gamma^\mu_{\kappa\nu}(x_1)
  \frac{\delta^3\mathcal W}{\delta g_{\rho\sigma}(x_2)\delta g_{\alpha\beta}(x_2)\delta g_{\kappa\nu}(x_1)}\bigg] = 0 \, ,
\label{WI3PF} 
\eeqa
where $\delta(x_1,x_2)\equiv \delta(x_1-x_2)$ and so on.

As we are interested in the flat space-time limit, we must evaluate \ref{WI2PFCoordinate} and
(\ref{WI3PF}) by letting the Christoffel symbols go to zero.
Another simplification is obtained by noticing that the Green's functions
\beq
\langle \frac{\delta \mathcal S}{\delta g_{\mu\nu}(x_1)}\rangle = - \frac{\delta \mathcal W}{\delta g_{\mu\nu}(x_1)}
\eeq
and
\beq\label{Tadpole2PF}
\langle \frac{\delta^2 \mathcal S}{\delta g_{\mu\nu}(x_1)\delta g_{\alpha\beta}(x_3)}\rangle
\eeq
are proportional to massless tadpoles, so that we can ignore them in the following expression
\beq
\frac{\d^2 \mathcal W}{\delta g_{\alpha\beta}(x_3) \delta g_{\mu\nu}(z)}
= \big\langle \frac{\delta \mathcal S}{\delta g_{\mu\nu}(x_1)}\frac{\delta \mathcal S}{\delta g_{\alpha\beta}(x_3)}\big\rangle
- \big\langle \frac{\d^2 \mathcal S}{\d g_{\alpha\beta}(x_3)\delta g_{\mu\nu}(x_1)} \big\rangle
= \big\langle \frac{\delta \mathcal S}{\delta g_{\mu\nu}(x_1)}\frac{\delta \mathcal S}{\delta g_{\alpha\beta}(x_3)}\big\rangle
\,.
\eeq
So the Ward identity for the 2-point function in flat coordinate space-time is immediately seen to be
\beq
\pd_{\nu} \langle T^{\mu\nu}(x_1) T^{\rho\sigma}(x_2) \rangle = 0 \, ,
\eeq
where, due to the vanishing of (\ref{Tadpole2PF}), we have set
\beq \label{2PF}
\langle T^{\mu\nu}(x_1) T^{\rho\sigma}(x_2) \rangle
\equiv
4\, \big\langle \frac{\delta\mathcal S}{\delta g_{\mu\nu}(x_1)}\frac{\delta\mathcal S}{\delta g_{\rho\sigma}(x_2)} \big\rangle \, .
\eeq
Obviously, its form in momentum space, exploiting (\ref{2PFMom}), is
\beq \label{WI2PFMom}
p_{\mu}\langle T^{\mu\nu}T^{\rho\sigma} \rangle(p) = 0 \, .
\eeq
The terms surviving in (\ref{WI3PF}) are those in the first line.
In order to make them explicit, we evaluate the functional derivative of the Christoffel symbols using
(\ref{Christoffel}), (\ref{Tricks}) and (\ref{Tricks2}), finding
\beqa\label{GammaDerivatives1}
\frac{\delta\Gamma^\mu_{\kappa\nu}(x_1)}{\delta g_{\rho\sigma}(x_2)}
&=&
\frac{1}{2}\delta^{\mu\alpha}\bigg[-s^{\rho\sigma}_{\,\,\,\,\,\,\kappa\nu}\pd_\alpha
+ s^{\rho\sigma}_{\,\,\,\,\,\,\alpha\nu}\pd_\k + s^{\rho\sigma}_{\,\,\,\,\,\,\alpha\kappa} \pd_\nu\bigg]\delta(x_1,x_2)\, ,
\eeqa
where the $s$ tensor is defined by eq. (\ref{Tricks2}) in the Appendix.
Plugging this into (\ref{WI3PF}) and using (\ref{2PF}), the second term becomes
\bea
8 \, \frac{\delta\Gamma^\mu_{\k\nu}(x_1)}{\delta g_{\r\s}(x_2)}
\frac{\delta^2 \mathcal W}{\delta g_{\alpha\beta}(x_3)\delta g_{\kappa\nu}(x_1)} 
&=&
\bigg[\delta^{\mu\r}\langle T^{\nu\sigma}(x_1)T^{\alpha\beta}(x_3)\rangle \, \pd_\nu +
\delta^{\mu\s}\langle T^{\nu\rho}(x_1)T^{\alpha\beta}(x_3)\rangle \, \pd_\nu 
\nonumber\\
&& \hspace{5mm}
- \, \langle T^{\rho\sigma}(x_1)T^{\alpha\beta}(x_3)\rangle \, \pd^\mu\bigg]\delta(x_1,x_2) \, .
\eea
A completely analogous relation holds for the exchanged term
$\big(g_{\alpha\beta}(x_3)\leftrightarrow g_{\rho\sigma}(x_2)\big)$. \\
Finally, we can recast the Ward identity (\ref{WI3PF}) in the form
\bea \label{WI3PFcoordinate}
\pd_\nu\langle T^{\mu\nu}(x_1)T^{\rho\sigma}(x_2)T^{\alpha\beta}(x_3) \rangle
&=&
\bigg[\langle T^{\rho\sigma}(x_1)T^{\alpha\beta}(x_3)\rangle\pd^\mu\d(x_1,x_2) + 
\langle T^{\alpha\beta}(x_1)T^{\rho\sigma}(x_2)\rangle\pd^\mu\d(x_1,x_3) \bigg]\nn \\
&-&
\bigg[\delta^{\mu\rho}\langle T^{\nu\sigma}(x_1)T^{\alpha\beta}(x_3)\rangle
+     \delta^{\mu\sigma}\langle T^{\nu\rho}(x_1)T^{\alpha\beta}(x_3)\rangle\bigg]\pd_\nu\d(x_1,x_2)\nn\\
&-&
\bigg[\delta^{\mu\alpha}\langle T^{\nu\beta}(x_1)T^{\rho\sigma}(x_2)\rangle
+ \delta^{\mu\beta}\langle T^{\nu\alpha}(x_1)T^{\rho\sigma}(x_2)\rangle\bigg]\pd_\nu\d(x_1,x_3)\, ,\nn\\
\eea
having used the definitions (\ref{NPF}) and (\ref{3PF}).\\
Fourier-transforming according to (\ref{3PFMom}) and (\ref{2PFMom}), we get the Ward identity in momentum space that we need, i.e.
\beqa\label{WI3PFmomenta2a}
&& k_\nu \langle T^{\mu\nu}T^{\alpha\beta}T^{\rho\sigma} \rangle(p,q) =
   p^\nu \langle T^{\alpha\beta}T^{\rho\sigma}\rangle(p) + q^\mu \langle T^{\rho\sigma}T^{\alpha\beta}\rangle(q) \nn \\
&&
-  p_\nu \bigg[\delta^{\mu\beta} \langle T^{\nu\alpha}T^{\rho\sigma} \rangle(q)
             + \delta^{\mu\alpha}\langle T^{\nu\beta}T^{\rho\sigma}  \rangle(q)\bigg]
-  q_\nu \bigg[\delta^{\mu\sigma}\langle T^{\nu\rho}T^{\alpha\beta}  \rangle(p)
+              \delta^{\mu\rho}  \langle T^{\nu\sigma}T^{\alpha\beta}\rangle(p)\bigg].
\eeqa
Similar Ward identities can be obtained when we contract with the momenta of the other lines. These are going to be essential in 
order to test the correctness of the computation once we turn to perturbation theory.

\subsection{The anomalous Ward identities for the TTT}
\label{DiagTTTWardAnom}

The anomalous Ward identities for the 3-graviton vertex is obtained after a lengthy computation, 
performing two functional variations of (\ref{TraceAnomalySymm}) and taking the flat-space limit, thereby obtaining
\beqa
\delta_{\mu\nu}\langle T^{\mu\nu}T^{\rho\sigma}T^{\alpha\beta} \rangle(p,q)
&=&
4 \, \mathcal A^{\alpha\beta\rho\sigma}(p,q)
- 2 \, \langle T^{\alpha\beta}T^{\rho\sigma} \rangle(p) - 2 \, \langle T^{\rho\sigma}T^{\alpha\beta} \rangle(q)\nn\\
&=&
4 \, \bigg[ \beta_a\,\big(\big[F\big]^{\alpha\beta\rho\sigma}(p,q)
- \frac{2}{3} \big[\sqrt{-g}\Box\,R\big]^{\alpha\beta\rho\sigma}(p,q)\big)
+ \beta_b\, \big[G\big]^{\alpha\beta\rho\sigma}(p,q) \bigg]\nn\\
&-&
2 \, \langle T^{\alpha\beta}T^{\rho\sigma} \rangle(p) 
- 2 \, \langle T^{\rho\sigma}T^{\alpha\beta} \rangle(q) \, ,\label{munu3PFanomaly}
\eeqa
where $\mathcal A^{\alpha\beta\rho\sigma}(p,q)$, $\mathcal A^{\mu\nu\rho\sigma}(-k,q)$ and
$\mathcal A^{\alpha\beta\mu\nu}(-k,p)$ are generated by the anomaly. We remark, if not obvious, that all the contractions with the 
metric tensor in the flat spacetime limit ($\delta_{\mu\nu}$) should be understood as being 4-dimensional. This is the case for all 
the anomaly equations.
The various contributions to the trace anomaly are given in terms of
the functional derivatives of quadratic invariants in appendix \ref{Functionals}. Analogous anomalous 
Ward identities can be obtained by tracing the other two pairs of indices. 

\section{Inverse mappings: the correlators $VVV$, $TOO$ and $TVV$ in position space using the Feynman expansion}

Having by now defined all the fundamental (anomalous and regular) Ward identities which allow to test the consistency 
of all the correlator which we are interested in, we now turn to provide the expression of these correlators in position space 
using their realization in free field theory. 

We remind that an important result of \cite{Osborn:1993cr} is the identification of the solution of the Ward identities in terms of a set of
constants and of certain linearly independent tensor structures in (Euclidean) position space. Since these same tensor structures must occur in direct computations of the same vertex functions in free field theories in momentum space, we can use the one-loop
computations of the vertex functions in momentum space to infer what those tensor structures must be, and find the exact correspondence between CFT amplitudes in position space and momentum space {\it a posteriori}, provided that we have enough linearly independent vertex functions for different free theories to determine the linear combinations uniquely. We call this procedure an "inverse mapping", as it allows to re-express the correlators of  \cite{Osborn:1993cr} in such a form that their Fourier integrability is explicit. This result is obtained by pulling out derivatives of the corresponding diagrams in such a way that integrability becomes trivial. More details on this procedure is contained in section 8. 

We start with the $VVV$ vertex function.
\begin{figure}[t]
\begin{center}
\includegraphics[scale=0.7]{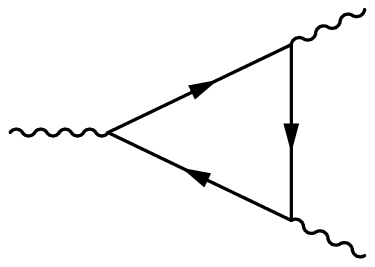}\qquad\qquad
\includegraphics[scale=0.7]{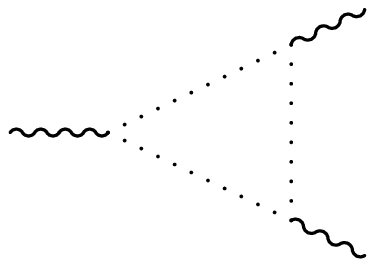}
\caption{The fermion and the scalar sectors contributing to the conformal VVV vertex in any dimension.}
\end{center}
\label{VVVV}
\end{figure}
The two types of diagrams contributing to the general conformal expression of the $VVV$ in any dimensions are shown in Fig. (\ref{VVVV}).
In  \cite{Osborn:1993cr} the $VVV$, as all the other correlators, are fixed by general CFT requirements. It takes the form  \cite{Osborn:1993cr}

\bea \label{VVVcoord}
\langle V_{\mu}^a(x_1) V_{\nu}^b (x_2) V_{\rho}^c(x_3) \rangle
&=&
\frac{f^{abc}}{(x^2_{12})^{d/2-1}\,(x^2_{23})^{d/2-1}\,(x^2_{31})^{d/2-1}}  \, \bigg\{ (a - 2b) \, X_{23\,\mu}\,X_{31\,\nu}\,X_{12\,\rho} \nonumber \\
&-&
b \left[\frac{1}{x^2_{23}} \, X_{23\,\mu} \, I_{\nu\rho}(x_{23})  + \frac{1}{x^2_{31}} \, X_{31\,\nu} \, I_{\mu\rho}(x_{31})
+ \frac{1}{x^2_{12}} \, X_{12\,\rho}\,I_{\mu\nu}(x_{12} ) \right]  \bigg\} \, ,
\eea
where $f^{abc}$ are the structure constants of the gauge group, $I_{\mu\nu}(x)$ is the inversion operator defined as
\beq
I^{\mu\nu}(x)=
\delta^{\mu\nu} - 2 \frac{x^\mu x^\nu}{x^2}
\eeq
and
\bea
x_{ij}
\equiv
x_i - x_j \, , \qquad
X_{ij} = - X_{ji}
\equiv
\frac{x_{ik}}{x^2_{ik}} - \frac{x_{jk}}{x^2_{jk}} \, ,  \quad  i,j,k = 1,2,3 \, .
\eea

The correlator is Fourier integrable, although this is not immediately evident from (\ref{VVVcoord}). The simplest way to prove this point consists in showing that (\ref{VVVcoord}) can be reproduced in $d$-dimensions by the combination of the scalar and the fermion sectors of a free field theory.
For this purpose we use two realizations of the vector current $V_{\mu}^a$, using scalar and fermion fields
\bea
V_{\mu}^a = \phi^{*} t^a \partial_{\mu} \phi - \partial_{\mu} \phi^{*} t^a \phi \,,  \qquad  V_{\mu}^a = \bar \psi t^a \gamma_{\mu} \psi \,.
\eea

The diagrammatic expansion of this correlator consists of two triangle diagrams, the direct and the exchanged, both in the scalar and fermion sectors. Using the Feynman rules in coordinate space we obtain, after some manipulations
\bea
\langle V_{\mu}^a(x_1) V_{\nu}^b (x_2) V_{\rho}^c(x_3) \rangle_{fermion} &=&
- \frac{ c_f \, f^{a b c} }{(d-2)^3} \Delta_{\mu \alpha \nu \beta \rho \gamma} \partial^{\alpha}_{12}\partial^{\beta}_{23}\partial^{\gamma}_{31}  \frac{1}{(x_{12}^2)^{d/2-1} (x_{23}^2)^{d/2-1} (x_{31}^2)^{d/2-1}}   \,,  \\
\langle V_{\mu}^a(x_1) V_{\nu}^b (x_2) V_{\rho}^c(x_3) \rangle_{scalar} &=&
\frac{c_s \, f^{a b c}}{(d-2)^2}   \left( \partial_{\mu}^{12} + \partial_{\mu}^{31} \right) \left( \partial_{\nu}^{23} + \partial_{\nu}^{12} \right) \left( \partial_{\rho}^{31} + \partial_{\rho}^{23} \right) \frac{1}{(x_{12}^2)^{d/2-1} (x_{23}^2)^{d/2-1} (x_{31}^2)^{d/2-1}}  \nn \\
\eea
where
\bea
\label{Delta6}
\Delta_{\mu \alpha \nu \beta \rho \gamma} = \frac{1}{4} Tr \left[ \gamma_{\mu} \gamma_{\alpha} \gamma_{\nu} \gamma_{\beta} \gamma_{\rho} \gamma_{\gamma} \right] \,,
\eea
and $c_f, c_s$ are normalization constants whose numerical values are irrelevant here.
 Written in these forms, the two expressions are manifestly integrable.
Tracing over the $\gamma$ matrices and applying the derivatives over all the denominators, we generate the result of  \cite{Osborn:1993cr} by taking a linear combination of these two sectors
\bea
\langle V_{\mu}^a(x_1) V_{\nu}^b (x_2) V_{\rho}^c(x_3) \rangle = \bigg( a \, t^a_{\mu\nu\rho} + b \, t^b_{\mu\nu\rho}  \bigg) \frac{ f^{abc}}{(x_{12}^2)^{d/2-1} (x_{23}^2)^{d/2-1} (x_{31}^2)^{d/2-1}}
\eea
where
\bea
t^a_{\mu\nu\rho} &=& \frac{1}{d(d-2)^2}   \left( \partial_{\mu}^{12} + \partial_{\mu}^{31} \right) \left( \partial_{\nu}^{23} + \partial_{\nu}^{12} \right) \left( \partial_{\rho}^{31} + \partial_{\rho}^{23} \right) - \frac{1}{d} t^b_{\mu\nu\rho} \,, \\
t^b_{\mu\nu\rho} &=&   - \frac{1}{(d-2)^3}  \Delta_{\mu \alpha \nu \beta \rho \gamma}\partial^{\alpha}_{12}\partial^{\beta}_{23}\partial^{\gamma}_{31} \,.
\eea
The equivalence between this expression and Eq. (\ref{VVVcoord}) can be verified explicitly.

\subsection{The TOO case}
The next correlator that we are going to investigate extensively is the $TOO$.
The structure of this function in coordinate space - for non coincident points - is given by \cite{Osborn:1993cr}
\beq
\label{TOOPO}
\langle T_{\mu\nu}(x_1) \, O(x_2) \, O(x_3) \rangle =
\frac{a}{(x^2_{12})^{d/2} \, (x^2_{23})^{\eta-d/2} \, (x^2_{31})^{d/2}} \, h^1_{\mu\nu}(\hat{X}_{23}) \, ,
\eeq
where $a$ is a constant, $\eta$ the dimension of the scalar field $O $  and where
\bea   \label{TOOstructures}
\hat{X}_\mu = \frac{X_\mu}{\sqrt{X^2}} \, , \nn \qquad
h^1_{\mu\nu}(\hat X) = \hat{X}_\mu \, \hat{X}_\nu - \frac{1}{d} \, \delta_{\mu\nu} \, .
\eea
In the short-distance limits of its external points this vertex is singular and needs regularization. In \cite{Osborn:1993cr} the authors, in their direct solutions of the Ward identites, introduce some extra terms which are given by 

\bea
\left[ \hat A_{\mu\nu}(x_{12}) - A_{\mu\nu}(x_{12}) + \hat A_{\mu\nu}(x_{31}) - A_{\mu\nu}(x_{31})\right] \frac{1}{(x_{23}^2)^{\eta}} \,,
\eea
where
\bea
A_{\mu\nu}(s) = \frac{a}{s^d} \left( \frac{s_\mu s_\nu}{s^2} - \frac{1}{d} \delta_{\mu\nu} \right) \,, \qquad
\hat A_{\mu\nu}(s) = \frac{a}{d} \left( \frac{\partial_{\mu\nu}}{d-2}\frac{1}{s^{d-2}} + \frac{\eta -d +1}{\eta} S_d \delta_{\mu\nu} \delta^d(s) \right) \,.
\eea
These are contact terms.
In the expression above $S_d$ denotes the volume of the $d$-dimensional sphere, $S_d = 2\,\pi^{\frac{1}{2}}/\Gamma(d/2)$.
The delta function term in $\hat A$ reflects the arbitrariness typical of any regularization scheme, and its coefficient is chosen to satisfy the Ward identities.\\
\subsubsection{Manifest integrability of the CFT result and comparisons with free field theory } 
Expanding the previous expression and bringing it in the derivative form we obtain
\bea
\langle T_{\mu\nu}(x_1) \, O(x_2) \, O(x_3) \rangle
&=&
\frac{a}{(d-2)^2}\bigg\{
 ( \partial_{\mu}^{12} \partial_{\nu}^{31} + \partial_{\nu}^{12} \partial_{\mu}^{31}  ) +
\frac{d-2}{d} ( \partial_{\mu \nu}^{12}  + \partial_{\mu\nu}^{31} )
\bigg\} \frac{1}{(x^2_{12})^{d/2-1}  (x^2_{23})^{\eta-d/2+1}  (x^2_{31})^{d/2-1}} \nn \\
&+&
a \frac{ x^2_{12}  x^2_{23} + x^2_{31} x^2_{23} - (x^2_{23})^2}{(x^2_{12})^{d/2}  (x^2_{23})^{\eta-d/2+1}  (x^2_{31})^{d/2}} \frac{\delta_{\mu\nu}}{d}
+  a \frac{\eta -d+1}{d \eta} S_d \delta_{\mu\nu} \frac{\delta^d(x_{12}) +  \delta^d(x_{31}) }{(x_{23}^2)^\eta}       \, .
\eea

Notice that the first term of the second line proportional to $\delta_{\mu\nu}$ is not manifestly integrable. As we have already mentioned,
one can use identities such as $x_{12}^2+ x_{13}^2 - x_{23}^2=2 x_{12}\cdot x_{13}$ in order to rewrite it in the form
\beq
\frac{ x^2_{12} \, x^2_{23} + x^2_{31} \, x^2_{23} - (x_{23}^2)^2}{(x^2_{12})^{d/2} \, (x^2_{23})^{d/2} \, (x^2_{31})^{d/2}}=
\frac{2 }{(d-2)^2} \partial^{12}_\mu \partial^{31 \,\mu} \frac{1}{ (x_{12}^2)^{d/2-1} (x_{31}^2)^{d/2-1} (x_{23}^2)^{\eta-d/2+1}}
\eeq
which shows its integrability when $\eta < d-1$.

In order to test the consistency of the result (\ref{TOOPO}) obtained from the application of the conformal Ward identities for the $TOO$,
we can consider a particular scalar free field theory. We suppose for instance that the scalar operator $O$ is given by $O = \phi^2$ with dimensions $\eta = d-2$, whose energy-momentum tensor $T$ is given by
\bea
T_{\mu\nu}
&=&
\pd_\mu \phi \, \pd_\nu\phi - \frac{1}{2} \, \delta_{\mu\nu}\,\pd_\alpha \phi \, \pd^\alpha \phi
+ \frac{1}{4}\,\frac{d-2}{d-1}\, \bigg[\delta_{\mu\nu} \pd^2 - \pd_\mu\,\pd_\nu\bigg]\, \phi^2 
\eea
which is conserved and traceless in $d$ dimensions. \\
Using the Feynman rules in coordinate space together with the expression of a scalar propagator we obtain the $T\phi^2\phi^2$ 
correlation function
\bea
\label{TphiphiDer}
\langle T_{\mu\nu}(x_1) \phi^2 (x_2) \phi^2 (x_3) \rangle 
&=&  
\frac{2 a (d-1)}{d (d-2)^2} \bigg[ \partial_{\mu}^{12} 
\partial_{\nu}^{31} + \partial_{\nu}^{12} \partial_{\mu}^{31}  - \delta_{\mu\nu} \partial^{12} \cdot \partial^{31}  - 
\frac{d-2}{2(d-1)} \bigg( - \partial_{\mu\nu}^{12} - \partial_{\mu\nu}^{31}   + \partial_{\mu}^{12} \partial_{\nu}^{31}  \nn \\
&+& 
\partial_{\nu}^{12} \partial_{\mu}^{31} + \delta_{\mu\nu} \left(\partial^2_{12} + \partial^2_{31} - 2 \partial^{12} \cdot 
\partial^{31} \right)  \bigg) \bigg]    \frac{1}{(x_{12}^2)^{d/2-1} (x_{23}^2)^{d/2-1} (x_{31}^2)^{d/2-1}} \nn \\
&-& 
a \frac{d-1}{d(d-2)} S_d \delta_{\mu\nu} \frac{\delta^d(x_{12}) +  \delta^d(x_{31}) }{(x_{23}^2)^{d-2}} \,.
\eea
The equivalence of this expression with the solution given in (\ref{TOOPO})  can be explicitly checked by performing the derivative of
(\ref{TphiphiDer}) and expanding the result. We remark that (\ref{TphiphiDer}) is clearly integrable and does not require any 
intermediate regularization. The first term in the previous expression comes from the triangle topology diagram while the last two, 
proportional to the delta functions, are contact terms with two-point topology.

\subsection{The TVV case: integrability and free field theory realization}

To identify the diagrammatic structure of the $TVV$ correlator we can proceed with an inverse mapping.
In fact, we know from \cite{Osborn:1993cr} that such solution is characterized by 2 constants when the 3 external
coordinates $(x_1,x_2, x_3)$ are separated. This homogeneous solution has to be modified by the additions of extra contact terms $(A 
- \hat{A})$ terms which have the topology of 2-point functions.

The homogeneous solution is then modified further by the addition of a 1/$\epsilon$ counterterm - in dimensional regularization -
to regulate its ultraviolet behaviour. This regularization procedure is crucial to obtaining the anomalous contribution. We will come to a discussion of this point once we move completely to momentum space. Before that let us provide a diagrammatic interpretation of the various contributions to this correlators, except for the contribution coming from the counterterm, using the information that in any dimension this can be constructed as a linear combination of two independent sectors, the fermion and the scalar.
Therefore we get
\bea
\langle T_{\mu\nu}(x_1) V^a_{\alpha} (x_2) V^b_{\beta} (x_3) \rangle
&=&
\sum_{I=f, s}\left( \langle T_{\mu\nu}[A](x_1) V^a_\alpha(x_2) V^b_\beta(x_3) \rangle^I_{A=0}  + \langle \frac{\delta 
T_{\mu\nu}[A](x_1)}{\delta A^{a\,\alpha}(x_2)}  V^b_\beta(x_3) \rangle^I_{A= 0} \right. \nonumber \\
&&
\left. \qquad \qquad + \langle \frac{\delta T_{\mu\nu}[A](x_1)}{\delta A^{b \,\beta}(x_3)}  V^a_\alpha(x_2) \rangle^I_{A= 0}
\right)
\label{contactTVV}
\eea
where the sum is over the fermion (f) and scalar (s) sectors.  
In a diagrammatic expansion, all the terms above have a diagrammatic interpretation, which will turn useful in order to derive an 
integrable expression of this vertex.

Using the Feynman rules in configuration space one can obtain the following parameterization 
of the $TVV$ vertex for fermions within the loop,
\bea
\langle T_{\mu\nu}[A](x_1) V^a_{\alpha} (x_2) V^b_{\beta} (x_3) \rangle_{A=0}^{f} =  
\frac{c \, \delta^{ab}}{d(d-2)^3} \, A_{\mu\nu\xi\eta}\, \Delta_{\xi\rho\alpha\si\beta\lambda}\,
(\partial_{\eta}^{12} + \partial_{\eta}^{31} ) \, \partial^{\rho}_{12} \partial^{\si}_{23} \partial^{\lambda}_{31} \nn\\
\times \frac{1}{(x_{12}^2)^{d/2-1} (x_{23}^2)^{d/2-1} (x_{31}^2)^{d/2-1}} \, ,
\label{fermionsector}
\eea
where $ \Delta_{\mu\rho\alpha\si\beta\lambda}$ is defined in eq. (\ref{Delta6}) and $A_{\mu\nu\rho\sigma}$ in Appendix \ref{VERTICES}.
This contribution alone is not sufficient to satisfy all the inhomogeneous Ward identities and we must consider also the 
contributions coming from the contact terms. In the framework of the analysis of \cite{Osborn:1993cr}, in which the correlation 
functions are obtained exploiting the symmetries without any reference to their perturbative structure, this is less evident. In fact 
in \cite{Osborn:1993cr}
the arbitrariness in the regularization procedure is exploited in order to impose the Ward identities by hand. This is achieved by 
introducing the differentially regulated expressions proportional to $A - \hat A$, which will be given below. These terms exactly 
correspond to the contributions proportional to 2-point functions discussed above, as we are going to show in a moment.
The two contact terms identified by the diagrammatic expansion are given by
\bea
\langle \frac{\delta T_{\mu\nu}[A](x_1)}{\delta A^{a\,\alpha}(x_2)}  V^b_\beta(x_3) \rangle_{A= 0}^{f}
&= & \frac{c \, \delta^{ab}}{d(d-2)^2} S_d \delta^d(x_{12}) \Delta^{(2)}_{\mu \nu \alpha \beta \rho \sigma}  \partial^{\rho}_{31} 
\frac{1}{(x_{31}^2)^{d/2-1}} \partial^{\si}_{31} \frac{1}{(x_{31}^2)^{d/2-1}} \\
%
\langle \frac{\delta T_{\mu\nu}[A](x_1)}{\delta A^{b \,\beta}(x_3)}  V^a_\alpha(x_2) \rangle_{A= 0}^{f}
&=& \frac{c \, \delta^{ab}}{d(d-2)^2} S_d \delta^d(x_{31}) \Delta^{(2)}_{\mu \nu \beta \alpha \rho \sigma}  \partial^{\rho}_{12} 
\frac{1}{(x_{12}^2)^{d/2-1}} \partial^{\si}_{31} \frac{1}{(x_{12}^2)^{d/2-1}}
\eea
with
\bea
\Delta^{(2)}_{\mu \nu \alpha \beta \rho \sigma} 
&=& 
  \delta_{\alpha\nu } \delta_{\beta\sigma}\delta_{\mu\rho }
+ \delta_{\alpha\mu } \delta_{\beta\sigma}\delta_{\nu\rho }
+ \delta_{\alpha\nu } \delta_{\beta\rho}  \delta_{\mu\sigma }
+ \delta_{\alpha\mu } \delta_{\beta\rho}  \delta_{\nu\sigma } 
- \delta_{\alpha\nu } \delta_{\beta\mu}   \delta_{\rho\sigma }
- \delta_{\alpha\mu } \delta_{\beta\nu}   \delta_{\rho\sigma } \nn \\
&-& 
2 \, \delta_{\mu\nu} \, \left(\delta_{\alpha\rho}\delta_{\beta\sigma} + \delta_{\alpha\sigma}\delta_{\beta\rho}
- \delta_{\alpha\beta}\delta_{\rho\sigma} \right)\, .
\eea
In the scalar sector the TVV correlation function can be recast in the manifestly integrable form as
\beqa
\langle T_{\mu\nu}[A](x_1) V^a_{\alpha} (x_2) V^b_{\beta} (x_3) \rangle_{A=0}^{s} = \nn
c  \, \delta^{ab} \frac{2(d-1)}{d (d-2)^3} \bigg[ \partial_{\mu}^{12} \partial_{\nu}^{31} + \partial_{\nu}^{12} \partial_{\mu}^{31}  
- \delta_{\mu\nu} \partial^{12} \cdot \partial^{31}  \nn \\
 - \frac{d-2}{2(d-1)} \bigg( - \partial_{\mu\nu}^{12} - \partial_{\mu\nu}^{31}
  + \partial_{\mu}^{12} \partial_{\nu}^{31} + \partial_{\nu}^{12} \partial_{\mu}^{31} + \delta_{\mu\nu} \left(\partial^2_{12} + 
  \partial^2_{31} - 2 \partial^{12} \cdot \partial^{31} \right)  \bigg) \bigg]\times \nn\\
 \times  \left( \partial_{\alpha}^{12} + \partial_{\alpha}^{23}\right)
\left( \partial_{\beta}^{31} + \partial_{\beta}^{23}\right)   \frac{1}{(x_{12}^2)^{d/2-1} (x_{23}^2)^{d/2-1} (x_{31}^2)^{d/2-1}}\, .
\label{scalarsector}
\eeqa
This contribution originates only from the triangle diagram. This term corresponds to the expression given in \cite{Osborn:1993cr} 
(for non coincident points) for the same correlator. The only differences are in the $\partial^2_{12}$ and $\partial^2_{31}$ terms 
which are proportional to $\delta_{\mu\nu}$, which vanish in the non-coincident point limit and are given by
\bea
&& 
- \frac{c \, \delta^{ab} }{d (d-2)^2} \delta_{\mu\nu} \left(\partial^2_{12} + \partial^2_{31} \right) \left( 
\partial_{\alpha}^{12} + \partial_{\alpha}^{23}\right)  \left( \partial_{\beta}^{31} + \partial_{\beta}^{23}\right)  
\frac{1}{(x_{12}^2)^{d/2-1} (x_{23}^2)^{d/2-1} (x_{31}^2)^{d/2-1}} \nn \\
&& 
= 
\frac{2 c \, \delta^{ab} }{d (d-2)} S_d \delta_{\mu\nu}   \bigg[\partial_{\alpha}^{23}  \left( \partial_{\beta}^{31} + 
\partial_{\beta}^{23}\right) \frac{\delta^d(x_{12}) }{ (x_{23}^2)^{d/2-1} (x_{31}^2)^{d/2-1}} + \partial_{\beta}^{23} \left( 
\partial_{\alpha}^{12} + \partial_{\alpha}^{23}\right)  \frac{ \delta^d(x_{31})}{(x_{12}^2)^{d/2-1} (x_{23}^2)^{d/2-1} } \bigg]\,. 
\label{top2a}
\eea
They have the topology of 2-point functions. These terms, together with those arising from the triangle diagrams, correspond exactly 
to those identified as $A-\hat A$ \cite{Osborn:1993cr}, which have been introduced  in order to satisfy the Ward identities (contact 
terms)
\bea
\langle \frac{\delta T_{\mu\nu}[A](x_1)}{\delta A^{a\,\alpha}(x_2)}  V^b_\beta(x_3) \rangle_{A= 0}^{s}
&=& \frac{c \, \delta^{ab} (d-1)}{d (d-2)^2} S_d \delta^d(x_{12}) \left((\partial^{23}_{\mu} + \partial^{31}_{\mu} ) \delta_{\nu \alpha} + (\partial^{23}_{\nu} + \partial^{31}_{\nu} ) \delta_{\mu \alpha}  - \delta_{\mu \nu}  (\partial^{23}_{\alpha} + \partial^{31}_{\alpha}) \right)\times \nn \\
&& \times  (\partial^{23}_{\beta} + \partial^{31}_{\beta} ) \frac{1}{(x_{31}^2)^{d/2-1} (x_{23}^2)^{d/2-1}} \label{top2b} \\
\langle \frac{\delta T_{\mu\nu}[A](x_1)}{\delta A^{b \,\beta}(x_3)}  V^a_\alpha(x_2) \rangle_{A= 0}^{s}
&=&
\frac{c \, \delta^{ab} (d-1)}{d (d-2)^2} S_d \delta^d(x_{31}) \left((\partial^{23}_{\mu} + \partial^{12}_{\mu} ) \delta_{\nu \beta} + (\partial^{23}_{\nu} + \partial^{12}_{\nu} ) \delta_{\mu \beta}  - \delta_{\mu \nu}  (\partial^{23}_{\alpha} + \partial^{12}_{\beta}) \right)\times \nn\\
&& \times  (\partial^{23}_{\alpha} + \partial^{12}_{\alpha} ) \frac{1}{(x_{12}^2)^{d/2-1} (x_{23}^2)^{d/2-1}} \label{top2c}.
\eea
This expression is in complete agreement with the solution given in \cite{Osborn:1993cr}, to which we refer for further details 

\bea
\label{TVVcoord}
\langle T_{\mu\nu}(x_1) V^a_\alpha(x_2)  V^b_\beta(x_3) \rangle &=&
\frac{\delta^{ab}}{(x^2_{12})^{d/2} \, (x^2_{31})^{d/2} \, (x^2_{23})^{d/2-1}} \, I_{\alpha\sigma}(x_{12}) \, I_{\beta\rho}(x_{31}) \, t_{\mu\nu\rho\sigma}(X_{23}) \nn \\
&&- \delta^{ab} \left[ A_{\mu\nu\alpha\rho}(x_{12}) - \hat A_{\mu\nu\alpha\rho}(x_{12}) \right] \frac{I_{\rho\beta}(x_{23})}{(x_{23}^2)^{d-1}} \nn\\
&&- \delta^{ab} \left[ A_{\mu\nu\sigma\beta}(x_{31}) - \hat A_{\mu\nu\sigma\beta}(x_{31}) \right] \frac{I_{\sigma\alpha}(x_{23})}{(x_{23}^2)^{d-1}}, \nn \\
\eea
which is expressed in terms of tensor structures whose coefficients, denoted as $a,b,c$ and $e$ in \cite{Osborn:1993cr}, 
satisfy two constraint equation, and of contact terms $A$  and $\hat{A} $ which are given in  \cite{Osborn:1993cr}. 
For this reason, only 2 independent constants are left free to parameterize any conformal correlator of this type in $d$ dimensions. In the notation of \cite{Osborn:1993cr} $e=0$ and hence $b=0$, so that there is only one independent 
structure. A final comment
concerns the issues of renormalization. These expressions are unrenormalized. The issue of renormalization will be addressed by discussing in parallel the position and the momentum space approaches, that we will do starting from the next section. For this reason we turn to specific realizations of theories containing scalars and fermions - which are conformal in any dimension - and vectors, which are conformal for $d=4$.

\section{The $TTT$ Amplitude}\label{TTT}

\subsection{The correlator}

Now we are ready to turn to the analysis of the 3-graviton vertex.
The general structure of the $<TTT>$ correlator in momentum space is \cite{Osborn:1993cr}
\beq \label{bareTTT}
\langle T^{\mu\nu}(x_1) \, T^{\rho\sigma}(x_2) \, T^{\alpha\beta}(x_3) \rangle =
\frac{1}{(x^2_{12})^{d/2} \, (x^2_{23})^{d/2} \, (x^2_{31})^{d/2}} \,
\mathcal{I}^{\mu\nu\mu'\nu'}\, \mathcal{I}^{\rho\sigma\rho'\sigma'} \, t^{\mu'\nu'\rho'\sigma'\alpha\beta}(X_{12}) \, \
\eeq
\beq \label{Inversion}
\mathcal{I}^{\mu \nu,\alpha \beta} (s) =
I^{\mu\rho}(s)I^{\nu\sigma}(s) {\epsilon_T}^{\rho\sigma,\alpha\beta} \, ,
\quad s=  x - y
\eeq
where
\beq\label{epsilon}
{\epsilon_T}^{\mu\nu,\alpha\beta} =
\frac{1}{2} \, (\delta^{\mu\alpha} \delta^{\nu \beta} + \delta^{\mu \beta} \delta^{\nu \alpha} \bigl)
- \frac{1}{d} \, \delta^{\mu \nu} \delta^{\alpha \beta}
\eeq
is the projector onto the space of symmetric traceless tensors.\\
We perform the computation of the 3-graviton vertex $TTT$ in free field theory, for $d=4$, in all its 3 relevant sectors, the 
conformally coupled scalar, the fermion and the
vector, since in this case the general solution of the Ward identities, for any CFT, is parameterized by 3 independent constants. 
This corresponds to the most general anomalous solution. For $d\neq 4$ the spin-1 sector is not conformally invariant and we can't 
build the general expression just by superposing the scalar and the fermion sectors. However, the combination of the scalar and the 
fermion sectors corresponds to an anomaly-free special solution also for generic $d$  \cite{Erdmenger:1996yc}.

As we have already mentioned above, the correctness of our results has been checked first by a complete test of all the Ward 
identities for each case, which is already a nontrivial test to pass, given the large complexity of the computations. At the same 
time we will show that the counterterm introduced in \cite{Osborn:1993cr, Erdmenger:1996yc} in position space, which is extracted 
from the general expression of the trace anomaly when $d=4$, coincides with that required in momentum space using dimensional 
regularization. The connection between the two methods will be discussed thoroughly.

\begin{figure}[t]
\centering
\includegraphics[scale=0.8]{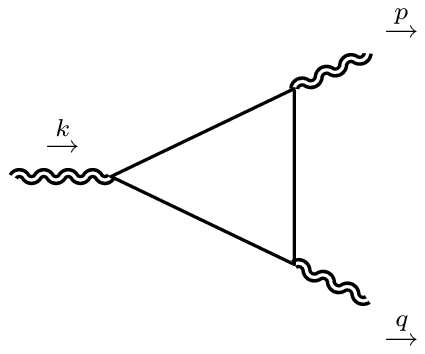}
\hspace{5mm}
\includegraphics[scale=0.8]{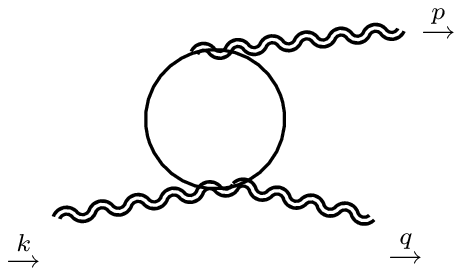}
\hspace{5mm}
\includegraphics[scale=0.8]{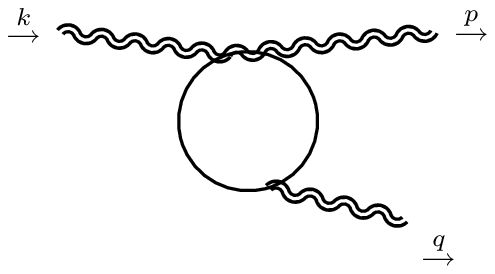}
\hspace{5mm}
\includegraphics[scale=0.8]{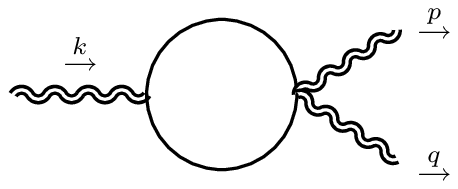}
\hspace{5mm}
\caption{One loop expansion of the 3-graviton vertex. Shown here are the general topologies,
i.e. the triangle and the self-energy type (T-bubble) contributions for the fermion case. The general correlator for any CFT in $d=4$ 
can be obtained by adding to these diagrams similar ones where the fermion is replaced by a scalar and a photon in the loops. Ghost 
corrections follow the same topologies.}
\label{Fig.diagramsTTT}
\end{figure}

\subsection{Inverse mapping for the $TTT$ Amplitude}
\label{InverseMappingTTT}

As done before for the $\langle VVV \rangle$, $\langle TOO \rangle$ and $\langle TVV \rangle$
correlators, here we check the result (\ref{bareTTT}) building explicitly the correlator from the diagrammatic expansion
in free field theory. This allows to come up with an expression for this vertex which is manifestly integrable.
We will be using the Feynman rules obtained from the Lagrangian descriptions for scalars, fermions and spin 1
in configuration space, given in section \ref{lags}.
We start testing the non-coincident case, for which we can omit the contact terms. This corresponds to the ``bulk" contribution to the
correlator, which involves only the triangle topology.
We give the d-dimensional expression for the scalar and the fermion cases, while - as already
remarked - we have to limit our analysis to $d=4$ for the spin-1 vector. Moreover, in the vector case
the gauge-fixing and ghost parts of the amplitude have to cancel since the vertex is obviously gauge invariant.
This has been explicitly verified in the computation in momentum space (see section \ref{ExplicitTTT}). So, in performing our
inverse mapping, we include in the interactions vertices only the Maxwell $\tilde{V}$ contributions, omitting ghosts and gauge-fixing terms. We have
\bea
\langle \frac{\delta\mathcal S}{\delta g_{\mu\nu}(x_1)}\,\frac{\delta\mathcal S}{\delta g_{\rho\sigma}(x_2)}
\frac{\delta\mathcal S}{\delta g_{\alpha\beta}(x_3)} \rangle^S
&=&
C^S_{TTT} V_{T\phi\phi}^{\mu\nu}(i\,\partial^{12},-i\,\partial^{31})\,V_{T\phi\phi}^{\rho\sigma}(i\,\partial^{23},-i\,\partial^{12})\,
V_{T\phi\phi}^{\alpha\beta}(i\,\partial^{31},-i\,\partial^{23})
\nonumber\\
&\times&
\frac{1}{(x_{12}^2)^{d/2-1}\,(x_{23}^2)^{d/2-1}\,(x_{31}^2)^{d/2-1}}
\, , \label{ScalarTriangle}\\
\langle \frac{\delta\mathcal S}{\delta g_{\mu\nu}(x_1)}\,\frac{\delta\mathcal S}{\delta g_{\rho\sigma}(x_2)}
\frac{\delta\mathcal S}{\delta g_{\alpha\beta}(x_3)} \rangle^F
&=&\nn
\eea
\bea
C^F_{TTT}\,(-1)\,\,\bigg( \textrm{Tr}\big[V_{T\bar\psi\psi}^{\mu\nu}(i\,\partial^{12},-i\,\partial^{31})\,
\,i\,\gamma\cdot\partial^{12}\,V_{T\bar\psi\psi}^{\rho\sigma}(i\,\partial^{23},-i\,\partial^{12})\,
\,i\,\gamma\cdot\partial^{23}\,V_{T\bar\psi\psi}^{\alpha\beta}(i\,\partial^{31},-i\,\partial^{23})\,
i\,\gamma\cdot\partial^{31}\big]&&\,
\nonumber\\
\hspace{-10mm}
+\textrm{Tr}\big[ V_{T\bar\psi\psi}^{\mu\nu}(i\,\partial^{31},-i\,\partial^{12})\,
i\,\gamma\cdot\,\partial^{31}\,V_{T\bar\psi\psi}^{\alpha\beta}(i\,\partial^{23},-i\,\partial^{31})\,
i\,\gamma\cdot\,\partial^{12}\,V_{T\bar\psi\psi}^{\rho\sigma}(i\,\partial^{12},-i\,\partial^{23})\,
i\,\gamma\cdot\partial^{12}\big]\bigg)\,
\nonumber\\
\times
\frac{1}{(x_{12}^2)^{d/2-1}\,(x_{23}^2)^{d/2-1}\,(x_{31}^2)^{d/2-1}} \, , &&
\nonumber \\
\label{FermionTriangle}
\eeqa
\bea
\langle \frac{\delta\mathcal S}{\delta g_{\mu\nu}(x_1)}\,\frac{\delta\mathcal S}{\delta g_{\rho\sigma}(x_2)}
\frac{\delta\mathcal S}{\delta g_{\alpha\beta}(x_3)} \rangle^V
&=&
C^V_{TTT}\,(-1)^3\,
\tilde{V}_{TAA}^{\mu\nu\gamma\delta}(i\,\partial^{12},-i\,\partial^{31})\,
\tilde{V}_{TAA}^{\rho\sigma\zeta\xi}(i\,\partial^{23},-i\,\partial^{12})\,
\tilde{V}_{TAA}^{\alpha\beta\chi\omega}(i\,\partial^{31},-i\,\partial^{23})\,
\nonumber\\
&\times&
\frac{\delta_{\gamma\xi}\,\delta_{\delta\chi}\delta_{\zeta\omega}}{x_{12}^2\,x_{23}^2\,x_{31}^2}\, ,
\label{VectorTriangle}
\eea
Notice that this last term enters only for $d=4$. 
Here and in the following, the dependences of the vertices on the coordinates are obtained by replacing the momenta
of (\ref{lags}) with appropriate derivatives respect to the external position variables. 
For instance  
\beq
V^{\mu\nu}_{T\phi\phi}(p,q)\to V^{\mu\nu}_{T\phi\phi}(\hat{p},\hat{q})=V^{\mu\nu}_{T\phi\phi}(i\, \partial^{12},- i\, \partial^{23})
\eeq
with 
\beq
\hat{p}\to i\, \partial^{12} \qquad\qquad \hat{q}\to - i\, \partial^{23}
\eeq
Explicitly
\bea
V^{\mu\nu}_{T\phi\phi}(i\, \partial^{12},- i\, \partial^{23})=
&=&
\frac{1}{2}\,(i\,\partial_{12\,\alpha}) \, (- i\,\partial_{23\,\beta}) \, C^{\mu\nu\alpha\beta}
\nonumber\\
&+&
\chi \bigg( \delta^{\mu\nu} \left( i\,\partial_{12} - i\,\partial_{23} \right)^2 
- \left( i\,\partial_{12}^{\mu} - i\,\partial_{23}^{\mu}\right)\,
\left( i\,\partial_{12}^{\nu} - i\,\partial_{23}^{\nu} \right) \bigg)
\, .
\eea
The replacements of $p,q$ and $l$, by the operatorial expressions $\hat{p},\hat{q}$ and $\hat{l}$ in \ref{DiagTTT}-\ref{DiagTTTWardAnom} are specific for each vertex. In appendix \ref{InverseTTT} we provide some more details on this procedure. Notice that we have chosen the
coupling parameter for the scalar field in $d$ dimensions at the corresponding conformal value $\chi = (d-2)/4(d-1)$.

Expanding the derivatives contained in each vertex, the expression given in (\ref{bareTTT}) is recovered by setting
\beq
C^S_{TTT} = -\frac{8}{S_d^3\,(d-2)^3}\, ,
\quad C^F_{TTT} = \frac{2^{d/2+1}}{S_d^3\,(d-2)^3}\, ,
\quad C^V_{TTT} = \frac{1}{S_4^3} \, .
\eeq
We compute next the contributions with the topology of 2-point functions, which are needed to account
for the behavior of the vertex in the short distance limit.
In coordinate space we can write them in a manifestly integrable form by pulling out derivatives in the same way as
for the triangle diagram. We replace the momenta with derivatives with respect to the corresponding
coordinates acting on propagators, obtaining very compact expressions for the vertex. We offer a few more details on this
computation in appendix \ref{InverseTTT}, quoting here the result.
In the scalar case we have
\bea
\langle \frac{\delta^2\mathcal{S}}{\delta g_{\mu\nu}(x_1)\delta g_{\alpha\beta}(x_3)}\,
\frac{\delta\mathcal{S}}{\delta g_{\rho\sigma}(x_2)} \rangle^S
&=&
\frac{C^S_{Q}}{2}\,
V^{\rho\sigma}_{T\phi\phi}(i\,\partial^{23},-i\,\partial^{12})\,
V^{\mu\nu\alpha\beta}_{TT\phi\phi}(i\,\partial^{12},-i\,\partial^{23},i\,\partial^{23}-i\,\partial^{31})\, \nn\\
&& \times \frac{\delta^{(d)}(x_{31})}{(x^2_{12})^{d/2-1}(x^2_{23})^{d/2-1}}
\nonumber
\eea
\bea
\langle \frac{\delta^2\mathcal{S}}{\delta g_{\mu\nu}(x_1)\delta g_{\rho\sigma}(x_2)}\,
\frac{\delta\mathcal{S}}{\delta g_{\alpha\beta}(x_3)} \rangle^S
&=&
\frac{C^S_{P}}{2}\,
V^{\alpha\beta}_{T\phi\phi}(i\,\partial^{31},-i\,\partial^{23})\,
V^{\mu\nu\alpha\beta}_{TT\phi\phi}(i\,\partial^{23},-i\,\partial^{31},-i\,\partial^{23}+i\,\partial^{12})
\nn\\
&&\times \frac{\delta^{(d)}(x_{12})}{(x^2_{23})^{d/2-1}(x^2_{31})^{d/2-1}}
\nonumber
\eea
\bea
\langle \frac{\delta^2\mathcal{S}}{\delta g_{\alpha\beta}(x_3)\delta g_{\rho\sigma}(x_2)}\,
\frac{\delta\mathcal{S}}{\delta g_{\mu\nu}(x_1)} \rangle^S
&=&
\frac{C^S_{K}}{2}\,
V^{\mu\nu}_{T\phi\phi}(i\,\partial^{12},-i\,\partial^{31})\,
V^{\alpha\beta\rho\sigma}_{TT\phi\phi}(i\,\partial^{31},-i\,\partial^{12},i\,\partial^{12}-i\,\partial^{23})\, \nn \\
&& \times \frac{\delta^{(d)}(x_{23})}{(x^2_{12})^{d/2-1}(x^2_{31})^{d/2-1}}.
\label{ScalarKBubble}
\eea
Notice that in the three contributions above, the $p,q,$ and $l$ dependence of the vertices correspond to mappings 
into $\hat{p}, \hat{q}$ and $\hat{l}$ which are specific for each T-bubble. 
Similarly, in the fermion sector we obtain
\bea
\langle \frac{\delta^2\mathcal{S}}{\delta g_{\mu\nu}(x_1)\delta g_{\alpha\beta}(x_3)}\,
\frac{\delta\mathcal{S}}{\delta g_{\rho\sigma}(x_2)} \rangle^F
&=&
- C^F_{Q}\,\delta^{(d)}(x_{31})\, \textrm{tr}\,
\big[
V^{\mu\nu\alpha\beta}_{TT\bar\psi\psi}(i\,\partial^{12},-i\,\partial^{23})\,
i\,\gamma\cdot\partial^{12}
V^{\rho\sigma}_{T\bar\psi\psi}(i\,\partial^{23},-i\,\partial^{12})\,
i\,\gamma\cdot\partial^{23}\big]
\nonumber \\
&\times&
\frac{1}{(x^2_{23})^{d/2-1}(x^2_{12})^{d/2-1}} \nn\, ,
\eeqa
and similar expressions for the $k-$ and $p$-bubles.
Finally, for the spin-1 vector field we obtain
\bea
\langle \frac{\delta^2\mathcal{S}}{\delta g_{\mu\nu}(x_1)\delta g_{\alpha\beta}(x_3)}\,
\frac{\delta\mathcal{S}}{\delta g_{\rho\sigma}(x_2)} \rangle^V
&=&
\frac{C^V_{Q}}{2}\, \,\delta^{(d)}(x_{31})\,
\tilde{V}^{\mu\nu\rho\alpha\beta\chi}_{TTAA}(i\,\partial^{12},-i\,\partial^{23})\,
\tilde{V}^{\rho\sigma\tau\omega}_{TAA}(i\,\partial^{23},-i\,\partial^{12})
\,\frac{\delta_{\zeta\tau}\,\delta_{\chi\omega}}{x^2_{12}\,x^2_{23}} \, ,
\eeqa
and similarly for the other bubble-type contributions.

Notice that this expression is affected by terms proportional to derivatives of $\delta$ functions.
We refer to appendix \ref{InverseTTT} for more details on the specific structures of these terms in momentum space, where we 
illustrate this point in a simple case. The complete structure of the $TTT$ vertex in position space is obtained by combining the 
triangle and the ``K",``P" and ``Q"-bubble topologies in the form
\bea
\langle T^{\mu\nu}(x_1)\,T^{\rho\sigma}(x_2)\,T^{\alpha\beta}(x_3) \rangle
&=&
\sum_{I=S,F,V} 8 \, 
\bigg[- \langle \frac{\delta \mathcal{S}}{\delta g_{\mu\nu}(x_1)}\,\frac{\delta \mathcal{S}}{\delta g_{\sigma\rho}(x_3)}\,
\frac{\delta \mathcal{S}}{\delta g_{\alpha\beta}(x_2)} \rangle^I
\nonumber\\
&&\hspace{-45mm}
+ \, \langle \frac{\delta^2\mathcal{S}}{\delta g_{\mu\nu}(x_1)\,\delta g_{\alpha\beta}(x_3)}\,
\frac{\delta\mathcal{S}}{\delta g_{\rho\sigma}(x_2)}\rangle^I  +
\langle \frac{\delta^2\mathcal{S}}{\delta g_{\mu\nu}(x_1)\,\delta g_{\rho\sigma}(x_2)}
\frac{\delta\mathcal{S}}{\delta g_{\alpha\beta}(x_3)}\rangle^I 
+ \langle \frac{\delta^2\mathcal{S}}{\delta g_{\alpha\beta}(x_3)\,\delta g_{\rho\sigma}(x_2)}\,
\frac{\delta\mathcal{S}}{\delta g_{\mu\nu}(x_1)}\rangle^I
\bigg] \, .
\eea
This expression is in agreement with the form of the energy-momentum tensor
three point function given in \cite{Osborn:1993cr}. The integrability of this result is manifest, due to the $(d/2-1)$ exponent of 
each propagator in position space, which corresponds, generically, to a $1/l^2$ behavior in momentum space. The vector terms, which 
exist in $d=4$ are, obviously, Fourier integrable.

\section{Moving to momentum space using Lagrangian realizations}
\label{lags}

At this point we use again the free field theory representation of these correlators to study their expression in momentum space. 
This will allow us to perform a direct comparison between position space and momentum space approaches for correlators affected
by the trace anomaly. We start by investigating the perturbative structure of these theories and derive the corresponding vertices.

The actions for the scalar and the fermion field are respectively
\bea\label{}
\mathcal{S}_{scalar}
&=&
\frac{1}{2} \, \int d^4 x \, \sqrt{-g}\,
\bigg[g^{\mu\nu}\,\nabla_\mu\phi\,\nabla_\nu\phi - \chi\,R\,\phi^2 \bigg]\, ,\label{scalarAction}\\
\mathcal{S}_{fermion}
&=&
\frac{i}{2} \, \int d^4 x \, V \, {V_\alpha}^\rho\,
\bigg[\bar{\psi}\,\gamma^\alpha\,(\mathcal{D}_\rho\,\psi)
- (\mathcal{D}_\rho\,\bar{\psi})\,\gamma^\alpha\,\psi \bigg] \, , \label{DiracAction}
\eea
where $\chi$ is the parameter corresponding to the ``improvement term", that we have chosen to be $1/6$ in the diagrammatic
calculation so to deal with the classically conformal invariant theory. ${V_\alpha}^\rho$ is the vielbein and $V$
$(= \sqrt{-g})$ its determinant, needed in such a way to embed
the fermion in the curved background, with its covariant derivative $\mathcal{D}_\mu$ as
\beq
\mathcal{D}_\mu = \pd_\mu + \Gamma_\mu =
\pd_\mu + \frac{1}{2} \, \Sigma^{\alpha\beta} \, {V_\alpha}^\sigma \, \nabla_\mu\,V_{\beta\sigma} \, .
\eeq
The $\Sigma^{\alpha\beta}$ are the generators of the Lorentz group  in the case of a spin $1/2$-field. \\
The action $\mathcal S$ for the photon field is given by
\beq
\mathcal{S}_{photon}=\mathcal{S}_M + \mathcal{S}_{gf} + \mathcal{S}_{gh}\, ,
\eeq
where the three contributions are the Maxwell action, the gauge fixing contribution and the ghost action
\beqa
\mathcal{S}_M    &=& - \frac{1}{4} \, \int d^4 x \, \sqrt{-g} \, F^{\a\b} F_{\a\b}\, ,\\
\mathcal{S}_{gf} &=& -\frac{1}{2 \xi} \, \int d^4 x \, \sqrt{-g} \, \left( \nabla_{\alpha}A^\alpha \right)^2\, \\
\mathcal{S}_{gh} &=& \int d^4 x \, \sqrt{-g}\, \partial^\a \bar{c} \, \partial_\a c\, .
\eeqa
We will be using Euclidean conventions for the generating functional, given by
\beq\label{GenFunct}
\mathcal{W} =
\frac{1}{\mathcal{N}}\int \mathcal D A_{\mu} \, \mathcal D \bar c \, \mathcal D c \, e^{- \mathcal S_E[A_\mu, \bar c, c]}\, .
\eeq
We will omit the "E" subscript from now on, as already done in (\ref{Generating}), to keep our notation easy.\\ 
The energy-momentum tensor is defined in (\ref{EMT}), which becomes, in the fermionic case,
\beq \label{FerTEI}
T^{\mu\nu} = \frac{1}{V} \, V^{\alpha\mu} \, \frac{\delta\mathcal{S}}{\delta {V^\alpha}_\nu}\, .
\eeq
This tensor is not symmetric in general, but its antisymmetric parts do not contribute to our calculations, 
so that, for our purposes, we can adopt the symmetric definition
\beq\label{SymmFerTEI}
T^{\mu\nu}
\stackrel{def}{\equiv} \frac{1}{2\,V}\bigg(
V^{\alpha\mu} \, \frac{\delta}{\delta {V^\alpha}_\nu} +
V^{\alpha\nu}\, \frac{\delta}{\delta {V^\alpha}_\mu}\bigg) \,
\eeq
as well. The energy-momentum tensors for the scalar and the fermion are
\bea
T^{\mu\nu}_{scalar}
&=&
\nabla^\mu \phi \, \nabla^\nu\phi - \frac{1}{2} \, g^{\mu\nu}\,g^{\alpha\beta}\,\nabla_\alpha \phi \, \nabla_\beta \phi
+ \chi \bigg[g^{\mu\nu} \Box - \nabla^\mu\,\nabla^\nu + \frac{1}{2}\,g^{\mu\nu}\,R - R^{\mu\nu} \bigg]\, \phi^2 \\
T^{\mu\nu}_{ferm}
&=&
\frac{i}{4} \, V \, \,
\bigg[ g^{\mu\rho}\,{V_\alpha}^\nu g^{\nu\rho}\,{V_\alpha}^\mu - 2\,g^{\mu\nu}\,{V_\alpha}^\rho \bigg]
\bigg[\bar{\psi} \, \gamma^{\alpha} \, \left(\mathcal{D}_\rho \,\psi\right) -
\left(\mathcal{D}_\rho \, \bar{\psi}\right) \, \gamma^{\alpha} \, \psi \bigg],
\eea
while the energy-momentum tensor for the photon field is given by the sum of three terms
\beq
T^{\mu\nu}_{QED} = T^{\mu\nu}_M + T^{\mu\nu}_{gf} + T^{\mu\nu}_{gh}\, ,
\eeq
with
\beqa
T^{\mu\nu}_M
&=&
\frac{1}{4}g^{\mu\nu}F^{\a\b}F_{\a\b} - F^{\mu\a}{F^\nu}_{\a} \, ,
\label{TEI M}
\\
T^{\mu\nu}_{gf}
&=&
\frac{1}{\xi}\{ A^\mu\nabla^\nu(\nabla_\r A^\r) + A^\nu\nabla^\mu(\nabla_\r A^\r ) -g^{\mu\nu}[ A^\r
\nabla_\r(\nabla_\s A^\s) + \frac{1}{2}(\nabla_\r A^\r)^2 ]\}\, ,
\label{TEI g.f.}
\\
T^{\mu\nu}_{gh}
&=&
\pd^\mu\bar{c}\, \pd^\nu c + \pd^\nu\bar{c}\, \pd^\mu c - g^{\mu\nu}\pd^{\r}\bar{c}\,\pd_{\r}c
\label{TEI gh}\, .
\eeqa
The computation of the vertices  can be done by taking (at most) two functional derivatives of the action
with respect to the metric, since the vacuum expectation values of the third order derivatives correspond to massless
tadpoles, which are zero in dimensional regularization. Given the complexity of the result and to avoid any error, we have checked 
that all the expressions obtained for the
1-loop vertices satisfy the corresponding Ward identities derived in the previous sections. They are given in 
Fig. \ref{listV} and their explicit expressions have been collected in Appendix \ref{VERTICES}.
\begin{figure}[t]
\begin{center}
\includegraphics[scale=0.7]{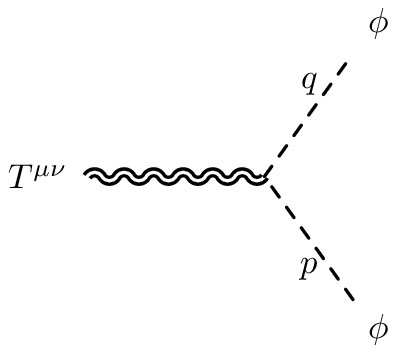}
\includegraphics[scale=0.7]{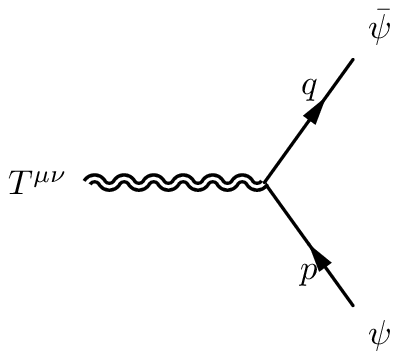}
\includegraphics[scale=0.7]{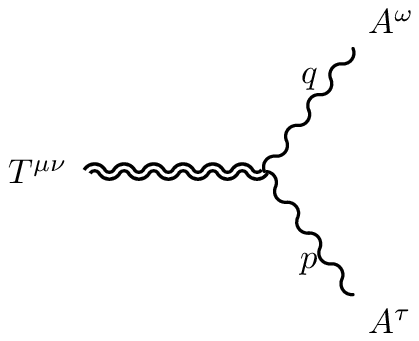}
\includegraphics[scale=0.7]{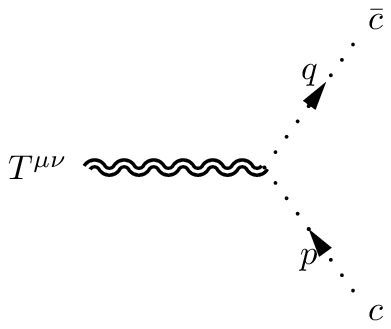} \\
\includegraphics[scale=0.7]{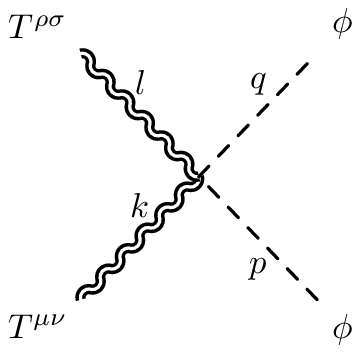}\qquad
\includegraphics[scale=0.7]{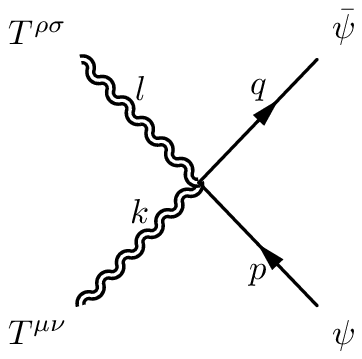}\qquad
\includegraphics[scale=0.7]{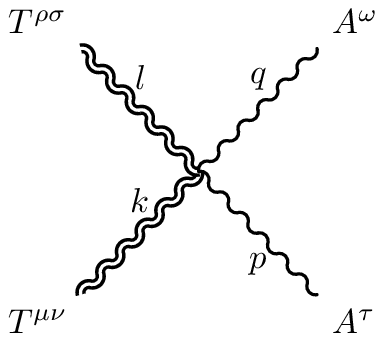}\qquad
\includegraphics[scale=0.7]{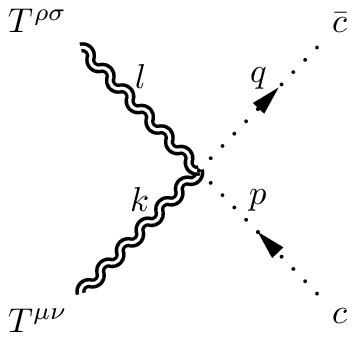}
\caption{List of the vertices used in the Lagrangian mapping of the conformal correlators}
\label{listV}
\end{center}
\end{figure}

\subsection{The interpretation of the counterterms:  the TT case}
\label{The TT case}

In this section we begin a discussion of the structure of anomalous correlators in momentum space, starting, for simplicity,
from the $TT$ case in the conformal limit. In the non-conformal case this correlator has been investigated in \cite{Bastianelli:2002qw,Bastianelli:2002fv} in the worldline approach.

This is a warm-up case before the more involved analysis of the 3-point functions that we will discuss afterwards. As we are
going to see, the interpretation of the anomaly and of its origin, in the process of renormalization, can be different in position 
and in momentum space. In fact, the anomaly can be attributed either to the specific structure of the counterterm in dimensional 
regularization, which violates conformal invariance in $d$ dimensions, while being traceless in $d=4$ or, alternatively, to the 
renormalized amplitude in d=4. In this second case the anomaly emerges as a feature of the $d=4$ renormalized amplitude and, 
specifically, of its 4-dimensional trace.

In the $TT$ case conformal symmetry fixes this correlator up to constant, and one can proceed with the
Fourier transform without resorting to a specific free field theory realization.
Using the inversion matrix in Euclidean space, we define the conformal energy-momentum tensor two-point function as
\beq\label{2PFOsborn}
\langle T^{\mu \nu}(x) \, T^{\alpha \beta} (y)\rangle =
\frac{C_T}{s^{2d}} \, \mathcal{I}^{\mu\nu ,\alpha\beta}(s) \, ,
\eeq
where $\mathcal{I}^{\mu\nu,\alpha\beta}(s)$ was defined in (\ref{Inversion}) and (\ref{epsilon}). \\
In order to move in the framework of differential regularization,  we pull out some derivatives and rewrite our correlator as
\beq
\label{2nd2PFOsborn}
\langle T^{\mu\nu}(x) \, T^{\alpha\beta}(0) \rangle
= \frac{C_T}{4(d-2)^2 d (d+1)} \, \hat{\Delta}^{(d)\,\mu\nu\alpha\beta}
\frac{1}{x^{2d - 4}} \, ,
\eeq
where
\bea \label{TransverseDeltaCoord}
\hat{\Delta}^{(d)\,\mu\nu\alpha\beta}
&=&
\frac{1}{2}\left( \hat{\Theta}^{\mu\alpha} \hat{\Theta}^{\nu\beta} +\hat{ \Theta}^{\mu\beta}\hat{ \Theta}^{\nu\alpha} \right)
- \frac{1}{d-1}\hat{ \Theta^{\mu\nu}} \hat{\Theta}^{\alpha\beta}\, ,
\quad
\text{with}
\quad
\hat{\Theta}^{\mu\nu} = \pd^\mu  \pd^\nu - \delta^{\mu\nu} \, \Box  \\
\pd_\mu \, \hat{\Delta}^{(d)\,\mu\nu\alpha\beta}
&=&
0 \, , \quad
\delta_{\mu\nu} \, \hat{\Delta}^{(d)\,\mu\nu\alpha\beta} = 0   \, .
\eea
For reasons that will be discussed in section \ref{direct}, this form of the $TT$ correlator is Fourier-integrable,
although it is characterized by a UV divergence as $x\to y$.
To move to momentum space we can split the ${1}/{(x^2)^{d-2}}$ term
into the product of two ${1}/{(x^2)^{d/2-1}}$ factors and apply straightforwardly the fundamental transform
({\it c.f.}  Eq. (\ref{fund})), obtaining
\bea \label{TransTT}
\langle T^{\mu\nu} \, T^{\alpha\beta} \rangle (p)
&\equiv&
\int \, d^d x \, \langle T^{\mu\nu}(x) T^{\alpha\beta}(0) \rangle
e^{- i\, p \cdot x}  \nn\\
&=&\frac{C_T}{4 (d-2)^2 d (d+1)} \, \int \, d^d x \, e^{- i\, p \cdot x}
\,\hat{ \Delta}^{(d)\,\mu\nu\alpha\beta}\, \frac{1}{(x^2)^{d/2-1}} \, \frac{1}{(x^2)^{d/2-1}} \nonumber \\
&=&
\frac{ (2\pi)^d \,  C(d/2 - 1)^2 \,  C_T}{4 (d-2)^2 d (d+1)}  \,
\Delta^{(d)\,\mu\nu\alpha\beta}(p) \, \int \, d^d l  \, \frac{1}{l^2 (l+p)^2} \, .
\eea
We have also defined
\bea \label{TransverseDeltaMom}
\Theta^{\mu\nu}(p)
&=&
\delta^{\mu\nu} \, p^2 - p^\mu \, p^\nu \, \\
\Delta^{(d)\,\mu\nu\alpha\beta}(p)
&=&
\frac{1}{2} \, \bigg(\Theta^{\mu\alpha}(p) \, \Theta^{\nu\beta}(p) + \Theta^{\mu\beta}(p) \, \Theta^{\nu\alpha}(p) \bigg)
- \frac{1}{d-1} \, \Theta^{\mu\nu}(p) \, \Theta^{\alpha\beta}(p)
\eea
as the momentum space counterparts of the two operators previously introduced.
In our notations $\Delta^{(4)\,\mu\nu\alpha\beta}$ is obtained from the expression above 
by setting $d=4$. The tensor indices, however, are still running from 0 to $d-1$.

Notice that in $d$ dimensions the $TT$ correlator is anomaly-free (i.e. traceless)
\beq
\delta_{\mu\nu} \langle T^{\mu\nu} \, T^{\alpha\beta} \rangle \, (p) =
\delta_{\alpha\beta}  \langle T^{\mu\nu} \, T^{\alpha\beta} \rangle \, (p) = 0 \, .
\eeq
As we move to $d=4$ the correlator in momentum space has a UV singularity, coming from the 2-point integral
\bea\label{B0}
\mathcal{B}_0(p^2)
&=&
\frac{1}{\pi^2} \, \int d^d l \, \frac{1}{l^2 \, (l + p )^2} =
\frac{\left[\Gamma(1 - \epsilon/2)\right]^2 \Gamma(\epsilon/2)}{\Gamma(2 - \epsilon)} \frac{1}{(\pi\, p^2)^{\epsilon/2}}
\nn  \\
&=&
\frac{2}{\bar{\epsilon}} + 2 + \ln \left(\frac{\mu^2}{p^2}\right) + O(\epsilon) \, ,
\label{Bzero}
\eea
where $\epsilon = 4 - d$ and we have introduced the quantity
$\frac{2}{\bar{\epsilon}} = \frac{2}{\epsilon} - \gamma - \ln \pi $, typical of the modified
minimal subtraction ($\overline{MS}$) scheme.
If we work in position space, renormalization is enforced by adding a local (i.e. $\sim \delta(x-y)$)
counterterm of the form $c_1/\bar{\epsilon} \, \hat{\Delta}^{(4)\mu\nu\alpha\beta} \, \delta(x-y)$.
The regulated correlator in $d=4$ is then defined as
\beq
\langle T^{\mu\nu}(x) \, T^{\alpha\beta}(0) \rangle
= \frac{C_T}{4(d-2)^2 d (d+1)} \, \hat{\Delta}^{(d)}_{\mu\nu\alpha\beta}
\frac{1}{x^{2d - 4}} + \frac{c_1}{\bar{\epsilon}} \, \hat{\Delta}^{(4)\,\mu\nu\alpha\beta} \, \delta^d(x-y) \, ,
\label{super}
\eeq
Notice that the counterterm is traceless for $d=4$ (i.e. contracting the indices with a 4-dimensional metric)
but not in general dimensions.
Therefore, if we split the $d$-dimensional
metric $(\delta^{(d)}_{\mu\nu})$ as a direct sum ($\oplus$) of a 4-dimensional ($\delta_{\mu\nu}\equiv \delta^{(4)}_{\mu\nu}$) and of 
a $(d-4)$-dimensional metrics acting on the subspaces $M_4$ and $M_{d-4}$
(i. e. $M_d=M_4 \oplus M_{d-4}$) we obtain
\beq \label{BrokenTrace}
\delta^{(d)}_{\mu\nu}\hat{ \Delta}^{(4)\mu\nu\alpha\beta} = \delta^{(4)}_{\mu\nu}\hat{ \Delta}^{(4)\mu\nu\alpha\beta}
+ \delta^{(d-4)}_{\mu\nu}\hat{ \Delta}^{(4)\mu\nu\alpha\beta} = \delta^{(d-4)}_{\mu\nu} \hat{\Delta}^{(4)\mu\nu\alpha\beta}
\eeq
and using the relation 
\beq
\delta^{(4)}_{\mu\nu}\hat{ \Delta}^{(4)\mu\nu\alpha\beta}=0
\eeq
we find that the $d$-dimensional trace of $\hat{\Delta}^{(4)}$ is $O(\epsilon)$
\beq
\delta^{(d)}_{\mu\nu}\, \hat{\Delta}^{(4)\,\mu\nu\alpha\beta} = -\frac{\epsilon}{3}\, \square\, \hat{\Theta}^{\alpha\beta} \, .
\eeq
If we now use the relation $\delta_{\mu\nu}\hat{\Delta}^{(d)\,\mu\nu\alpha\beta}=0$, it is clear that 
the trace of renormalized $TT$ correlator gives the correct anomaly. In particular, the trace operation cancels 
the ${1}/{\epsilon}$ pole of the counterterm
\beqa
\delta^{(d)}_{\mu\nu}\langle T^{\mu\nu}(x) \, T^{\alpha\beta}(0) \rangle &=&
c_1 \, \frac{1}{\bar{\epsilon}} \, \delta^{(d-4)}_{\mu\nu} \, \hat{\Delta}^{(4)\,\mu\nu\alpha\beta} \, \delta^d(x-y) \nn\\
&=& -\left[ \frac{c_1}{3} \square\hat{\Theta}^{\alpha\beta} + \frac{\epsilon}{2} (\gamma + \ln\pi ) \right] \,\delta^d(x-y)
\label{count}
\eeqa
which is finite as $\epsilon\to 0$ and reproduces the expected anomaly.
The selection of the counterterm is in agreement with the anomalous Ward identity of the 2-point function in momentum space.
This can be checked directly from Eq. (\ref{TraceAnomaly}), by computing its first functional derivative around flat space,
which leaves $ \Box R $ as the only contribution to the $TT$ anomaly
\beq\label{TraceAnomaly2PF}
\delta^{(4)}_{\mu\nu} \, \langle T^{\mu\nu}\,T^{\alpha\beta} \rangle(p) =
2 \, \beta_c\,\big[\Box R \big]^{\alpha\beta}(p) =
2 \, \beta_c \, p^2 \, \Theta^{\alpha\beta}(p) \, .
\eeq
Below we will be omitting the subscript $(4)$ when referring to a 4-dimensional kronecker $\delta_{\mu\nu}$, 
unless it is strictly necessary for clarity.

In the expression above, we have introduced the notation $\big[\Box R\big]^{\alpha\beta}(p)$ to indicate the Fourier-transformed
functional derivative of the box $(\Box)$ of the scalar curvature evaluated in the limit of flat spacetime.
The last two equations allow us to fix the final structure of the fully renormalized correlator in the form
\beq\label{Ren2PF1}
  \langle T^{\mu\nu} \, T^{\alpha\beta} \rangle_{ren}(p)
= \langle T^{\mu\nu}\,T^{\alpha\beta}\rangle_{bare}(p)
+ 6 \, \frac{\beta_c}{\bar{\epsilon}} \, \Delta^{(4)\,\mu\nu\alpha\beta}(p)
=
\langle T^{\mu\nu} \, T^{\alpha\beta}\rangle_{bare}(p)
- 4 \, \frac{\beta_a}{\bar{\epsilon}} \, \Delta^{(4)\,\mu\nu\alpha\beta}(p)\, ,
\eeq
where we have used in the last step Eq. (\ref{constraints}).

In position space, as clear from (\ref{count}), the anomaly can be attributed to the counterterm.
This approach allows to write down the solution of the Ward identities
as an anomaly free solution (for $x\neq y$) superimposed to the inhomogenous terms, exactly as stated in Eq. (\ref{super}).
This procedure is general, and can be applied to any correlator.

It is instructive, for comparison, to comment on the same approach in dimensional regularization working in momentum space.
One can start from a field theory realization of the same (unrenormalized) correlator obtaining
\bea \label{2PFp}
\langle T^{\mu\nu} \, T^{\alpha\beta} \rangle(p)
&=&
\left\{\frac{1}{2} \, \big[ \Theta^{\mu\alpha}(p) \, \Theta^{\nu\beta}(p) + \Theta^{\mu\beta}(p) \, \Theta^{\nu\alpha}(p)\big]
- \frac{1}{3} \, \Theta^{\mu\nu}(p) \, \Theta^{\alpha\beta}(p)\right\} \, C_1(p) \nn \\
&+&
\frac{1}{3}\,\Theta^{\mu\nu}(p) \, \Theta^{\alpha\beta}(p) \, C_2(p)\nn\\
&\equiv&
\Delta^{(4)\,\mu\nu\alpha\beta}(p)  \, C_1(p) +
\frac{1}{3}\Theta^{\mu\nu}(p) \, \Theta^{\alpha\beta}(p) \, C_2(p) \, ,
\eea
where the form factors are given, in the cases of a conformally coupled scalar, a Dirac fermion and a photon, by
\bea
C_1(p)\bigg|_{conf. scalar}
&=&
\frac{16 + 15 \, \mathcal{B}_0(p^2)}{14400\,\pi^2}\,
\qquad C_2(p)\bigg|_{conf. scalar}
=
- \frac{1}{1440\,\pi^2}\, , \\
C_1(p)\bigg|_{Dir. fermion}
&=&
\frac{2 + 5 \,\mathcal{B}_0(p^2)}{800\,\pi^2} \, ,
\qquad C_2(p)\bigg|_{Dir. fermion}
=
- \frac{1}{240\,\pi^2}\, , \\
C_1(p)\bigg|_{photon}
&=&
\frac{-11 + 10\, \mathcal{B}_0(p^2)}{800\,\pi^2} \, ,
\qquad C_2(p)\bigg|_{photon}
=
- \frac{1}{120\,\pi^2}\, .
\eea

Notice that the singularity of Eq. (\ref{2PFp}) is contained in the expressions of $C_1(p)$ due to the presence of the scalar 2-point 
function $\mathcal{B}_0$ which needs to be renormalized. The constant terms in these coefficients are due to the mass-independent
renormalization of the correlator, here performed in dimensional regularization, which, for each separate case, conformal scalar, 
fermion and photon, can be absorbed into a redefined renormalization scale $\mu$.
The two structures in the last line of (\ref{2PFp}) separately respect the energy-momentum conservation
Ward identity for the 2-point function \ref{WI2PFMom}, but only the first one, $\Delta^{(4)\,\alpha\beta\rho\sigma}(p)$, is traceless 
in $d=4$, while tracing the second we obtain the anomalous relation
\beq
\delta_{\mu\nu} \frac{1}{3}\,\Theta^{\mu\nu}(p)\,\Theta^{\alpha\beta}(p) = p^2 \, \Theta^{\alpha\beta}(p)\,.
\eeq
The singular contribution in Eq. (\ref{2PFp}) can be eliminated by the ordinary
renormalization procedure, leaving a result that is finite and whose trace can be taken {\em directly in 4 dimensions}.
In this approach the anomaly can be attributed to the regularization procedure and not directly to the counterterm,
which is traceless (compare (\ref{BrokenTrace}) for d = 4), while it is the finite part of the correlator,
going like $C_2(p)$, to be anomalous. 

The complete TT correlation function and its positive spectral functions were calculated in both the tensor and scalar 
sectors for a scalar field of arbitrary mass and curvature coupling $\xi$ in 4-dimensions in \cite{AndMolMott:2003}. In the case
of general mass and $\xi$, conformal invariance does not hold and the second tensor structure in (\ref{2PFp}) is
always present. By taking $\xi= 1/6$ and the limit of zero mass, one can also see from the spectral function
approach in \cite{AndMolMott:2003} how the trace anomaly appears.

As in the case of the chiral anomaly, a
dispersive analysis shows that the spectral density of an anomalous correlator is affected, under certain circumstances, by
typical contributions which amount to {\em anomaly poles}. Anomaly poles emerge from a collinear configuration of a certain
amplitude interpreted as a real space-time (on-shell) process. Similar poles have been found in the $TT$ case in 2-dimensions 
\cite{Bertlmann:2000da}. In higher dimensions because of the kinematics explained in \cite{Coleman:1982yg} one must go at least to 
triangle amplitudes at least as complicated as TVV or TTT in order to find these pole terms.

We have stressed this point to emphasize that the approach followed in
position space, which consists in the addition of a contact counterterm to regulate the anomaly, is not in contradiction with
the ordinary diagrammatic picture. It simply doesn't give the complete kinematical understanding of the origin of
the anomaly, which the spectral function dispersive approach attributes to the existence of a collinear region in the (anomalous) 
diagrams of the perturbative expansion. In the following, we will try to match these two quite different descriptions by
discussing more complex correlators.

\section{The counterterm for the $TVV$ in position and in momentum space}

We now turn to the question of the renormalization of TVV correlator in $d=4$ dimensions. This can be performed either
1) by solving the renormalized Ward identities in position space or
2) by a
perturbative computation in momentum space of all the diagrams in dimensional regularization.
The two methods are obviously quite different and the goal of this section is to test their correspondence,
given the results of \cite{Osborn:1993cr}.

As already emphasized in section (\ref{The TT case}),
the renormalized 3-point functions have to satisfy the requirement of general covariance as well as renormalized anomalous Ward
identities. The solution of these identities can be directly found by rewriting them in momentum space.
For the $\langle TVV \rangle$ case, the requirement of general covariance is also supplememented with gauge current
conservation.
If we denote our counterterm by $D_{\mu\nu\alpha\beta}(p,q)$, the algebraic conditions satisfied
by the counterterm are given by
\bea    \label{DivWardTVV}
(p + q)^\mu \,  D_{\mu\nu\alpha\beta} (p,q)
&=&
q_\nu \, \Theta_{\alpha\beta}(p) - \delta_{\nu\beta} \, q^\mu \, \Theta_{\mu\alpha}(p) +
p_\nu \,  \Theta_{\alpha\beta}(q) - \delta_{\nu\alpha} \,  p^\mu \,  \Theta_{\mu\beta}(q)  \, ,  \nonumber \\
p^\alpha \, D_{\mu\nu\alpha\beta}(p,q)
&=&
q^\beta \, D_{\mu\nu\alpha\beta}(p,q)  = 0 \, ,
\eea
with $\Theta_{\alpha\beta}(p)$ being the counterterm for the vector-vector 2-point function.
In fact, the equations above are just the divergent parts of the general covariance and gauge invariance Ward identities
for our three point function,
\bea \label{WardTVV}
(p+q)^\mu \, \langle T_{\mu\nu} \, {V^a}_\alpha \, {V^b}_\beta \rangle(p,q)
&=&
q_\nu \,  \langle {V^a}_\alpha \, {V^b}_\beta \rangle(p) - \delta_{\nu\beta} \,   q^\mu \,  \langle {V^a}_\mu \, {V^b}_\alpha
\rangle(p) +
p_\nu \,  \langle {V^a}_\alpha \, {V^b}_\beta \rangle(q) \nn\\
&&- \delta_{\nu\alpha} \,  p^\mu \, \langle {V^a}_\mu \, {V^b}_\beta
\rangle(q)  \, ,  \nonumber \\
p^\alpha \, \langle T_{\mu\nu} \, {V^a}_\alpha \, {V^b}_\beta \rangle(p,q)
&=&
q^\beta \, \langle T_{\mu\nu} \, {V^a}_\alpha \, {V^b}_\beta \rangle(p,q)  = 0 \, .
\eea
To see how they arise, we introduce the counterterms for the two correlators at hand, modulo two constants
\bea
\langle {V^a}_\alpha \, {V^b}_\beta \rangle_{ren}(p)
&=&
\langle {V^a}_\alpha \, {V^b}_\beta \rangle_{bare}(p) + \frac{1}{\epsilon} \, C_{VV} \, \Theta_{\alpha\beta}(p) \, , \nonumber
\\
\langle T_{\mu\nu} \, {V^a}_\alpha \, {V^b}_\beta \rangle_{ren}(p,q)
&=&
\langle T_{\mu\nu} \, {V^a}_\alpha \, {V^b}_\beta \rangle_{bare}(p,q) + \frac{1}{\epsilon} \, C_{TVV} \,
D_{\mu\nu\alpha\beta}(p,q) \, .
\eea
Replacing them in (\ref{WardTVV}) and equating the coefficients of the ${1}/{\epsilon}$ terms we immediately obtain
(\ref{DivWardTVV}) and the condition $C_{VV} = C_{TVV}$. These constraints are sufficient to state that the counterterm is
\bea
D_{\mu\nu\alpha\beta}(p,q)
&=&
\delta_{\alpha\beta} \,  ( p_\mu \, q_\nu + q_\mu \, p_\nu )
+ p \cdot q \,  (\delta_{\mu\beta}\, \delta_{\nu\alpha}  + \delta_{\mu\alpha} \, \delta_{\nu\beta} )  \nonumber \\
&-&
(\delta_{\beta\nu} \, p_\mu + \delta_{\beta\mu} \,  p_\nu ) \,  q_\alpha -
( \delta_{\mu\alpha} \, q_\nu + \delta_{\alpha\nu} \, q_\mu ) \, p_\beta
- \delta_{\mu\nu} \, ( p \cdot q \, \delta_{\alpha\beta} -   q_\alpha \,  p_\beta )  \, .
\eea
A consistency condition on this tensor, which is easily seen to be satisfied, is that the trace anomaly constraint in $d$ dimensions,
\beq
\delta^{\mu\nu} \, D_{\mu\nu\alpha\beta}(p,q) =   (4-d)  \, ( p \cdot q \, \delta_{\alpha\beta} - q_\alpha \, p_\beta )
\equiv  \epsilon  \, ( p \cdot q \, \delta_{\alpha\beta} - q_\alpha \, p_\beta ) \,
\eeq
reproduces the anomaly.

It is instructive to see how the same operation can be performed diagrammatically.
For this purpose we just recall that the general form of the $TVV$ amplitude can be expanded in a basis of 13 tensor structures  
$t_i^{\mu\nu\alpha\beta}(p,q)$ defined in \cite{Giannotti:2008cv}
\bea
\Gamma_{\mu\nu\alpha\beta}(p,q) =  \, \sum_{i=1}^{13} F_i (k^2; p^2,q^2)\ t^i_{\mu\nu\alpha\beta}(p,q)\,,
\label{Gamt}
\eea
where we have defined the tensors
\bea
&&
u^{\alpha\beta}(p,q) \equiv (p\cdot q) \,  \delta^{\alpha\beta} - q^{\alpha} \, p^{\beta}\,,\\
&&
w^{\alpha\beta}(p,q) \equiv p^2 \, q^2 \, \delta^{\alpha\beta} + (p\cdot q) \, p^{\alpha} \, q^{\beta}
- q^2 \,  p^{\alpha} \, p^{\beta} - p^2 \, q^{\alpha} \, q^{\beta}\,,
\label{uwdef}
\eea
which are Bose symmetric,
\bea
&&u^{\alpha\beta}(p,q) = u^{\beta\alpha}(q,p)\,,\\
&&w^{\alpha\beta}(p,q) = w^{\beta\alpha}(q,p)\,.
\eea
Gauge invariance is respected due to the conditions
\bea
&&p_{\alpha} \, u^{\alpha\beta}(p,q)  = q_{\beta} \, u^{\alpha\beta}(p,q) = 0\,,\\
&&p_{\alpha} \, w^{\alpha\beta}(p,q)  = q_{\beta} \, w^{\alpha\beta}(p,q) = 0\,.
\eea
A complete perturbative analysis shows that the only tensor structure which is affected by the renormalization
procedure is $t_{13}$, which coincides with the $D_{\mu\nu\alpha\beta}$ counterterm introduced above.
 As discussed in
\cite{Giannotti:2008cv} for QED and in \cite{Armillis:2009pq, Armillis:2010qk} for QED and QCD by direct computations,
renormalization of the $TVV$ vertex affects only this tensor structure. Given the complexity of the computations and the
wide difference between the general CFT approach and the ordinary diagrammatic one, this agreement is obviously nontrivial. 
As in the $TT$ case, the anomaly is generated by the ($d - 4$)-dimensional part of the trace, which
simplifies with the $1/(d-4)$  factor in front of the counterterm. In particular, all our previous comments concerning the 
renomalization of the $TT$ case remain valid also here, since in our approach the anomaly is computed
after subtracting the infinities, by taking the 4-dimensional trace of the renormalized $TVV$ vertex.
In particular, one can check that of the 13 structures  $t_i$ only $t_1$ has a non-vanishing trace, 
while the remaining ones are traceless. 
As discussed in \cite{Giannotti:2008cv} for the fermion case, $t_1$ carries all the information about the anomaly 
and its corresponding form factor $(F_1)$ contains an anomaly pole. The extraction of this additional information about the $TVV$ 
correlator indeed requires a complete analysis of the same in momentum space.

\subsection{TVV on-shell in $d=4$ and the anomaly poles}

As we have mentioned, the complete $TVV$ correlator can be obtained in any dimension as a superposition of a scalar and of
a fermion sectors. Obviously, this result holds for any CFT, and the explicit evaluation that we provide is completely general. In 
the off-shell case the fermion loop has been analyzed in \cite{Giannotti:2008cv} \cite{Armillis:2009pq}. Explicit resuls for this 
sector can be found in  \cite{Armillis:2009pq}. In this section we extend the computation to the scalar sector, focusing on the 
on-shell case for the two external vectors, since the expressions in the general case are far lengthier.

In the on-shell case the 13 structures $t^i$ simplify drastically.
We use three structures $A^1$, $A^2$ and $D$, with $D$ being the counterterm discussed above, to describe the parameterization of 
this vertex. In terms of the momenta of the two outgoing gauge bosons $(p,q)$, with $p^2=q^2=0$  and $p\cdot q=k^2/2 $ we have
\bea
{\Gamma_{\mu\nu\alpha\beta}^{ab}(p,q)}^{f/s}
=
{ F^{ab}_1(p \cdot q)}^{f/s}
A^1_{\mu\nu\alpha\beta}(p,q) + { F^{ab}_2(p \cdot q)}^{f/s}
A^2_{\mu\nu\alpha\beta}(p,q) + { F^{ab}_3(p \cdot q)}^{f/s}
D_{\mu\nu\alpha\beta}(p,q)
\label{fourD}
\eea
with
\bea
A^1_{\mu \nu \a \b}
&=&
(2 p \cdot q \, \delta^{\mu\nu} - k^{\mu}k^{\nu}) \, u^{\a \b} (p,q) \, , \\
 A^2_{\mu \nu \a \b}
&=&
- 2 \, u^{\a \b} (p,q) \left( 2 p \cdot q \, \delta^{\mu \nu} + 2 (p^\mu \, p^\nu + q^\mu \, q^\nu )
- 4 \, (p^\mu \, q^\nu + q^\mu \, p^\nu) \right)  \,,
\eea
with form factors given by 
\bea
{F_1^{a b}(p \cdot q)}^f &=&  \delta^{ab} \, \frac{1}{72\, \pi^2 \, p \cdot q} \,, \\
{F_2^{a b}(p \cdot q)}^f &=&  \delta^{ab} \, \frac{1}{576 \pi^2 \, p \cdot q} \,, \\
{F_3^{a b}(p \cdot q)}^f &=& -\delta^{ab} \, \frac{1}{288 \pi^2} \bigg[ 12 \mathcal B_0(2p\cdot q,0,0) + 11 \bigg] \,,
\eea
for the fermion sector and
\bea
{F_1^{a b}(p \cdot q)}^s &=&   \delta^{ab},\frac{1}{144 \pi^2 \, p \cdot q} \,, \\
{F_2^{a b}(p \cdot q)}^s &=& - \delta^{ab}\,\frac{1}{576 \pi^2 \, p \cdot q} \,, \\
{F_3^{a b}(p \cdot q)}^s &=& - \delta^{ab}\,\frac{1}{576 \pi^2} \bigg[ 6 \mathcal B_0(2p\cdot q,0,0) + 7\bigg] \,,
\eea
for the scalar sector. Notice that both the scalar (s) and the fermion (f) sectors have anomaly poles. The anomaly
is attributed to the tensor structure $A_1$ which has a nonzero trace. As we have clarified above, the anomaly is not attributed to 
$D$ (i.e. $t_{13}$), which is the counterterm
found in position space, but to the tensor structure $A_1$, after renormalization. The remaining structures $A_2$ and $D$ are, in 
fact, traceless in 4-dimensions. This structure coincides with the form factor
$t_1$ of \cite{Giannotti:2008cv}, which has a nonzero trace. As remarked before, the dynamical origin of the trace anomaly has 
necessarily to be found in momentum space.
\begin{figure}[t]
\label{TVVfigure}
\begin{center}
\includegraphics[scale=0.7]{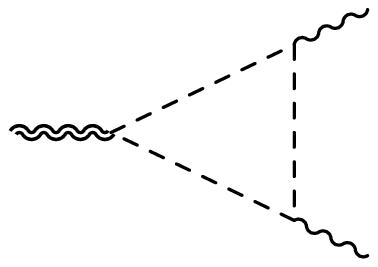}\qquad
\includegraphics[scale=0.7]{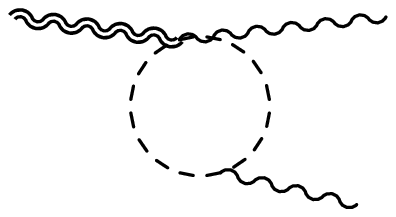}\qquad
\includegraphics[scale=0.7]{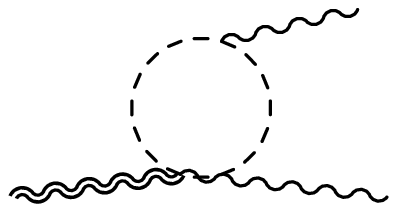}
\caption{The fermion/scalar sectors in the TVV vertex.}
\end{center}
\end{figure}

\subsection{TVV in $d$ dimension}

These results can be generalized, with some extra effort, to $d$ dimensions. By our inverse mapping procedure the result of the 
computation in this case remains valid for any conformal theory, since the two sectors, scalar and fermion, are sufficient to 
describe the general solution of the Ward identities.
The result can be given in a form which is quite similar to those in (\ref{fourD}). 
We obtain
\bea
\Gamma_{\mu\nu\alpha\beta}^{ab}(p,q)_f 
&=&  
f_1^{ab}(p \cdot q) C^f_{\mu\nu\alpha\beta}(p,q) + f_2^{ab}(p \cdot q) D_{\mu\nu\alpha\beta}(p,q) \,, \nn \\
\Gamma_{\mu\nu\alpha\beta}^{ab}(p,q)_s &=&  s_1^{ab}(p \cdot q)
C^s_{\mu\nu\alpha\beta}(p,q) + s_2^{ab}(p \cdot q)
D_{\mu\nu\alpha\beta}(p,q) \, .
\eea
The form factors are found to be
\bea
f_1^{ab}(p \cdot q) 
&=& 
\frac{1 \, \delta^{ab}}{(2 \pi)^d p \cdot q} \frac{d-4}{d (d-1)(d-2)} \pi^2 \mathcal B_0(2 p \cdot q ,0 , 0) \nn
\eea
\bea
f_2^{ab}(p \cdot q) 
&=& 
-\frac{2 \, \delta^{a b}}{(2\pi)^d}\frac{d(d-3)+4}{d(d-1)(d-2)} \pi^2 \mathcal B_0(2 p \cdot q ,0 ,0) \,, \nn
\eea
\bea
s_1^{ab}(p \cdot q) 
&=& 
\frac{4\delta^{ab}}{(2 \pi)^d} \frac{d-4}{d(d-1)(d-2) p\cdot q}  \pi^2 \mathcal B_0(2 p \cdot q ,0 , 0)   \, ,\nn
\eea
\bea
s_2^{ab}(p \cdot q) 
&=&  
-\frac{2\,\delta^{ab}}{(2 \pi)^d} \frac{1}{d(d-1)} \pi^2 \mathcal B_0(2 p \cdot q ,0 , 0)   \, ,
\eea
where the tensors in the basis are given by
\bea
C^f_{\mu\nu\alpha\beta}(p,q)
&=&
 ( p \cdot q \, \delta_{\alpha\beta} -   q_\alpha \,  p_\beta ) \left( d (
p_\mu \, p_\nu +  q_\mu \, q_\nu)  +  (d-4)  ( p_\mu \, q_\nu +  q_\mu \,
p_\nu) -2 (d-2) p \cdot q \, \delta_{\mu\nu}   \right) \,, \nn \\
C^s_{\mu\nu\alpha\beta}(p,q)
&=&
 ( p \cdot q \, \delta_{\alpha\beta} -   q_\alpha \,  p_\beta ) \left(
p_\mu \, q_\nu +  q_\mu \, p_\nu -  p \cdot q \, \delta_{\mu\nu}
\right) \,, \nn \\
D_{\mu\nu\alpha\beta}(p,q)
&=&
\delta_{\alpha\beta} \,  ( p_\mu \, q_\nu + q_\mu \, p_\nu )
+ p \cdot q \,  (\delta_{\mu\beta}\, \delta_{\nu\alpha}  +
\delta_{\mu\alpha} \, \delta_{\nu\beta} )  \nonumber \\
&-&
(\delta_{\beta\nu} \, p_\mu + \delta_{\beta\mu} \,  p_\nu ) \,  q_\alpha -
( \delta_{\mu\alpha} \, q_\nu + \delta_{\alpha\nu} \, q_\mu ) \, p_\beta
- \delta_{\mu\nu} \, ( p \cdot q \, \delta_{\alpha\beta} -   q_\alpha \,
p_\beta )  \, .
\eea

Notice that in this case all the structures (C, D) are traceless since there is no anomaly.
As a final observation, we remark that in the on-shell case, the only topology that survives in the expansion of this correlator 
corresponds to a master integral of type $\mathcal{B}_0$ which corresponds to a massless 2-point function. The other master integral 
which also heavily appears in the perturbative expansion, $\mathcal{C}_0$, which corresponds to the scalar triangle diagram, drops 
out in the on-shell limit.

\subsection{Renormalization of the $TTT$}
\label{Renormalization}

In this section we address the problem of the renormalization of the 3-graviton vertex and compare the standard Lagrangian
approach with the deductive method of \cite{Osborn:1993cr}, which is developed for the analysis in $d$ dimensions.
Since our interest, for this vertex, is sharply focused on the $d=4$ case, we need to clarify a few points. Notice that one
of the two counterterms that appear at Lagrangian level, $G$, is a total divergence in 4 but not in $d$ dimensions.
In particular, $G$ generates a counterterm which is effectively a projector on the extra $(d-4)$-dimensional space and as such,
gives a contribution which needs to be included in order to perform a correct renormalization of the vertex.
This has been verified by an explicit computation in dimensional regularization.

We recall that in perturbation theory the one loop counterterm Lagrangian is
\beq\label{CounterAction}
S_{counter} = - \frac{1}{\epsilon}\sum_{I=f,s,V}n_I \int d^d x \sqrt{-g} \bigg( \beta_a(I) F + \beta_b(I) G\bigg)
\eeq
We have used the 4-dimensional realization of $F$
\beq\label{RecallF}
F =
R^{\alpha\beta\gamma\delta}R_{\alpha\beta\gamma\delta} - 2\,R^{\alpha\beta}R_{\alpha\beta} + \frac{1}{3} \, R^2
\eeq
which is obtained from (\ref{Geometry1}) with $d\to 4$.
$G$, obviously does not contribute to every correlator. For instance, in the case of the $TT$, the counterterm is obtained by
functional differentiation twice of $S_{counter}$, but one can easily check (see Eq. (\ref{Magic})) that the second variation of
$G$ vanishes in the flat limit. Hence, the only counterterm is given by
\beq
\label{2PFCounterterm}
D_F^{\alpha\beta\rho\sigma}(x_1,x_2) =
4 \, \frac{\delta^2}{\delta g_{\alpha\beta}(x_1)\delta g_{\rho\sigma}(x_2)} \, \int\,d^d w\,\sqrt{-g} \, F\, .
\eeq
Its form in momentum space is given by
\beq
D_F^{\alpha\beta\rho\sigma}(p) = 4 \,\Delta^{(4)\,\alpha\beta\rho\sigma}(p)\, ,
\eeq
and we recover the renormalized 2-point function in (\ref{Ren2PF1}) just with its inclusion, i.e.
\beq\label{Ren2PF2}
\langle T^{\alpha\beta}\,T^{\rho\sigma} \rangle_{ren}(p)
= \langle T^{\alpha\beta}\,T^{\rho\sigma} \rangle(p)
- \frac{\beta_a}{\,{\bar\epsilon}} \, D_F^{\alpha\beta\rho\sigma}(p)\, .
\eeq
In the case of the 3-graviton vertex the counterterm action (\ref{CounterAction}) generates the vertices 
\beq
-\frac{1}{\epsilon}\, \bigg(\beta_a D_F^{\mu\nu\rho\sigma\alpha\beta}(z,x,y)
+ \beta_b \, D_G^{\mu\nu\rho\sigma\alpha\beta}(z,x,y)\bigg)\, ,
\eeq
where
\bea
D_F^{\mu\nu\rho\sigma\alpha\beta}(x_1,x_2,x_3)
&=&
8 \, \frac{\delta^3}{\delta g_{\mu\nu}(x_1) \delta g_{\rho\sigma}(x_2)\delta g_{\alpha\beta}(x_3)}
\int\,d^d w\,\sqrt{-g}\, F\, ,\label{DF}\\
D_G^{\mu\nu\rho\sigma\alpha\beta}(z,x,y)
&=&
8 \, \frac{\delta^3}{\delta g_{\mu\nu}(x_1) \delta g_{\rho\sigma}(x_2)\delta g_{\alpha\beta}(x_3)}
\int\,d^d w\,\sqrt{-g}\, G \label{DG}\, .
\eea
(\ref{DG}) and (\ref{DF}) are obtained by functionally deriving three times the general functional
\beq
\mathcal{I}(a,b,c) \equiv \int\,d^4 x\,\sqrt{-g}\,
\big(a\,R^{abcd}R_{abcd} + b\,R^{ab}R_{ab} + c\, R^2 \big)\, ,
\eeq
with respect to the metric for appropriate $a, b$ and $c$, i.e.
\bea
a = 1 \, ,\quad b = -2 \, ,\quad c = \frac{1}{3} \, ,
\nonumber \\
a = 1 \, ,\quad b = -4 \, ,\quad c = 1 \, . \nonumber
\eea
Some of the computations are, for convenience, reproduced
in appendix \ref{FunctionalIntegral}.\\

It is known that $D_G^{\mu\nu\alpha\beta\rho\sigma}(p,q)$ is found to vanish identically in four dimensions.
In fact, its explicit form is
\beq\label{ExplicitDG}
D_G^{\mu\nu\alpha\beta\rho\sigma}(p,q) =
-240 \big(E^{\mu\sigma\alpha\gamma\kappa,\nu\rho\beta\delta\lambda} +
E^{\mu\rho\alpha\gamma\kappa,\nu\sigma\beta\delta\lambda} + \alpha\leftrightarrow \beta\big)
\,q_\gamma\, q_\delta\, p_\kappa\, p_\lambda\, ,
\eeq
where $E^{\mu\sigma\alpha\gamma\kappa,\nu\rho\beta\delta\lambda}$ is a projector onto
completely antisymmetric tensors with five indices, so that it would yield zero in four dimensions,
reflecting the fact that the integral of the Euler density is a topological invariant in such dimensions.
We have explicitly checked
by an explicit computation that, given the structure of the counterterm Lagrangian in (\ref{CounterAction}), one needs
necessarily to include the contribution from the $G$ part of the functional, in the form given by $D_G$, in order to remove all
the divergences. This choice brings us to a counterterm contribution which regulates $TTT$ which is slightly different from the
approach followed in \cite{Osborn:1993cr}. The two approaches, in fact, differ by a finite renormalization, since in our case
we reproduce the entire anomaly, including the local contribution ($\beta_c \neq 0$).
The fully renormalized 3-point correlator in momentum space can be written down as
\beq\label{Ren3PF}
\langle T^{\mu\nu}T^{\rho\sigma}T^{\alpha\beta} \rangle_{ren}(p,q) =
\langle T^{\mu\nu}T^{\rho\sigma}T^{\alpha\beta} \rangle_{bare}(p,q) -
\frac{1}{{\epsilon}}\,\bigg(\beta_a\,D_F^{\mu\nu\alpha\beta\rho\sigma}(p,q)
+ \beta_b\,D_G^{\mu\nu\alpha\beta\rho\sigma}(p,q)\bigg)\,
\eeq
and the goal is to proceed with an identification both of $D_F$ and $D_G$ from the
diagrammatic expansion in momentum space.
The cancellation of all of the ultraviolet poles, for suitable expressions of $D_F$ and $D_G$, has been thoroughly checked
from our explicit results. As we have already discussed in the previous cases, after renormalization, we can take the trace of
(\ref{Ren3PF}) (in four dimensions) and obtain the entire trace anomaly.

In parallel, it is instructive to see how one can derive the analogue of (\ref{Ren3PF}), using our expression of $F$, which is
4-dimensional, but following the same approach of \cite{Osborn:1993cr}, i.e. by using the Ward identities.
In this case we are bound to introduce the generic counterterms to the $TTT$ vertex
\beq \label{Ren3PFansatz}
\langle T^{\mu\nu}T^{\rho\sigma}T^{\alpha\beta} \rangle_{ren}(p,q) =
\langle T^{\mu\nu}T^{\rho\sigma}T^{\alpha\beta} \rangle_{bare}(p,q)
+ \frac{1}{\epsilon} \,
\bigg(C_F \, D_F^{\mu\nu\alpha\beta\rho\sigma}(p,q) + C_G \, D_G^{\mu\nu\alpha\beta\rho\sigma}(p,q)\bigg), \,
\eeq
written in terms of arbitrary coefficients $C_F$ and $C_G$. Notice that, for convenience, we have formulated
(\ref{Ren3PFansatz}) in momentum space, but the $1/\epsilon$ corrections are supported only at the coincidence point
($x_1=x_2=x_3$), for appropriate $D_F$ and $D_G$, as one could check by performing a transform of this expression.

With the addition of the new contact terms which guarantee the regularization of the correlator, the new renormalized
vertex must satisfy (\ref{WI3PFcoordinate}) and two similar identities which follow exchanging indices and momenta properly.\\
One can check that $D_G^{\mu\nu\alpha\beta\rho\sigma}(p,q)$ is transverse, as (\ref{ExplicitDG}) shows clearly,
\beq \label{DGConstraints1}
k_{\nu}D_G^{\mu\nu\alpha\beta\rho\sigma}(p,q) = 0 \, , \quad
p_{\alpha}D_G^{\mu\nu\alpha\beta\rho\sigma}(p,q) = 0 \, \quad
q_\sigma D_G^{\mu\nu\alpha\beta\rho\sigma}(p,q) =0\, ,
\eeq
so that by inserting the expressions (\ref{Ren2PF2}) and (\ref{Ren3PFansatz}) into these Ward identities and taking
(\ref{DGConstraints1}) into account, one obtains three conditions on the F-contribution to the counterterm, the first being
\beqa \label{DFConstraints1}
&&
C_F \, k_\nu D_F^{\mu\nu\alpha\beta\rho\sigma}(p,q) =
- 4 \, \beta_a \bigg\{q^\mu \Delta^{(4)\,
\rho\sigma\alpha\beta}(p) + p^\mu \Delta^{(4)\,\alpha\beta\rho\sigma}(q) \nn\\
&&
-  q_\nu \bigg[\delta^{\mu\rho}\Delta^{(4)\,\nu\sigma\alpha\beta}(p)
+ \delta^{\mu\sigma}\Delta^{(4)\,\nu\rho\alpha\beta}(p)\bigg]
-  p_\nu \bigg[\delta^{\mu\alpha}\Delta^{(4)\,\nu\beta\rho\sigma}(q)
+ \delta^{\mu\beta}\Delta^{(4)\,\nu\alpha\rho\sigma}(q)\bigg]\bigg\}\, .
\eeqa
\begin{figure}[t]
\centering
\includegraphics[scale=0.7]{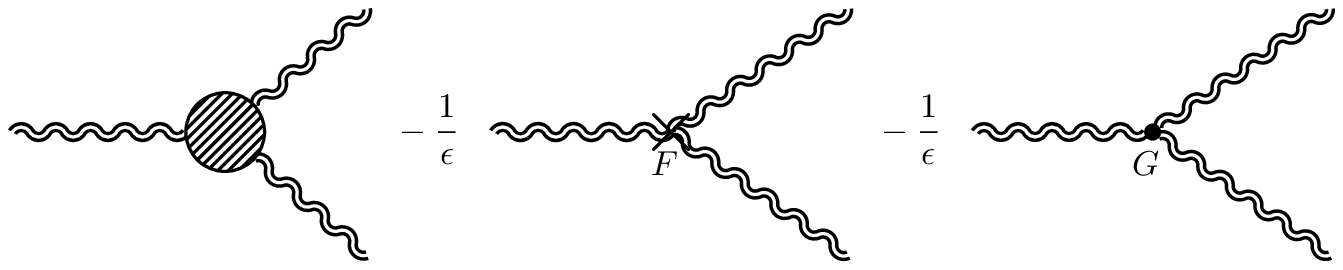}
\hspace{5mm}
\caption{$TTT$ and its counterterms generated with the choice of the square of the Weyl ($F$) tensor in 4 dimensions and the Euler density ($G$).}
\includegraphics[scale=0.7]{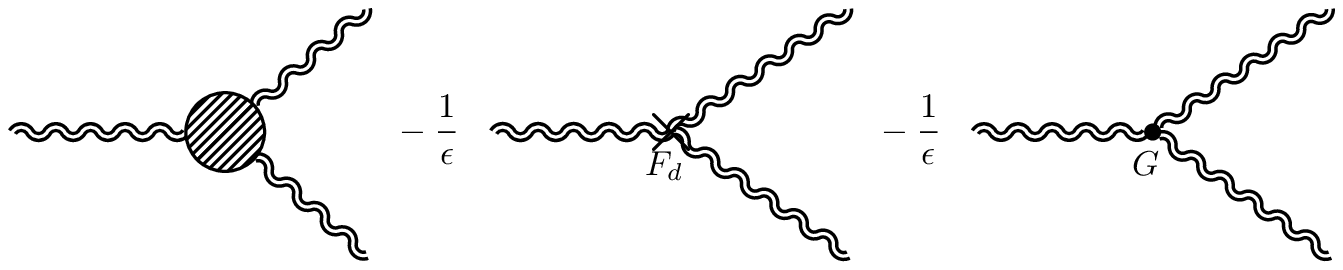}
\hspace{5mm}
\caption{The contributions to the renormalized $TTT$ vertex from the square of the Weyl tensor in $d$-dimensions ($F^d$) and the Euler density ($G$).}
\label{Fig.diagrams3grav}
\end{figure}
and the other two coming from a permutation of the indices and of the momenta.
They are seen to be satisfied if $C_F = -\beta_a$.\\
Exactly the same argument can be applied to the three anomalous trace identities in $d = 4 + \epsilon$
dimensions in order to fix $C_G$. Notice that, in this approach, the anomaly is reproduced by taking the traces
of $D_F^{\mu\nu\alpha\beta}(p,q)$ and $D_G^{\mu\nu\alpha\beta}(p,q)$ in $d$ dimensions, obtaining
\bea\label{CTTraces}
\delta_{\mu\nu}D_F^{\mu\nu\alpha\beta\rho\sigma}(p,q)
&=&
-4 \,\epsilon \,\bigg( \big[F\big]^{\alpha\beta\rho\sigma}(p,q)
- \frac{2}{3} \big[\sqrt{-g}\Box\,R\big]^{\alpha\beta\rho\sigma}(p,q) \bigg)
-  8 \, \bigg(\Delta^{(4)\,\alpha\beta\rho\sigma}(p) + \Delta^{(4)\,\alpha\beta\rho\sigma}(q)\bigg) \nn\\
\label{DFConstraints2}\,
\delta_{\mu\nu}D_G^{\mu\nu\alpha\beta\rho\sigma}(p,q)
&=&
- 4 \, \epsilon \,\big[G\big]^{\alpha\beta\rho\sigma}(p,q)
\label{DGConstraints2}\, .
\eea
According to the previously established notation, $\big[F\big]^{\alpha\beta\rho\sigma}(p,q)$ and
$\big[G\big]^{\alpha\beta\rho\sigma}(p,q)$ are the Fourier-transformed second functional derivatives
of the squared Weyl tensor and the Euler density respectively.
Requiring (\ref{munu3PFanomaly}) to be satisfied by the renormalized 2 and 3-point correlators we get
\bea \label{Trace}
\delta_{\mu\nu} \, \bigg( - \beta_a \,\ D_F^{\mu\nu\alpha\beta\rho\sigma}(p,q)
+ C_G \, D_G^{\mu\nu\alpha\beta\rho\sigma}(p,q) \bigg)
&=&
4 \, \epsilon \,
\bigg[ \beta_a \,
\bigg(\big[F\big]^{\alpha\beta\rho\sigma}(p,q) - \frac{2}{3}\big[\sqrt{-g}\Box\,R\big]^{\alpha\beta\rho\sigma}(p,q)\bigg)
\nonumber\\
&+&
\beta_b \, \big[G\big]^{\alpha\beta\rho\sigma}(p,q)
\bigg]
- 8 \, \bigg(\Delta^{(4)\,\alpha\beta\rho\sigma}(p) + \Delta^{(4)\,\alpha\beta\rho\sigma}(q)\bigg)\, ,\nn\\
\eea
and other two similar equations, obtained by shuffling indices and momenta as for the general covariance Ward identites.\\
In this way the conditions (\ref{Ren2PF2}), (\ref{DFConstraints2}), (\ref{DGConstraints2})
and (\ref{constraints}) allow us to obtain the relation $C_G = -{\beta_b}$, as expected. We have verified by
direct computation for scalar, fermion and vector fields that the approach followed in ref. \cite{Osborn:1993cr}
of solving the Ward identities by adding contact terms to the homogenous expression of vertex (obtained for separate 
points) matches precisely the renormalization procedure above in momentum space.

Notice that in \cite{Osborn:1993cr} the choice of $F$ is slightly different from ours, since the authors essentially define
a counterterm which at a Lagrangian level would be of the form
\beq
\mathcal{\tilde{S}}_{counter} = 
- \frac{1}{\epsilon} \int d^4x \sqrt{-g} \, \big( \beta_a \, F^{d} + \beta_b \, G \big) \,
\label{newren}
\eeq
based on the $d$-dimensional expression of the squared of the Weyl tensor ($F^d$).
Such a choice does not generate a local anomaly contribution proportional to $\square R$ as $d\to 4$. In fact the authors 
choose to work with $\beta_c=0$ from the beginning, since the inclusion of the local anomaly contribution amounts just 
to a finite renormalization with respect to (\ref{newren}). Notice that in $d$ dimensions, if we take the trace of the functional
derivative in (\ref{Magic}) for $a=1$, $b = -{4}/({d-2})$, $c = {2}/((d-1)(d-2))$, which are the $d$-dimensional coefficients
appearing in $F^d$, one can explicitly check that the contribution proportional to $\square R$ in the anomalous trace cancels.
For this purpose we can expand the integrand of (\ref{newren}) around $d=4$ (in $\epsilon = 4 - d$) up to $O(\epsilon)$,
obtaining that the counterterm action can be separated in a pole plus a finite part, i.e.
\beq
\mathcal{\tilde{S}}_{counter} =
\mathcal S_{counter} + \mathcal S_{fin.\,ren.} = \mathcal S_{counter} 
+ \beta_a \, \int d^4x\, \sqrt{-g} \, \bigg(R^{\alpha\beta}\,R_{\alpha\beta} - \frac{5}{18} \, R^2\bigg)  + O(\epsilon)\, .
\eeq
Recalling the definition (\ref{VEVEMT}) and using (\ref{Magic}),
we see that the contribution of this finite part to the vev of the energy-momentum tensor is
\beq
g_{\mu\nu} \langle \, T^{\mu\nu} \, \rangle_{fin.ren.}
=
-\beta_c \, \square R
\, .
\eeq

\begin{figure}
\begin{center}
\includegraphics[scale=0.7]{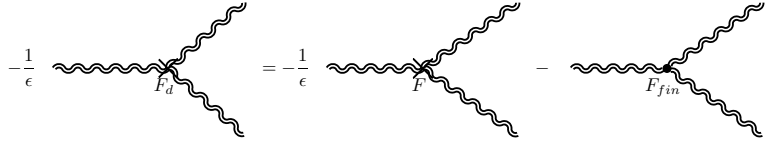}
\hspace{5mm}

\caption{The relation between the counterterm generated by $F^d$ and the same obtained from $F$. The difference is a finite
renormalization ($F_{fin}$) generated by the $\sqrt{-g} R^2$ term in the counterterm Lagrangian, which generates the local
contirbution to the trace anomaly.}
\end{center}
\end{figure}

Comparing this with (\ref{TraceAnomaly}), we see that this extra contribution will cancel
the local anomaly.\\
So this approach is equivalent, for what concerns the anomaly, to supplying the action of the theory with
the finite renormalization usually met in the literature, i.e.
\beq
\mathcal S^{(2)}_{fin.\,ren.} \equiv - \frac{\beta_c}{12} \, \int d^4x\, \sqrt{-g} \, R^2\, ,
\eeq
which is known to cancel the local anomaly, due to the similar relation
\beq
g_{\mu\nu} \, \frac{2}{\sqrt{-g}} \, \frac{\delta \mathcal S^2_{fin.\,ren.}}{\delta g_{\mu\nu}}
=
- \beta_c \,\square R\, ,
\eeq
which holds in $d=4$ as well.\\

\subsection{The renormalized on-shell 3-graviton vertex in 4 dimensions }
\label{ExplicitTTT}

In all of the three cases examined, the vertex $\Gamma^{\mu\nu\alpha\beta\rho\sigma}(p,q)$
can be expanded on a basis made up of thirteen tensors, if we go on shell on the two outgoing
gravitons, which amounts to contract the amplitude with polarization tensors which are transverse and traceless
\beq\label{helicity}
e^s_{\lambda\kappa}(p)\, , \quad {(e^s)^\lambda}_\lambda = 0\, , \quad p^\lambda\,e^s_{\lambda\kappa} = 0 \, ,
\eeq
where the superscript denotes the helicity state.\\
It is easy to see that the contraction of the amplitude with the polarization tensors with the properties (\ref{helicity})
for the two outgoing gravitons is equivalent to the replacements
\beq\label{OnShell}
p^2       \rightarrow 0  \, , \quad q^2      \rightarrow 0 \, , \quad
p^\alpha  \rightarrow 0  \, , \quad p^\beta  \rightarrow 0 \, , \quad
q^\rho    \rightarrow 0  \, , \quad q^\sigma \rightarrow 0 \, ,
\eeq
so that we will give the amplitude in terms of tensors which are non-vanishing after this limit is taken.\\
The expansion of our Green's function for a theory with $n_S$ scalars, $n_F$ fermions and $n_V$ vector bosons
can be written in general as
\beq
\langle T^{\mu\nu}T^{\rho\sigma}T^{\alpha\beta} \rangle(p,q)\bigg|_{On-Shell}
= \sum_{n_I=n_S,n_F,n_V}\, n_I\,\sum_{i=1}^{13}\, \Omega^I_i(s) \, t_i^{\mu\nu\alpha\beta\rho\sigma}(p,q)\, ,
\quad s = k^2 = (p+q)^2 = 2 \, p \cdot q \, .
\eeq
The form factors  for the three theories at hand are listed in Table \ref{FormFactors},
modulo the three overall factors, in the first row.
\begin{table}
$$
\begin{array}{|c|c|c|c|}
\hline
i
&   \Omega^S_{i}(s) \times {720\,\pi^2}
&   \Omega^F_{i}(s) \times {240\,\pi^2}
&   \Omega^V_{i}(s) \times {1152\,\pi^2}
\\
\hline
\hline
1
&  - \frac{1}{2 \, s}
&  - \frac{1}{s}
&    \frac{72}{5 \, s}  \\
\hline
2
&  - \frac{1}{s}
&  - \frac{1}{3 \, s}
&    \frac{64}{5 \, s}
\\
\hline
3
&  - \frac{7 + 30 \, \mathcal{B}_0(s)}{120}
&    \frac{13 - 30 \, \mathcal{B}_0(s)}{60}
&    \frac{82 - 120 \, \mathcal{B}_0(s)}{25}
\\
\hline
4
&  -  \frac{2 + 5 \, \mathcal{B}_0(s)}{10}
&     \frac{7 - 70 \, \mathcal{B}_0(s)}{120}
&     \frac{2 \, (482 + 130 \, \mathcal{B}_0(s))}{25}
\\
\hline
5
&     \frac{1}{6}
&   - \frac{-1 + 10 \, \mathcal{B}_0(s)}{48}
&   - \frac{79 + 50 \, \mathcal{B}_0(s)}{5}
\\
\hline
6
&    \frac{23 + 20 \, \mathcal{B}_0(s)}{20}
&    \frac{33 + 70 \, \mathcal{B}_0(s)}{60}
&  - \frac{104 \, (22 + 5 \, \mathcal{B}_0(s))}{25}
\\
\hline
7
&  - \frac{s\,(16 + 15 \, \mathcal{B}_0(s))}{20}
&  - \frac{3\,s\,(2 + 5 \, \mathcal{B}_0(s))}{10}
&  - \frac{s\,(-11 + 10 \, \mathcal{B}_0(s))}{80}
\\
\hline
8
&  - \frac{s\,(47 + 30 \, \mathcal{B}_0(s))}{80}
&  - \frac{3\, s\,(9 + 10 \, \mathcal{B}_0(s))}{40}
&    \frac{s\,(2 + 5 \, \mathcal{B}_0(s))}{40}
\\
\hline
9
&    \frac{s\,(2 + 5 \, \mathcal{B}_0(s))}{40}
&  - \frac{7 s\,(1 - 10 \, \mathcal{B}_0(s))}{480}
&  - \frac{s\,(487 + 130 \, \mathcal{B}_0(s))}{50}
\\
\hline
10
&    \frac{s \, (9 + 10 \, \mathcal{B}_0(s))}{20}
&    \frac{s \, (137 + 430\,  \mathcal{B}_0(s))}{480}
&  - \frac{s \, (883 - 230 \, \mathcal{B}_0(s))}{50}
\\
\hline
11
&  - \frac{s \, (7 + 5 \, \mathcal{B}_0(s))}{20}
&  - \frac{7 \, s \, (9 + 10 \, \mathcal{B}_0(s))}{240}
&    \frac{s \, (467 + 130 \, \mathcal{B}_0(s))}{25}
\\
\hline
12
&  - \frac{s \, (121 + 90 \, \mathcal{B}_0(s))}{240}
&  - \frac{s \, (97 + 130 \, \mathcal{B}_0(s))}{240}
&    \frac{2 \, s \, (299 + 35 \, \mathcal{B}_0(s))}{25}
\\
\hline
13
&   \frac{5 \, s^2 \, (3 + 2  \, \mathcal{B}_0(s))}{32}
&   \frac{5 \, s^2 \, (9 + 10 \, \mathcal{B}_0(s))}{96}
&  - s^2 \, (13 - \mathcal{B}_0(s))
\\
\hline
\end{array}
$$
\caption{Form factors for the vertex $\Gamma^{\mu\nu\alpha\beta\rho\sigma}(p,q)$ in the on-shell limit.}
\label{FormFactors}
\end{table}
The 13 tensors $t_i^{\mu\nu\alpha\beta\rho\sigma}(p,q)$ are listed below. They are given by
\bea
t_1^{\mu \nu \alpha \beta \rho \sigma}(p,q)
&=&
\big( p^{\mu} p^{\nu} + q^{\mu} q^{\nu}\big) \, p^{\rho} p^{\sigma} q^{\alpha} q^{\beta} \nn\\
t_2^{\mu \nu \alpha \beta \rho \sigma}(p,q)
&=&
\big( p^{\mu} q^{\nu} + p^{\nu} q^{\mu} \big) \, p^{\rho} p^{\sigma} q^{\alpha} q^{\beta}
\nn
\eeqa
\bea
t_3^{\mu \nu \alpha \beta \rho \sigma}(p,q)
&=&
\big(p^{\mu} p^{\nu}  + q^{\mu} q^{\nu} \big) \,
\big( p^{\sigma} q^{\beta} \delta^{\alpha \rho} + p^{\sigma} q^{\alpha} \delta^{\beta \rho}
+ p^{\rho} q^{\beta} \delta^{\alpha \sigma} + p^{\rho} q^{\alpha}\delta^{\beta \sigma} \big) \nn \\
t_4^{\mu \nu \alpha \beta \rho \sigma}(p,q)
&=&
p^{\rho} p^{\sigma} \, \big(
q^{\beta} q^{\nu} \delta^{\alpha \mu} + q^{\beta} q^{\mu} \delta^{\alpha \nu}
+ q^{\alpha} q^{\nu} \delta^{\beta \mu} + q^{\alpha} q^{\mu} \delta^{\beta \nu}
\big) \nn \\
&+&
q^{\alpha} q^{\beta} \, \big(
p^{\nu} p^{\sigma} \delta^{\mu \rho} + p^{\nu} p^{\rho} \delta^{\mu \sigma}
+ p^{\mu} p^{\sigma} \delta^{\nu \rho} + p^{\mu} p^{\rho} \delta^{\nu \sigma}
\big)
\nn \\
t_5^{\mu \nu \alpha \beta \rho \sigma}(p,q)
&=&
\big(p^{\mu} q^{\nu} + q^{\mu} p^{\nu}\big) \, \bigg(
  p^{\rho}   \big(q^{\alpha} \delta^{\beta\sigma} + q^{\beta} \delta^{\alpha\sigma} \big)
+ p^{\sigma} \big(q^{\alpha}\delta^{\beta \rho}   + q^{\beta} \delta^{\alpha\rho} \big) \bigg) \nn
\eeqa
\bea
t_6^{\mu \nu \alpha \beta \rho \sigma}(p,q)
&=&
\delta^{\mu \nu} p^{\rho} p^{\sigma} q^{\alpha} q^{\beta}
\nn
\eea
\bea
t_7^{\mu \nu \alpha \beta \rho \sigma}(p,q)
&=&
  p^{\rho}p^{\sigma}  \, \big(\delta^{\mu\alpha}\delta^{\nu\beta} + \delta^{\mu\beta}\delta^{\nu\alpha}\big)
+ q^{\alpha}q^{\beta} \, \big(\delta^{\mu\rho}\delta^{\nu\sigma} + \delta^{\mu\sigma}\delta^{\nu\rho}\big) \nn\\
&-&
\frac{1}{2} \, \bigg(
  p^{\mu}p^{\rho}   \big( \delta^{\alpha\sigma}\delta^{\nu\beta} + \delta^{\beta\sigma}\delta^{\nu\alpha} \big)
+ p^{\nu}p^{\rho}   \big( \delta^{\alpha\sigma}\delta^{\mu\beta} + \delta^{\beta\sigma}\delta^{\mu\alpha} \big) \nn \\
&+&
  p^{\mu}p^{\sigma} \big( \delta^{\alpha\rho}\delta^{\nu\beta}   + \delta^{\beta\rho}\delta^{\nu\alpha} \big)
+ p^{\nu}p^{\sigma} \big( \delta^{\alpha\rho}\delta^{\mu\beta}   + \delta^{\beta\rho}\delta^{\mu\alpha}  \big) \nn \\
&+&
  q^{\mu}q^{\alpha} \big( \delta^{\beta\sigma}\delta^{\nu\rho}   + \delta^{\beta\rho}\delta^{\nu\sigma}  \big)
+ q^{\nu}q^{\alpha} \big( \delta^{\beta\sigma}\delta^{\mu\rho}   + \delta^{\beta\rho}\delta^{\mu\sigma}  \big)\nn \\
&+&
  q^{\mu}q^{\beta}  \big( \delta^{\alpha\sigma}\delta^{\nu\rho}  + \delta^{\alpha\rho}\delta^{\nu\sigma}  \big)
+ q^{\nu}q^{\beta}  \big( \delta^{\alpha\sigma}\delta^{\mu\rho}  + \delta^{\alpha\rho}\delta^{\mu\sigma} \big)
\bigg) \nn
\eea
\bea
t_8^{\mu \nu \alpha \beta \rho \sigma}(p,q)
&=&
\big(p^{\mu} p^{\nu}+q^{\mu} q^{\nu}\big) \,
\big(\delta^{\alpha \sigma} \delta^{\beta \rho} + \delta^{\alpha \rho} \delta^{\beta \sigma})
\nn\\
t_9^{\mu \nu \alpha \beta \rho \sigma}(p,q)
&=&
  p^{\rho} \, \bigg(
  q^{\mu} (\delta^{\alpha \sigma} \delta^{\beta \nu}+\delta^{\alpha \nu} \delta^{\beta \sigma})
+ q^{\nu} (\delta^{\alpha \sigma} \delta^{\beta \mu}+\delta^{\alpha \mu} \delta^{\beta \sigma})
\bigg) \nn \\
&+&
p^{\sigma} \, \bigg(
   q^{\mu} (\delta^{\alpha \rho} \delta^{\beta \nu}+\delta^{\alpha \nu} \delta^{\beta \rho})
+  q^{\nu} (\delta^{\alpha \rho} \delta^{\beta \mu} + \delta^{\alpha \mu} \delta^{\beta \rho})
\bigg) \nn \\
&+&
q^{\alpha} \, \bigg(
  p^{\mu}  (\delta^{\beta \sigma} \delta^{\nu \rho} +  \delta^{\beta \rho} \delta^{\nu \sigma})
+ p^{\nu}  (\delta^{\beta \sigma} \delta^{\mu\rho}  +  \delta^{\beta \rho} g_{\mu \sigma})
\bigg) \nn \\
&+&
q^{\beta} \, \bigg(
  p^{\mu}  ( \delta^{\alpha\sigma}\delta^{\nu\rho} + \delta^{\alpha\rho} \delta^{\nu\sigma})
+ p^{\nu}  ( \delta^{\alpha\sigma}\delta^{\mu\rho} + \delta^{\alpha\rho}\delta^{\mu\sigma})
\bigg)
\nn
\eea
\bea
t_{10}^{\mu \nu \alpha \beta \rho \sigma}(p,q)
&=&
p^{\rho} \, \bigg(
  q^{\alpha} (\delta^{\beta \nu} \delta^{\mu \sigma} + \delta^{\beta \mu} \delta^{\nu \sigma})
+ q^{\beta}  (\delta^{\alpha \nu} \delta^{\mu \sigma} + \delta^{\alpha \mu} \delta^{\nu\sigma})
\big) \nn \\
&+&
p^{\sigma}  \, \bigg(
   q^{\alpha} (\delta^{\beta \nu} \delta^{\mu\rho} + \delta^{\beta \mu} \delta^{\nu \rho})
+  q^{\beta}  (\delta^{\alpha \nu} \delta^{\mu \rho} + \delta^{\alpha \mu} \delta^{\nu \rho}) \bigg) \nn \\
&-& p.q \, \bigg(
  \delta^{\alpha \rho}    (\delta^{\beta \nu} \delta^{\mu \sigma} + \delta^{\beta \mu} \delta^{\nu \sigma})
+ \delta^{\alpha \nu}     (\delta^{\beta \sigma} \delta^{\mu \rho} + \delta^{\beta \rho} \delta^{\mu \sigma}) \nn \\
&+&
  \delta^{\alpha \mu}     (\delta^{\beta \sigma} \delta^{\nu \rho} + \delta^{\beta \rho} \delta^{\nu \sigma} )
+ \delta^{\alpha \sigma}  (\delta^{\beta \nu} \delta^{\mu \rho}+\delta^{\beta \mu} \delta^{\nu\rho})
\bigg)
\nn \\
t_{11}^{\mu \nu \alpha \beta \rho \sigma}(p,q)
&=&
\big(p^{\nu} q^{\mu} + p^{\mu} q^{\nu}\big) \,
\big(\delta^{\alpha \sigma} \delta^{\beta \rho} + \delta^{\alpha \rho} \delta^{\beta \sigma}\big)
\nn \\
t_{12}^{\mu \nu \alpha \beta \rho \sigma}(p,q)
&=&
 \delta^{\mu \nu} \,
\bigg( p^{\rho} \big( q^{\beta} \delta^{\alpha \sigma} + q^{\alpha} \delta^{\beta\sigma} \big)
+ p^{\sigma}  \big( q^{\beta} \delta^{\alpha \rho} + q^{\alpha} \delta^{\beta \rho} \big)
 \bigg)
\nn\\
t_{13}^{\mu \nu \alpha \beta \rho \sigma}(p,q)
&=&
\delta^{\mu \nu} \, \big(\delta^{\alpha \sigma} \delta^{\beta \rho}+\delta^{\alpha \rho} \delta^{\beta \sigma}\big).
\eea
The correlator is affected by ultraviolet divergences coming from the two-point integrals $\mathcal{B}_0(s)$ (see Eq. (\ref{B0})).
This is true in the off-shell case too, as all the other contributions to the scalar coefficients
of its tensor expansion are made up of the three invariants $p^2, q^2$ and $p \cdot q$ plus the scalar 3-point integral
\beq
\mathcal {C}_0(s,s_1,s_2) =
\frac {1}{\pi^2} \int d^dl\, \frac{1}{l^2\,(l + p_1)^2\, (l + p_2)^2} \, , s = (p_1+p_2)^2 \, , s_i= p_i^2\, i=1,2 \, \quad
\eeq
which is finite for $d=4$. In the $\overline{MS}$ scheme the renormalized two-point integral is defined as
\beq
\mathcal{B}_0^{\overline{MS}}(p^2) = 2 + \ln\left(\frac{\mu^2}{p^2}\right) \, .
\eeq
which simply replaces the unrenormalized expression ${B}_0(p^2)$ (\ref{Bzero}) given in Tab. \ref{FormFactors}, after using the
renormalization procedure discussed above.
We have checked that by taking the trace of these 13 tensors one reproduces the Weyl, Euler and local contributions
to the trace anomaly satisfied by the vertex which in this on-shell case are given by
\beqa
\delta_{\mu\nu}\langle T^{\mu\nu}T^{\alpha\beta}T^{\mu\nu}\rangle(p,q)\bigg|_{On-Shell}
&=&
4 \, \bigg\{\beta_a\,\bigg(\big[F\big]^{\alpha\beta\rho\sigma}(p,q)
- \frac{2}{3} \big[\sqrt{-g}\Box\,R\big]^{\alpha\beta\rho\sigma}(p,q)\bigg) 
\nonumber\\
&+& \beta_b\, \big[G\big]^{\alpha\beta\rho\sigma}(p,q)\bigg\}\bigg|_{On-Shell}
\label{munu3PFanomaly1}
\eeqa
\beqa
\delta_{\alpha\beta}\langle T^{\mu\nu}T^{\alpha\beta}T^{\mu\nu}\rangle(p,q)\bigg|_{On-Shell}
&=&
4 \, \bigg\{\beta_a\,\bigg(\big[F\big]^{\mu\nu\rho\sigma}(-k,q)
- \frac{2}{3} \big[\sqrt{-g}\Box\,R\big]^{\mu\nu\rho\sigma}(-k,q)\bigg)
\nonumber\\
&+& 
\beta_b\, \big[G\big]^{\mu\nu\rho\sigma}(-k,q)
- \frac{1}{2}\,\langle T^{\mu\nu}T^{\rho\sigma}\rangle(k)\bigg\}\bigg|_{On-Shell}
\label{albe3PFanomaly1}
\eeqa
\beqa
\delta_{\rho\sigma}\langle T^{\mu\nu}T^{\alpha\beta}T^{\mu\nu}\rangle(p,q)\bigg|_{On-Shell}
&=&
4 \, \bigg\{\beta_a\,\bigg(\big[F\big]^{\mu\nu\alpha\beta}(-k,p)
- \frac{2}{3} \big[\sqrt{-g}\Box\,R\big]^{\mu\nu\alpha\beta}(-k,p)\bigg)
\nonumber\\
&+& 
\beta_b\, \big[G\big]^{\mu\nu\alpha\beta}(-k,p) 
- \frac{1}{2} \,\langle T^{\mu\nu}T^{\alpha\beta}\rangle(k)\bigg\}\bigg|_{On-Shell}\, ,
\nonumber\\
\label{rosi3PFanomaly1}
\eeqa
with
\bea
\left[F\right]^{\alpha\beta\rho\sigma}(p,q)\bigg|_{On-Shell}
&=&
2\, p^{\rho} \, p^{\sigma} q^{\alpha} q^{\beta}
- p\cdot q\,\bigg( p^{\sigma}  q^{\beta} \delta^{\alpha\rho} - p^{\rho} q^{\beta} \delta^{\alpha\sigma}
- p^{\sigma} q^{\alpha} \delta^{\beta\rho} - p^{\rho} q^{\alpha} \delta^{\beta\sigma}\bigg)
\nonumber \\
&+&
(p\cdot q)^2\, \bigg(\delta^{\alpha\sigma} \delta^{\beta\rho} + \delta^{\alpha\rho} \delta^{\beta\sigma}\bigg)
\\
\left[G\right]^{\alpha\beta\rho\sigma}(p,q)\bigg|_{On-Shell}
&=&
2\, p^{\rho} p^{\sigma} q^{\alpha} q^{\beta}
- p\cdot q\,\bigg( p^{\sigma} q^{\beta} \delta^{\alpha\rho} -  p^{\rho} q^{\beta} \delta^{\alpha \sigma}
- p^{\sigma} q^{\alpha} \delta^{\beta\rho} - p^{\rho} q^{\alpha} \delta^{\beta \sigma}\bigg)
\nonumber\\
&+&
(p\cdot q)^2\, \bigg(\delta^{\alpha \sigma} \delta^{\beta \rho} + \delta^{\alpha \rho} \delta^{\beta \sigma}\bigg)
\\
\left[\sqrt{-g}\,\square R\right]^{\alpha\beta\rho\sigma}(p,q)\bigg|_{On-Shell}
&=&
\frac{1}{2}\, p\cdot q\,\bigg(p^{\sigma} q^{\beta} \delta^{\alpha\rho}   + p^{\rho} q^{\beta} \delta^{\alpha \sigma}
+ p^{\sigma} q^{\alpha} \delta^{\beta\rho} + p^{\rho} q^{\alpha} \delta^{\beta \sigma}\bigg)
\nonumber\\
&-&
\frac{3}{2}\, (p\cdot q)^2 \bigg(g^{\alpha \sigma} \delta^{\beta \rho} - \delta^{\alpha \rho} \delta^{\beta \sigma}\bigg).
\eea
The on-shell limits of the two point functions appearing in the r.h.s. of (\ref{albe3PFanomaly1})
and (\ref{rosi3PFanomaly1}) are obtained from (\ref{2PFp}) replacing $p\rightarrow k$
and using (\ref{OnShell}) in (\ref{TransverseDeltaMom}).

\section{The $T\phi^2 \phi^2$ and $VVV$ correlators in momentum space}

A similar analysis allows to obtain the expression in momentum space of the $T\phi^2 \phi^2$, discussed before
in position space. We give the complete $d$-dimensional off-shell expression. It can be decomposed into four
independent tensor structures
\bea
\Gamma^{T\phi^2 \phi^2}_{\mu\nu}(p,q) &=& F_1(p,q) \left( p_{\mu}p_{\nu}-
\frac{p^2}{d} \delta_{\mu\nu} \right) + F_1(q,p) \left( q_{\mu}q_{\nu}-
\frac{q^2}{d} \delta_{\mu\nu} \right) \nn \\
&+& F_2(p,q) \left( p_{\mu}q_{\nu} + p_{\nu}q_{\mu} - \frac{2 \, p \cdot q}{d}
\delta_{\mu\nu} \right)  + F_3(p,q) \frac{1}{d} \delta_{\mu\nu}
\eea
where the first three tensors are traceless while the last one has a non-vanishing trace. \\
The three form factors are given by
\bea
F_1(p,q) 
&=& 
\frac{1}{(2\pi)^d}\,\frac{\pi^2}{2(d-2) (p\cdot q^2 -p^2 q^2)^2} \bigg\{
(d-1) p \cdot q (p \cdot q + q^2) (p+q)^2 \, \mathcal B_0((p+q)^2) \nn \\
&-& 
\mathcal B_0(p^2) \bigg[  p^2 ((d-1) p \cdot q^2 + 2 \, p \cdot q
\, q^2 + q^2 ) + (d-2) p \cdot q^2 (2 \, p \cdot q + q^2) \bigg] \nn \\
&-& 
\mathcal B_0(q^2) \bigg[ q^2 \, p \cdot q ((3d-5) p \cdot q + p^2)
- q^4 ((d-3) p^2 - (d-1) p \cdot q) + (d-2) p \cdot q^3 \bigg]  \nn \\
&+& 
(p \cdot q + q^2) (p+q)^2 ((d-2) p \cdot q^2 + p^2 q^2 ) \mathcal C_0(p^2, (p+q)^2, q^2)
\bigg\} \,, \nn
\eea
\bea
F_2(p,q) 
&=& 
\frac{1}{(2\pi)^d}\frac{\pi^2}{2(d-2) (p\cdot q^2 -p^2 q^2)^2} \bigg\{
\mathcal B_0(p^2) \bigg[ (d-1) p^2 \, p \cdot q^2 + (d-2) p \cdot q^3
+ p^2 q^2 \,p \cdot q \bigg] \nn \\
&+&  \mathcal B_0(q^2) \bigg[ (d-1) q^2 \, p \cdot q^2 + (d-2) p \cdot
q^3 + p^2 q^2 \, p \cdot q \bigg] \nn \\
&-&  
p \cdot q \, \mathcal B_0((p+q)^2) \bigg[ (d-1) p \cdot q (p^2+q^2) + (d-2) p^2 q^2 + d \, p \cdot q^2\bigg] \nn \\
&-&  
\frac{(d-2) p\cdot q^2 + p^2 q^2}{d-1} \mathcal C_0(p^2, (p+q)^2, q^2) 
\bigg[ (d-1)p \cdot q (p^2+q^2) + (d-2) p^2 q^2 + d\, p\cdot q^2\bigg]
\bigg\} \,,  \nn \\
F_3(p,q) 
&=&  
\frac{1}{(2\pi)^d}\pi^2 \left( \mathcal B_0(p^2) + \mathcal B_0(q^2)  \right) \,.
\eea
Finally, we present here the expression of the conformal contributions to the $VVV$ with two external legs on mass-shell. This
limit is achieved contracting with the two polarization vectors ($e_{\alpha}(p)\, , e_{\beta}(q)$) and sending the invariants
$p^2$, $q^2$ to zero. The fermion sector, for instance gives
\bea
\Gamma^{VVV_{ferm}}_{\alpha\beta\lambda}(p,q) 
&=&
\frac{1}{(2\pi)^d}\,
\frac{f^{abc}}{(d-2)(d-1)} Ê\bigg\{ d(d-3) \delta_{\alpha\beta}(p-q)_{\lambda} + 2 (d-2)^2 \left( \delta_{\beta \lambda} q_{\alpha} -
Ê\delta_{\alpha \lambda} p_{\beta} \right) \nn \\
&-& \frac{d-4}{p \cdot q} (p-q)_{\lambda} p_{\beta} q_{\alpha}
\bigg\} \pi^2 \, \mathcal B_0(2 p\cdot q, 0,0),
\eeqa
while the scalar sector gives
\bea
\Gamma^{VVV_{scalar}}_{\alpha\beta\lambda}(p,q) 
&=&
- \frac{1}{(2\pi)^d}
\frac{f^{abc}}{(d-2)(d-1)} \bigg\{ \delta_{\alpha\beta}(p-q)_{\lambda} + (d-2)
\left( \delta_{\beta \lambda} q_{\alpha} -  \delta_{\alpha \lambda}
p_{\beta} \right) \nn \\
&+& \frac{d-4}{2\,p \cdot q} (p-q)_{\lambda} p_{\beta} q_{\alpha}
\bigg\} \pi^2  \, \mathcal B_0(2 p\cdot q, 0,0).
\eeqa

\section{Handling any massless correlator: a direct approach in $d$ dimensions} \label{direct}

In the previous sections we have tried to compare perturbative results in free field theory with general ones coming from the
requirement of conformal symmetry imposed on certain correlators. We have also seen that in this case, working backward from
the explicit field theory representation of the
lowest order realization of these correlators, one can match the general solutions. This is the case of the $VVV$, $TVV$ and
$TOO$ correlators in general dimensions, while for the $TTT$ the 4-dimensional solution of the Ward identities is completely
matched by a combination of scalar, vector and fermion sectors. As we consider the same 3-graviton vertex in d-dimensions, the
vector contribution is not conformally invariant, and therefore the combination of the scalar and the fermion sectors does not
match the most general $d$-dimensional solution. Checking the finiteness of the general solution and formulating it entirely in
momentum space are not obvious steps, since a correlator such as the $TTT$, once expanded, contains several hundreds of terms.
For this reason we are going to illustrate an algorithm that allows to compute vertices of such a complexity using
a direct approach. Our analysis
will be formulated in general but illustrated with few examples only up to correlators of rank-4. The algorithm has been
implemented for the same vertex in a symbolic manipulation program, which is available upon request.

\begin{itemize}
\item{\bf The steps}
\end{itemize}
As we have already mentioned in the previous sections, given any correlator, we can formulate a general procedure which allows
us to transform its expression to momentum space, with the following steps: \\
\begin{itemize}
\item{\bf 1)} expansion of the correlator into its single tensor components;

\item{\bf 2)} rewriting of each component in terms of some ``R-substitutions", that we will define below;

\item{\bf 3)} application of the dimensional shift $d\to d- 2 \omega$ which can be performed generically in the expression
    resulting from point 2); and
\item{\bf 4)}  implementation of the transform. The transform is implemented by Eq. (\ref{fund}) for each single difference
    $x_{ij}= x_i - x_j$. For correlators of higher orders, say of rank $n$ ($n > 3$), the transform is used $n$ times.
\end{itemize}
As we are going to describe below, this method and the regularization imposed by the dimensional shift allows to test quite
straighforwardly the integrability of any correlator,  a point already emphasized in \cite{Osborn:1993cr} where this
regularization has been first introduced. The transform can be applied in several independent ways. These features share
some similarities with the so called ``method of uniqueness" (see for instance \cite{Kazakov:1986mu}) used for massless
integrals in momentum or in configuration space.

\subsection{Pulling out derivatives}\label{pull}

One of the main steps that we will follow in the computation of the transform of the x-space expression of the correlators
consists in the rewriting of a given x-space tensor in terms of derivatives of other terms. We call this rule a ``derivative
relation." It allows one to reduce the degree of singularity of a given tensor structure, when the variables are coincident, in the
spirit of differential regularization. Differently from the standard approach given by differential regularization, which is
4-dimensional, we will be working in $d$ dimensions. We will be using the term ``integrable" to refer to expressions
for which the Fourier transform exists and that are well defined in $d$-dimensions, although they may be singular in d=4.
Derivative relations, combined with the basic transform
\bea \label{fund}
\frac{1}{(x^2)^\alpha}&=& \frac{1}{4^\alpha \pi^{d/2}}\,\frac{\Gamma(d/2 - \alpha)}{\Gamma(\alpha)} \,\int d^d l \, \frac{e^{i
l\cdot x}}{(l^2)^{d/2 - \alpha}}
\equiv C(\alpha) \, \int d^d l \, \frac{e^{i l\cdot x}}{(l^2)^{d/2 - \alpha}} \nonumber \\
C(\alpha) &=&\frac{1}{4^{\alpha}\,\pi^{d/2}} \frac{\Gamma(d/2 - \alpha)}{\Gamma(\alpha)}
\eea
allow one to perform a direct mapping of these correlators to momentum space.
We proceed with a few examples to show how the lowering of the singularity takes place.

We start from tensors of rank-1.
At this rank we use the relation
\bea
\frac{x^\mu}{(x^2)^\alpha}
&=&
-\frac{1}{2 (\alpha -1)} \partial_\mu \frac{1}{(x^2)^{\alpha -1}} \nonumber \\
&=&
- \frac{i}{2^{2\alpha-1} \pi^{d/2}} \, \frac{\Gamma(d/2 +1 - \alpha) }{\Gamma(\alpha) } \, \int d^d l \, e^{i l\cdot x}\frac{
l_\mu}{(l^2)^{d/2 - \alpha +1}} \,
\eea
to extract the derivative, where in the last step we have used (\ref{fund}).
Notice that by using (\ref{fund}) with $\alpha=d/2-1$ one can immediately obtain the equation
\beq
\Box \frac{1}{(x^2)^{d/2-1}}= - \frac{4\,\pi^{d/2 }}{\Gamma(d/2-1)} \, \delta^{(d)}(x)
\label{deltaeq}
\eeq
which otherwise needs Gauss' theorem to be derived.

Scalar 2-point functions describing loops in x-space are next in difficulty.
As an illustration, consider the generalized 2-point function
\beq
\frac{1}{[(x-y)^2]^\alpha [(x-y)^2]^\beta}\, .
\eeq
%
%
Using (\ref{fund}) separately for the $1/[(x-y)^2]^\alpha$ and the $1/[(x-y)^2]^\beta$ factors,
the Fourier transform ($\mathcal{F T}$) of this expression is found to be
\bea\label{FourierB0General}
\mathcal{FT} \left[ \frac{1}{[(x-y)^2]^\alpha [(x-y)^2]^\beta} \right]
&\equiv&
\int\, d^d x \, d^d y \, \frac{e^{- i ( p\cdot x + q \cdot y )}}{[(x-y)^2]^\alpha [(x-y)^2]^\beta}
\nonumber \\
&=&
(2\pi)^{2d} \, C(\alpha) \, C(\beta)\, \int d^d l \, \frac{1}{[l^2]^\alpha [( l+p)^2 ]^\beta}\, .
\eea
Uniqueness allows to reformulate the transform by combining the powers of the propagators into a single factor
\beq
\mathcal{FT}\left[\frac{1}{[(x-y)^2]^{\alpha + \beta}}\right] =
(2\pi)^{2d} \, \frac{C(\alpha + \beta)}{(p^2)^{d/2 - \alpha - \beta}},
\label{unsplit}
\eeq
giving, for consistency, a functional relation for the integral in (\ref{FourierB0General})
\bea \label{B0general}
\int\,d^dl \, \frac{1}{[l^2]^\alpha[(l+p)^2]^\beta} =
\frac{C(\alpha + \beta)}{C(\alpha) \, C(\beta)} \frac{1}{(p^2)^{d/2 - \alpha - \beta}} =
\pi^{d/2}\, \frac{\Gamma(d/2-\alpha)\Gamma(d/2-\beta)\Gamma(\alpha+\beta-d/2)}{\Gamma(\alpha)\Gamma(\beta)\Gamma(d-\alpha-\beta)}\, 
\frac{1}{(p^2)^{\alpha+\beta-d/2}}.\nonumber \\
\eea
In the $TT$ and $TVV$ cases, x-space expressions such as $x^{\mu_1}...x^{\mu_n}/ (x^2)^\alpha$ up to rank-4 are common, and the
use of derivative relations - before proceeding with their final transform to momentum space - can be done in several ways.
Also in this case, as for the scalar functions, uniqueness shows that the result does not depend on the way we combine the factors
at the denominators with the corresponding numerators.

To deal with tensor expressions in position space we introduce some notation.
We denote by
\beq
{R^n}_{\mu_1,\dots, \mu_n}(x,\alpha)
\equiv
\frac{x_{\mu_1}, \dots x_{\mu_n}}{(x^2)^\alpha} \, ,
\eeq
the ratio between a generic tensor monomial in the vector $x$ and a power of $x^2$.
We do so to denote in a compact way the tensor structures that appear in the expansion of any correlator.
We call these expressions ``R-terms". \\
After some differential and algebraic manipulation we can easily derive the first four R-terms as
\bea
{R^1}_\mu(x,\alpha)
&=&
-\frac{1}{2 \, (\alpha-1)} \, \pd_\mu \, \frac{1}{(x^2)^{\alpha-1}} \, , \nonumber \\
{R^2}_{\mu\nu}(x,\alpha)
&=&
\frac{1}{4 \, (\alpha-2) \, (\alpha-1)} \, \pd_\mu\,\pd_\nu \, \frac{1}{(x^2)^{\alpha-2}}
+ \frac{\delta_{\mu\nu}}{2\,(\alpha-1)} \,
\frac{1}{(x^2)^{\alpha-1}} \, , \nonumber \\
{R^3}_{\mu\nu\rho}(x,\alpha)
&=&
- \frac{1}{8 (\alpha-3)(\alpha-2)(\alpha-1)} \,\pd_\mu\,\pd_\nu \, \pd_\rho \frac{1}{(x^2)^{\alpha-3}} +
\frac{1}{2 (\alpha-1)}\, \big[ \delta_{\mu\nu}{R^1}_\rho
+ \delta_{\mu\rho}{R^1}_\nu + \delta_{\nu\rho}{R^1}_\mu \big](x,\alpha-1)\, ,
\nonumber \\
{R^4}_{\mu\nu\rho\sigma}(x,\alpha)
&=&
\frac{1}{16(\alpha-4)(\alpha-3)(\alpha-2)(\alpha-1)} \,
\pd_\mu \, \pd_\nu \, \pd_\rho \, \pd_\sigma \, \frac{1}{(x^2)^{\alpha-4}} \nonumber \\
&+&
\frac{1}{2(\alpha-1)} \, \big[
\delta_{\mu\nu} {R^2}_{\rho\sigma} + \delta_{\rho\sigma} {R^2}_{\mu\nu} +
\delta_{\mu\rho} {R^2}_{\nu\sigma} + \delta_{\nu\sigma} {R^2}_{\mu\rho} +
\delta_{\mu\sigma} {R^2}_{\nu\rho} + \delta_{\nu\rho} {R^2}_{\mu\sigma}
\big](x,\alpha-1)
\nonumber \\
&-&
\frac{1}{4(\alpha-2)(\alpha-1)} \, (\delta_{\mu\nu}\delta_{\rho\sigma} + \delta_{\mu\rho}\delta_{\nu\sigma}
+ \delta_{\mu\sigma}\delta_{\nu\rho})\, \frac{1}{(x^2)^{\alpha-2}} \label{Rterms} \, .
\eea
The use of R-terms allows to extract immediately the leading singularities of the correlators, as we show below.
One can use several different forms of R-substitutions for a given tensor component and the procedure is in fact not unique.
For example, a second rank tensor can be rewritten in R-form in several ways
\bea
\frac{(x -y)_\mu(x-y)_\nu}{[(x-y)^2]^{d+1}}
&=&
{R^2}_{\mu\nu}(x-y,d+1)
\nonumber \\
&=&
{R^1}_\mu(x-y,d/2+1)\, {R^1}_\nu(x-y,d/2)
\nonumber \\
&=&
\frac{1}{(x-y)^2} \, {R^1}_\mu(x-y,d/2) \, {R^1}_\nu(x-y,d/2) \, .
\eea
The derivative relations in the three cases shown above are obviously different, but the transform is unique.
One can also artificially rewrite the numerators at will by introducing trivial identities in position space,
without affecting the final expression of the mapping.
We will be using this method in order to extract some of the logarithmic integrals generated by this procedure.
Obviously, this is possible only if we guarantee an intermediate regularization.
We implement it by a dimensional shift of the exponents of the propagators.
The regulator will allow to smooth out the singularity of the correlators around
the value $\alpha=d/2$, which is the critical value beyond which a function such as $1/[x^2]^{\alpha}$ is not integrable.

The structure of the singularities in x-space of the corresponding scalars and tensor correlators can be
identified using the basic transform. For instance, using (\ref{fund}) for $\alpha=d/2$ one encounters a pole
in the expression of the transform. For this reason we regulate dimensionally in x-space
such a singularity by shifting $d\rightarrow d- 2 \omega$. At the same time we compensate with a regularization scale $\mu$ to
preserve the dimension of the redefined correlator. A similar approach has been discussed in \cite{Dunne:1992ws}, in an attempt to relate differential and dimensional regularization. In our case as in \cite{Osborn:1993cr}, however, $\omega$ is an independent regulator which serves to test integrability in momentum space, and for this reason is combined with a fundamental transform which is given by
\beq
\frac{\mu^{2 \omega}}{[x^2]^{d/2- \omega}} =
\frac{\mu^{2 \omega}}{4^{d/2- \omega} \pi^{d/2}} \, \frac{\Gamma(\omega)}{\Gamma(d/2-\omega)} \, \int d^d l \,
\frac{e^{i l\cdot x}}{[l^2]^{\omega}}
\eeq
that we can expand around $\omega\sim 0$ to obtain
\beq \label{second}
\frac{\mu^{2\omega}}{[x^2]^{d/2- \omega}} = \frac{\pi^{d/2}}{\Gamma(d/2)} \, \delta^{(d)}(x) \, \left[\frac{1}{\omega}
 - \gamma  + \log 4 + \psi(d/2) \right] -
\frac{1}{(4\pi )^{d/2 } \, \Gamma(d/2)} \, \int d^d l \,{e^{i l\cdot x}} \log\left(\frac{l^2}{\mu^2}\right) + O(\omega)\,.
\eeq
The subtraction of this pole in $d$ dimensions is obviously related to the need of redefining correlators
which are not integrable, in analogy with the approach followed in differential regularization.
The most popular example is
$1/[x^2]^2$, which has no transform for $d=4$, but is rewritten in the derivative form as \cite{Freedman:1991tk}
\beq \label{defin}
\frac{1}{x^4}=\Box \, G(x^2),
\eeq
where $G(x^2)$ is defined by
\beq
G(x^2)= \frac{\log x^2 M^2}{x^2} + c \, ,
\eeq
with $c$ being a constant. This second approach can be easily generalized to $d$ dimensions.
One can use derivative relations such as
\beq
\frac{1}{[x^2]^\alpha}= \frac{1}{2(\alpha -1)(2 \alpha - d)} \, \Box \, \frac{1}{[x^2]^{\alpha -1}}
\label{squareeq}
\eeq
which is correct as far as $\alpha\neq d/2$.
For $\alpha=d/2 $ this relation misses the singularity at $x=0$, which is apparent from (\ref{deltaeq}).
For this reason, as far as $\alpha=d/2 - \omega$ Eq. (\ref{squareeq}) remains valid and it can be used
together with (\ref{deltaeq}) and an expansion in $\omega$ to give
\bea \label{third}
\frac{\mu^{2 \omega}}{[x^2]^{d/2- \omega}}
&=&
- \frac{1}{2\,\omega}\frac{\mu^{2 \omega}}{d - 2 - 2 \omega} \, \Box
\frac{1}{[x^2]^{d/2-1-\omega}}
\nonumber \\
&=&
\frac{1}{4 - 2\,d}\left(\frac{1}{\omega} + \frac{2}{d-2} \right)\Box \frac{1}{[x^2]^{d/2-1}}-
\frac{1}{2 (d-2)}\Box\frac{ \log (\mu^2 x^2)}{[x^2]^{d/2-1}} \nonumber \\
&=&
\frac{\pi^{d/2}}{\Gamma(d/2)} \, \left(\frac{1}{\omega}
+ \frac{2}{d-2}\right) \, \delta^{(d)}(x) - \frac{1}{2(d-2)} \, \Box\frac{\log (\mu^2 x^2)}{(x^2)^{d/2-1}}.
\eea
The d-dimensional version of differential regularization ($D f R$) can be obtained by requiring the subtraction of all
the terms in (\ref{third}) which are proportional to $\delta^d(x)$, giving
\beq
\frac{1}{[x^2]^{d/2}}\vline_{DfR}\equiv -\frac{1}{2 (d-2)}\Box\frac{\log (\mu^2 x^2)}{(x^2)^{d/2-1}}.
\eeq
This procedure clearly agrees with the traditional version of differential regularization in $d=4$
\cite{Freedman:1991tk}
\beq
\frac{1}{x^4}\equiv-\frac{1}{4}\Box\frac{
\log(x^2 \mu^2)}{x^2}.
\eeq
Notice that this analysis shows that, according to (\ref{third}), the logarithmic integral in (\ref{second}) is given by
\bea    \label{LogInt}
\int d^d l e^{i l \cdot x}\log \left(\frac{l^2}{\mu^2}\right)
&=&
(2\pi)^d \, \left[ - \gamma + \log 4 + \psi(d/2) - \frac{2}{d-2}  \right] \, \delta^{(d)}(x)
+ \frac{(4\pi)^{d/2}}{2(d-2)}\Gamma(d/2) \, \Box \, \frac{ \log(\mu^2 x^2)}{[x^2]^{d/2-1}}
\nonumber \\
&=&
\frac{(4\pi)^{d/2}}{2(d-2)}\Gamma(d/2) \, \Box \, \frac{ \log(\bar{\mu}^2 x^2)}{[x^2]^{d/2-1}} \, ,
\eea
having redefined the regularization scale properly
\beq \label{massscale}
\log \bar{\mu}^2 = \log \mu^2  + \gamma - \log 4 - \psi(d/2) + \frac{2}{d-2} \, .
\eeq
Notice that a regulated (but singular) correlator can be mapped in several ways into momentum space, with identical results.
For instance, we can take $1/[x^2]^{d/2}$ and use on it Eq. (\ref{fund}) once
\bea \label{transf}
&&
\int d^d x \, e^{i k \cdot x}\frac{1}{[x^2]^{d/2}}
\to
\int d^d x \, e^{i k \cdot x}\frac{\mu^{2\omega}}{[x^2]^{d/2-\omega}} =
\frac{1}{4^{d/2-\omega}\, \pi^{d/2}}\, \frac{\Gamma(\omega)}{\Gamma(d/2- \omega)}\,
\int d^d x\, d^d l\, e^{i (k + l) \cdot x}\frac{\mu^{2\omega}}{[l^2]^{\omega}}
\nonumber \\
&&
= 4^{\omega}\, \pi^{d/2}\, \frac{\Gamma(\omega)}{\Gamma(d/2- \omega)}\,
\frac{\mu^{2\omega}}{[k^2]^{\omega}}\, ,
\eea
twice
\bea
&&
\int d^dx \, \frac{\mu^{2\omega}}{x^2 [x^2]^{d/2-1-\omega}}
= \frac{1}{4^{d/2-\omega} \pi^d}\, \frac{\Gamma(d/2 - 1)\, \Gamma(1+\omega)}{\Gamma(d/2-1-\omega)}
\int d^dx\, d^d l_1\, d^d l_2\,  e^{i (k+ l_1 + l_2) \cdot x} \, \frac{\mu^{2\omega}}{[l_1^2]^{d/2-1} [l_2^2]^{1+\omega}}
\nonumber \\
&=& 4^{\omega}\, \pi^{d/2}\, \frac{\Gamma(\omega)}{\Gamma(d/2- \omega)}\,
\frac{\mu^{2\omega}}{[k^2]^{\omega}}\, ,
\eea
(where in the last step (\ref{B0general}) was used) or any number of times, obtaining the same transform.

As one can easily work out, the use of the dimensional regulator generates, after a Laurent expansion in $\omega$, some
logarithmic integrals
in momentum space. As we shall show, if the $1/\omega$ poles cancel, then these integrals can be avoided, in the sense
that it will be possible to rewrite the correlator in such a way that they are absent. This means that in this case one has
to go back and try to rewrite the correlator in such a way
that it takes a finite form already in position space. In this case the mapping of the correlators to momentum space is similar
to the usual Feynman expansion typical of perturbation theory. The condition of Fourier transformability is in fact necessary
in order to have, eventually, a Lagrangian description of the correlator.
On the other hand, if the same poles do not cancel, then the logarithms are a significant aspect of the correlator which,
for sure, can't be reproduced by a local field theory Lagrangian in any simple way, in particular not by a free field theory.
We have left to appendix \ref{Distributional} a few more examples on the correct handling of these distributional identities.

\subsection{Regularization of tensors}

The regularization of other tensor contributions using this extension of differential regularization can be handled
in a similar and straightforward way.
The use of the derivative relations on the R-terms, that map the tensor structures into derivative of less singular terms,
combined at the last stage with the basic transform, allows to get full control of any correlator and guarantee
their consistent mappings into momentum space.
We provide a few examples to illustrate the procedure.

Consider for instance the tensor structure
\beq
t_{\mu} = \frac{(x-y)_\mu}{[(x-y)^2]^{d/2 +1}} \, ,
\eeq
whose R-form is, trivially,
\beq
t_{\mu}
=
{R^1}_\mu\left( x-y,\frac{d}{2}+1\right)
=
- \frac{1}{d} \, \pd_\mu \, \frac{1}{[(x-y)^2]^{d/2}} \, ,
\eeq
where the derivative is intended with respect to $x-y$.
Now we send $d\to d - 2 \omega$ in the exponent of the denominator, since $d/2$ is a critical value
for the integrability of the exponent, introducing the proper mass scale.
This allows us to use the basic transform (\ref{fund}), getting
\beq
t_\mu(\omega) =
- \frac{i \, \mu^{2\omega}}{(d- 2\omega) \, 4^{d/2-\omega} \, \pi^{d/2}} \, \frac{\Gamma(\omega)}{\Gamma(d/2-\omega)} \,
\int \, d^d l \, \frac{l_\mu}{[l^2]^\omega} \, e^{i l\cdot (x-y)} \, .
\eeq
We can expand in $\omega$ obtaining
\bea
t_\mu(\omega)
&=&
\frac{i}{d \, 2^d \pi^{d/2} \, \Gamma(d/2)}
\bigg[
- \bigg(\frac{1}{\omega} + \frac{2}{d} - \gamma + \log 4 + \psi(d/2)\bigg) \,
\int d^d l\, e^{i l\cdot (x-y)}\,  l_{\mu}
\nonumber\\
&&\hspace{25mm}
+ \int d^d l\, e^{i l\cdot (x-y)} \, l_{\mu}\, \log\left(\frac{l^2}{\mu^2}\right)
\bigg]
+ O(\omega) \nonumber \\
&=& \frac{\pi^{d/2}}{d\,\Gamma(d/2)} \, \pd_\mu
\bigg[
- \bigg( \frac{1}{\omega} + \frac{4(d-1)}{d(d-2)} \bigg) \,
\delta^{(d)}(x-y)
+ \frac{\Gamma(d/2)}{2(d-2)\pi^{d/2}} \, \Box \frac{\log(\bar{\mu}^2(x-y)^2)}{[(x-y)^2]^{d/2-1}}
\bigg] \, ,
\eea
where in the last step we have used (\ref{LogInt}).
Notice that the strength of the singularity has increased from $\delta(x)/\omega$ to $\partial_\mu\delta(x)/\omega$,
due to the higher power $(d/2)$ of the denominator in position space. It is clear that for finite correlators
these singular contributions must cancel.
In general, the introduction of the regulator $\omega$ allows to perform algorithmically all the computations of any lengthy
expression leaving its implementation to a symbolic manipulation program. Obviously, for finite correlators this approach might
look redundant, but it can be extremely useful in order to check the cancellation of all the multiple and single
pole singularities in a very efficient way. We will present more examples of this approach in the next sections.

A more involved example is given by
\beq
t_{\mu\nu} = \frac{(x-y)_\mu (x-y)_\nu}{[(x-y)^2]^{d/2+1}}
\eeq
to which corresponds the regulated expression
\beq
t_{\mu\nu}(\omega)= \frac{\mu^{2\omega}(x-y)_\mu (x-y)_\nu}{[(x-y)^2]^{d/2+1 - \omega}}
\eeq
and a minimal R-form which is given by
\beq
t_{\mu\nu}(\omega) = \mu^{2\omega} \, {R^2}_{\mu\nu}\left(x - y,\frac{d}{2} + 1 - \omega\right).
\eeq
Using the list of replacements given in (\ref{Rterms}), the derivative form of $t_{\mu\nu}$ is given by
\beq
t_{\mu\nu }(\omega)= \frac{\mu^{2\omega}}{(d - 2 - 2 \omega) \, (d - 2\omega)} \, \pd_\mu \, \pd_\nu \,
\frac{1}{[(x-y)^2]^{d/2 - \omega -1}}
+ \frac{\delta_{\mu\nu}}{d + 2 - 2\omega} \, \frac{\mu^{2\omega}}{[(x-y)^2]^{d/2 - \omega}}
\eeq
whose singularities are all contained in the second term, whose Fourier transform is given by
\beq
\mathcal{FT}\bigg[\frac{\delta_{\mu\nu}}{d + 2 - 2\omega} \, \frac{\mu^{2\omega}}{[(x-y)^2]^{d/2 - \omega}}\bigg]
= \frac{1}{\omega} \, \frac{\delta_{\mu\nu}}{2^d \, \pi^{d/2} \, (d+2) \, \Gamma(d/2)}  + O(\omega^0)
\eeq
where we have omitted the regular terms. The procedure therefore allows to identify quite straightforwardly
the leading singularities of any tensor in x-space, giving, in this specific case
\beq
\frac{(x-y)_\mu (x-y)_\nu}{[(x-y)^2]^{d/2+1 - \omega}} \sim
\frac{1}{\omega} \, \frac{\delta_{\mu\nu}}{2^d \, \pi^{d/2} \, (d+2) \, \Gamma(d/2)} \, .
\eeq
We can repeat the procedure for correlators of higher rank. The singularities, after performing all the substitutions,
are proportional to the non-derivative terms isolated by the repeated replacement of Eq. (\ref{Rterms}).

\subsection{Regularization of 3-point functions}

In the case of 3-point functions the analysis of the corresponding singularities can be extracted quite simply.
Let's consider, for instance, the identity
\bea \label{fundFor3}
&&
\mathcal{FT}\bigg[\frac{1}{[(x-y)^2]^{\alpha_1}[(z-x)^2]^{\alpha_2}[(y-z)^2]^{\alpha_3 }}\bigg]
\equiv
\int \, d^d x \, d^d y \, d^d z \,
\frac{e^{- i(k\cdot z+  p\cdot x + q\cdot y)}}{[(x-y)^2]^{\alpha_1}[(z-x)^2]^{\alpha_2}[(y-z)^2]^{\alpha_3 }}
\nonumber \\
&=&
(2\pi)^{3d} \, \prod_{i=1}^{3}\left( \frac{\Gamma(d/2 - \alpha_i)}{4^{\alpha_i} \pi^{d /2}\Gamma(\alpha_i)}\right) \,
\delta^{(d)}(k+p+q)\,
\int \frac{d^dl}{[l^2]^{d/2-\alpha_1} [(l+p)^2]^{d/2-\alpha_2} [(l-q)^2]^{d/2-\alpha_3}}\, ,\nonumber \\
\eea
obtained using the fundamental transform (\ref{fund}), where all the physical momenta $(k,p,q)$ are treated as incoming.
The convention for matching the momenta in (\ref{fund}) with the couples of coordinate is
\beq
l_1 \leftrightarrow x-y\, , \qquad l_2 \leftrightarrow z-x\, , \qquad l_3 \leftrightarrow y-z \, ,
\eeq
and the shift $l \rightarrow l-q$ (which is always possible in a regularized expression) has been performed at the end.\\
It is clear that the prefactor on the r.h.s. of this relation has poles for $\alpha_i = d/2 + n$, with $n\geq 0$.
At the same time the loop integral is asymptotically divergent if $d=\sum_i
\alpha_i $, where it develops a logarithmic singularity. In dimensional regularization such a singularity corresponds to a
single pole in $\epsilon= d -\sum_i \alpha_i$. One can be more specific by discussing further examples
of typical 3-point functions.\\
For instance, consider the tensor structure
\beq
{\mathcal{Q}^1}_{\alpha\beta\mu\nu} =
\frac{(y-z)_\alpha \, (y-z)_\beta \, (y-z)_\mu \, (y-z)_\nu}{[(x-y)^2]^{d/2+1} \, [(z-x)^2]^{d/2-1} \, [(y-z)^2]^{d/2+1}} \, ,
\eeq
which appears in the $TVV$ correlator and can be reduced to its R-form in several ways.
We use a minimal substitution and have
\beq
{\mathcal{Q}^1}_{\alpha\beta\mu\nu} = \frac{1}{[(x-y)^2]^{d/2+1}} \, \frac{1}{[(z-x)^2]^{d/2-1}} \,
{R^4}_{\alpha\beta\mu\nu}\left(y-z,\frac{d}{2}+1\right)
\eeq
and application of the derivative reductions in (\ref{Rterms}) gives
\bea
{\mathcal{Q}^1}_{\alpha\beta\mu\nu}
&=&
\frac{1}{(d-6)\,(d-4)\,(d-2)\,d}\, \frac{1}{[(x-y)^2]^{d/2+1}} \, \frac{1}{[(x-z)^2]^{d/2 - 1}}
\nonumber \\
&\times&
\bigg\{
\pd_\alpha \, \pd_\beta \, \pd_\mu \, \pd_\nu \, \frac{1}{[(y-z)^2]^{d/2 - 3}}
+ (d-6)\,(d-4) \,
\frac{\delta_{\mu\nu}\,\delta_{\alpha\beta} + \delta_{\mu\alpha}\,\delta_{\nu\beta} + \delta_{\mu\beta}\,\delta_{\nu\alpha}}
{[(y - z)^2]^{d/2 - 1}}
\nonumber \\
&+&
(d-6)  \, \left(
  \delta_{\mu\nu}    \, \pd_\alpha \,  \pd_\beta  + \delta_{\alpha\beta} \, \pd_\mu  \, \pd_\nu
+ \delta_{\mu\alpha} \, \pd_\nu    \,  \pd_\beta  + \delta_{\nu\beta}    \, \pd_\mu  \, \pd_\alpha
+ \delta_{\nu\alpha} \, \pd_\mu    \,  \pd_\beta  + \delta_{\mu\beta}    \, \pd_\nu  \, \pd_\alpha
\right) \, \frac{1}{[(y-z)^2]^{d/2-2}}
\bigg\}  \, .
\nonumber\\
\eea
Before moving to momentum space, a quick glance at this equation shows that its transform does not exist.
This appears obvious from the presence of the overall factor ${1}/({[(x-y)^2]^{d/2+1}})$ which needs regularization.
The mapping can be performed using the rules defined above, which give, for instance, for the coefficient of
$\delta_{\mu\nu}\, \delta_{\alpha\beta} + \delta_{\mu\alpha}\, \delta_{\nu\beta} + \delta_{\mu\beta}\, \delta_{\nu\alpha}$,
\bea
&&
\mathcal{F T}
\bigg[
\frac{1}{d\,(d-2)} \,
\frac{\mu^{2\omega}}{[(x-y)^2]^{d/2 + 1 - \omega} \, [(z-x)^2]^{d/2 - 1} \, [(y-z)^2]^{d/2 - 1}}
\bigg]
\nonumber \\
&=&
\frac{(2\pi)^{3d} \, \delta^{(d)}(k+p+q)}{d\,(d-2)} \, \frac{4^{1+\omega}}{(4\pi)^{3d/2}}\,
\frac{\Gamma(\omega-1)}{\Gamma(d/2-1)^2 \,\Gamma(d/2-1-\omega)} \,
\int \, d^d l \, \frac{\mu^{2\omega}}{(l^2)^{\omega-1}\,(l+p)^2\,(l-q)^2}\nonumber
\eea
\bea
&=&
\frac{\delta^{d}(k+p+q)}{d(d-2)}\, \frac{4\,\pi^{3d/2}}{\Gamma(d/2-1)^3}\,
\bigg[
-\frac{1}{\omega}  \int d^dl\, \frac{l^2}{(l+p)^2(l-q)^2}
+\,   \int d^dl\, \frac{l^2\, \log\left(l^2/\bar{\mu}^2\right)}{(l+p)^2(l-q)^2}
\bigg]
+  O(\omega). \nonumber \\
\eea
In a similar way the Fourier transform of the first term is
\bea
&&
\mathcal{F T}\bigg[\frac{1}{(d-6)\,(d-4)\,(d-2)\,d}\, \frac{\mu^{2\omega}}{[(x-y)^2]^{d/2+1-\omega}} \,
\frac{1}{[(z-x)^2]^{d/2-1}} \, \pd_\mu\,\pd_\nu\,\pd_\alpha\,\pd_\beta \, \frac{1}{[(y-z)^2]^{d/2-3}}\bigg] \nonumber \\
&=&
\frac{(2\pi)^{3d}\,\delta^{(d)(k+p+q)}}{(d-6)\,(d-4)\,(d-2)\,d}\, \frac{4^{3+\omega}}{(4\pi)^{3d/2}}\,
\frac{2\,\Gamma(\omega-1)}{\Gamma(d/2-3)\,\Gamma(d/2-1)\,\Gamma(d/2+1-\omega)}
\nonumber \\
&\times&
\int d^dl\, \frac{(l-q)_\alpha\,(l-q)_\beta\,(l-q)_\mu\,(l-q)_\nu}{(l^2)^{\omega-1}\,(l+p)^2\,[(l-q)]^3}
\nonumber \\
&=&
\frac{\delta^{(d)(k+p+q)}}{d\,(d-2)}\,\frac{32\,\pi^{3d/2}}{\Gamma(d/2-1)^3}\,
\bigg[
- \frac{1}{\omega}\, \int d^dl\, \frac{l^2\,(l-q)_\alpha\,(l-q)_\beta\,(l-q)_\mu\,(l-q)_\nu}{(l+p)^2[(l-q)^2]^3}
\bigg]
\nonumber\\
&& \hspace{40mm}
+\, \int d^dl\, \frac{\log\left(l^2/\bar{\mu}^2\right)\, (l-q)_\alpha\,(l-q)_\beta\,(l-q)_\mu\,(l-q)_\nu}
{(l+p)^2[(l-q)^2]^3} + O(\omega)\, ,
\eea
illustrating quite clearly how the general procedure can be implemented.

At this point we pause for some comments. The regularization can be performed by sending $d\to d - 2 \omega$ - with no distinction among the various terms - or, alternatively, one can regulate only the non integrable terms. The two approaches, in a generic
computation, will differ only at
$O(\omega)$ and as such they are equivalent. One can obviously check this by an explicit computation.

Another important point concerns the possibility of performing an explicit computation of the logarithmic integrals.
They are indeed calculable in terms of generalized hypergeometric functions (for general $\omega$), but the small $\omega$
expansion of these functions is rather difficult to re-express as a combination of ordinary functions and polylogs.
This is due to the need of performing a double expansion (in $\epsilon$ and in $\omega$) if we move to $d=4$ and insist, as we
should, on the use of dimensional regularization in the computation of the momentum integrals. This difficulty is attributed to
the absence of simple expansions of hypergeometric functions (ordinary and generalized) about non integer (real)
values of their indices.
However, if the $1/\omega$ terms for a combination of terms similar to those shown above cancel, there are some steps which can
be taken in order to simplify this final part of the computation.

To set the stage for the explicit examples of three point functions treated with this procedure,
we introduce here a systematic short-hand notation to denote the momentum-space integrals.
We define
\bea \label{StandInt}
I_{\mu_1,\dots,\mu_n}(p)
&=&
\int d^d l \, \frac{l_{\mu_1}\, \dots \, l_{\mu_n}}{l^2\,(l+p)^2}\, ,
\nonumber\\
J_{\mu_1,\dots,\mu_n}(p_1,p_2)
&=&
\int d^d l \, \frac{l_{\mu_1}\, \dots \, l_{\mu_n}}{l^2\,(l+p_1)^2(l+p_2)^2}\, ,
\nonumber \\
IL_{\mu_{1}\dots\mu_{n}}(p_1,p_2,p_3)
&=&
\int d^dl \, \frac { l_{\mu_1}\, \dots l_{\mu_n} \, \log \left((l+p_1)^2/\mu^2\right)}{(l+p_2)^2(l+p_3)^2} \, ,
\nonumber\\
ILL_{\mu_{1}\dots\mu_{n}}(p_1,p_2,p_3,p_4)
&=&
\int d^d l \frac{l_{\mu_1}\, \dots l_{\mu_n} \,
\log \left((l+p_1)^2/\mu^2\right)\log \, \left((l+p_2)^2/\mu^2\right)}{(l+p_3)^2(l+p_4)^2}\, .
\eea
For correlators which are finite, the double logarithmic contributions will appear in combinations
that can be re-expressed in terms of ordinary Feynman integrals.

\subsection{Application to the $VVV$ case}

To illustrate the way to proceed in general, we reconsider the $VVV$ case, that we know to be integrable, but treated this time
with the general algorithm. We expand the correlator and perform the R-substitutions (\ref{Rterms}). The direct algorithm gives
an expression which is not immediately recognized as being integrable

\bea f^{abc} \, \bigg\{\frac{(a - 2 \, b)}{(d-2)^3} \,
&\times&
\bigg[ \pd^{31}_\mu \, \frac{1}{(x^2_{31})^{d/2-1}}
\, \pd^{12}_\nu \, \frac{1}{(x^2_{12})^{d/2-1}} \, \pd^{23}_\rho \, \frac{1}{(x^2_{23})^{d/2-1}} \nonumber \\
&+&
\pd^{12}_\mu \, \frac{1}{(x^2_{12})^{d/2-1}} \, \pd^{23}_\nu \, \frac{1}{(x^2_{23})^{d/2-1}} \, \pd^{31}_\rho \,
\frac{1}{(x^2_{31})^{d/2-1}}    \bigg] \nonumber \\
+  \frac{a}{d \, (d-2)^2} \,
&\times&
\bigg[ \frac{1}{(x^2_{12})^{d/2-1}} \left(
\pd^{31}_\mu \, \frac{1}{(x^2_{31})^{d/2-1}} \, \pd^{23}_\nu \, \pd^{23}_\rho \, \frac{1}{(x^2_{23})^{d/2-1}}  +
\pd^{23}_\nu \, \frac{1}{(x^2_{23})^{d/2-1}} \, \pd^{31}_\mu \, \pd^{31}_\rho \, \frac{1}{(x^2_{31})^{d/2-1}}  \right)
\nonumber \\
&+&
\frac{1}{(x^2_{23})^{d/2-1}} \left(
\pd^{31}_\rho \, \frac{1}{(x^2_{31})^{d/2-1}} \, \pd^{12}_\mu \, \pd^{12}_\nu \, \frac{1}{(x^2_{12})^{d/2-1}} +
\pd^{12}_\nu \, \frac{1}{(x^2_{12})^{d/2-1}} \, \pd^{31}_\mu \, \pd^{31}_\rho \, \frac{1}{(x^2_{31})^{d/2-1}}  \right)
\nonumber \\ &+&
\frac{1}{(x^2_{31})^{d/2-1}} \left(
\pd^{23}_\rho \, \frac{1}{(x^2_{23})^{d/2-1}} \, \pd^{12}_\mu \, \pd^{12}_\nu \, \frac{1}{(x^2_{12})^{d/2-1}} +
\pd^{12}_\mu \, \frac{1}{(x^2_{12})^{d/2-1}} \, \pd^{23}_\nu \, \pd^{23}_\rho \, \frac{1}{(x^2_{23})^{d/2-1}}  \right) \bigg]
\nonumber \\
- \frac{1}{d-2} \, \left( b - \frac{a}{d+2} \right) \,
&\times&
\bigg[  \frac{1}{(x^2_{31})^{d/2-1}} \, \left(
\frac{\delta_{\mu\nu}}{(x^2_{12})^{d/2}} \, \pd^{23}_\rho \, \frac{1}{(x^2_{23})^{d/2-1}} +
\frac{ \delta_{\nu\rho}}{(x^2_{23})^{d/2}} \, \pd^{12}_\mu\, \frac{1}{(x^2_{12})^{d/2-1}}\right) \nonumber \\
&+&
\frac{1}{(x^2_{23})^{d/2-1}} \, \left(
\frac{\delta_{\mu\nu}}{(x^2_{12})^{d/2}} \, \pd^{31}_\rho \, \frac{1}{(x^2_{31})^{d/2-1}} +
\frac{\delta_{\mu\rho}}{(x^2_{31})^{d/2}} \, \pd^{12}_\nu\, \frac{1}{(x^2_{12})^{d/2-1}} \right) \nonumber \\ &+&
\frac{1}{(x^2_{12})^{d/2-1}} \, \left(
\frac{\delta_{\mu\rho}}{(x^2_{31})^{d/2}} \, \pd^{23}_\nu \, \frac{1}{(x^2_{23})^{d/2-1}} +
\frac{\delta_{\nu\rho}}{(x^2_{23})^{d/2}} \, \pd^{31}_\mu\, \frac{1}{(x^2_{31})^{d/2-1}}\right)  \bigg] \bigg\} \, . \eea
The apparent non-integrability is due to terms of the form ${1}/{(x^2_{ij})^{d/2}}$ in the last addend. For this reason,
ignoring any further information, to test the approach we proceed with a regularization of the non-integrable terms.
The expression in momentum space is obtained by sending
$d\to d - 2\omega$ in all the terms of the form ${1}/{(x^2_{ij})^{d/2}}$. Expanding in $\omega$ the result, one can show that,
as expected, the $1/\omega$ terms cancel, proving its integrability. We fill in few more details to clarify this point.
A typical not manifestly integrable term in $VVV$ is
\beq
\frac{1}{(x^2_{31})^{d/2-1}} \,
\frac{1}{(x^2_{12})^{d/2}} \, \pd^{23}_\rho \, \frac{1}{(x^2_{23})^{d/2-1}} +
\frac{1}{(x^2_{23})^{d/2-1}} \,
\frac{1}{(x^2_{12})^{d/2}} \, \pd^{31}_\rho \, \frac{1}{(x^2_{31})^{d/2-1}}
\eeq
which in momentum space after $\omega$ regularization gives (omitting an irrelevant constant)
\beq
\mu^{2 \omega}\, \Gamma(\omega)\, \int d^d l\, \frac{ 2 l^\rho - q^\rho}{(l^2) (l-q)^2 [(l+p)^2]^{\omega}}.
\eeq
Expanding in $\omega$, the residue at the pole is given by the integral
\beq
\int d^d l\frac{2 l^\rho - q^\rho}{l^2 (l-q)^2}
\eeq
which vanishes in dimensional regularization.
The finite term is logarithmic and it is given by
\beq
\int d^d l\frac{\log \left((l+p)^2/\mu^2\right)\, (2 l^\rho - q^\rho)}{l^2 (l-q)^2}.
\eeq
The scale dependence also disappears, since the $\log\mu^2$ term is also multiplied by the same vanishing integral. Obviously, the
nontrivial part of the computation is in the appearance of a finite logarithmic integral which, due to the finiteness of the
correlator, has to be re-expressed in terms of other non-logarithmic contributions, i.e. of ordinary Feynman integrals.
There is no simple way to relate one single integral to an ordinary non-logarithmic contribution unless one performs the entire
computation and expresses the result in terms of special polylogarithmic functions, using consistency.  For correlators which are integrable, however, it is possible to relate two log integrals to regular Feynman integrals.
Single log integrals, at least in this case, can also be evaluated explicitly, as we illustrate in an appendix.

By applying the algorithm we get
\bea
&&
\langle {V^a}_{\mu} \, {V^b}_{\nu} \, {V^c}_{\rho} \rangle \, (p,q) =
(2\pi)^{3d} \, \delta^{(d)}(k+p+q) \, i \, f^{abc}
\nn \\
&\times&
\bigg\{
C(d/2-1)^3 \,
\bigg[
\frac{a\,(6-4d)+2\,b\,d}{d(d-2)^3}
\bigg(2 \, J_{\mu\nu\rho}(p,-q) + (p+q)_{\mu} \, J_{\nu\rho}(p,-q) + p_{\nu} \, J_{\mu\rho}(p,-q)
\nonumber \\
&& \hspace{50mm}
- \, q_{\rho} \, J_{\mu\nu}(p,-q) - p_{\nu}\,q_{\mu}\,J_{\rho}(p,-q) - p_{\mu}\,q_{\rho}\,J_{\nu}(p,-q)\bigg)
\bigg]
\nonumber \\
&+&
\frac{a}{d(d - 2)^2} \,
\bigg(- 2 \, \left(p_\mu +  q_\mu\right) \, \big( p_\nu\,J_\rho(p,-q) + q_\rho\,J_\nu(p,-q)\big)
+ q_\rho p_\nu \big(2 \, J_\mu(p,-q) +  (p-q)_\mu \, J(p,-q) \big) \bigg)
\nonumber \\
&-&
\frac{C(d/2-1)^2}{(4\pi)^{d/2}\,\Gamma(d/2)\,(d-2)} \,
\bigg(\frac{a}{d+2} - b \bigg) \,
\bigg[
\delta_{\mu\nu}  \, \bigg( 2 \, IL_\rho(p,0,-q) - q_\rho \, IL(p,0,-q)\bigg) \nonumber \\
&&
\hspace{54mm}
+ \, \delta_{\mu\rho} \, \bigg( 2 \, IL_\nu(-q,0,p)  + p_\nu  \, IL(-q,0,p)\bigg)
\nonumber \\
&&
\hspace{54mm}
+ \, \delta_{\nu\rho} \, \bigg( 2 \, IL_\mu(q,0,k)  + k_\mu  \, IL(q,0,k)\bigg)
\bigg]
\bigg\} \, .
\eea
One can easily show the scale independence of the result, which is related to the finiteness of the expressions
and to the fact that the logarithmic contributions, in this case, are an artifact of the approach.
For this reason, when the scale independence of the regulated expressions has been proved, then one can go back and try to
rewrite the correlator in such a way that it is manifestly integrable. Obviously this may not be a straightforward thing to do,
especially if the expression is given by hundreds of terms in configuration space. If, even after proving the finiteness of
the expression, one is unable to rewrite it in an integrable form, one can always continue applying the algorithm that we have
presented, generating the logarithmic integrals. Pairs of log integrals can be related to ordinary Feynman integrals by applying appropriate
tricks.
We have illustrated in an appendix an example where we discuss the computation of the single log-integral appearing in $VVV$
as an example. In the $TOO$ case one encounters both single and double-log integrals. For non-conformal correlators
these second type of integrals are, in general, expected and turn out to be a characteristic feature of these correlators in momentum space.

\subsection{Direct methods for the $TOO$ case and double logs}

A similar analysis can be pursued in the $TOO$ case. Also for this correlator we can apply a direct approach in order to show
the way to proceed in the test of its regularity. Using our basic transform (\ref{fundFor3}) and introducing the regulator
$\omega$ to regulate the intermediate singularities we can easily transform it to momentum space
\bea
&&
\mathcal{FT}\bigg[ \langle T_{\mu\nu}(x_1) \, O(x_2) ,\ O(x_3) \rangle  \bigg]
\equiv
\langle T_{\mu\nu} \, O \, O  \rangle(p,q) =
(2\pi)^{3d} \, \delta^{(d)}(k+p+q) \, a
\nonumber\\
&\times&
\bigg\{
\frac{C(d/2-1)^3}{d\,(d-2)^2}\,
\bigg[
- 4\, (d-1)\, J_{\mu\nu}(p,-q) - 2 \, (d-1) \, \bigg( (q_\nu - p_\nu)\, J_\mu(p,-q)  + (q_\mu - p_\mu)\, J_\nu(p,-q) \bigg)
\nonumber \\
&+&
\bigg(d\,(p_\mu q_\nu + p_\nu q_\mu ) - (d-2)\, (p_\mu p_\nu + q_\mu q_\nu) \bigg) \, J(p,-q)
\bigg]
\nonumber \\
&+&
\frac{C(d/2-1)^2\, C(d/2-\omega)}{d}\, \delta_{\mu\nu}\,
\bigg(\int\,d^dl\,\frac{\mu^{2\omega}}{l^2[(l+p)^2]^{\omega}(l-q)^2}
    + \int\,d^dl\,\frac{\mu^{2\omega}}{l^2(l+p)^2[(l-q)^2]^{\omega}} \bigg)
\nonumber \\
&-&
\frac{C(d/2-1)\, C(d/2-\omega)^2}{d}\, \delta_{\mu\nu}\,
\int\,d^dl\,\frac{(\mu^{2\omega})^2}{[l^2]^2[(l+p)^2]^{\omega}[(l-q)^2]^{\omega}}
\bigg\} \, .
\eea
The expression above is affected by double and single poles in $\omega$ once we perform an expansion in this parameter,
which are expected to vanish in order to guarantee a finite result.

The coefficient of the double pole is easily seen to take the form
\beq
- \delta_{\mu\nu}\, \frac{a\, (2\,\pi^2)^d\, d\, C(d/2-1)}{\Gamma(d/2)^2} \, I(0) \, ,
\eeq
where the integral vanishes in dimensional regularization, being a massless tadpole. \\
The coefficient of the simple pole is instead given by
\bea \label{simple}
&&
\delta_{\mu\nu}\, \frac{4^d\, \pi^{5d/2}\, C(d/2-1)^2}{d\, \Gamma(d/2)}\,
\bigg\{
\frac{1}{\Gamma(d/2-2) \, \Gamma(d/2)^2} \,
\bigg[
2 \, \bigg( \gamma - \log 4 - \psi(d/2) \bigg) \, I(0)
\nonumber \\
&+&
\bigg( IL(p,0,0) + IL(-q,0,0)  \bigg) \bigg]  +
\frac{1}{\Gamma(d/2-1)^2 \, \Gamma(d/2)} \,
\bigg[
I(p) + I(q)
\bigg]
\bigg\}.
\eea
The first term of (\ref{simple}) vanishes as in the case of the double pole,
while for the remaining contributions we use the relation
\bea\label{IntegralTrick}
IL(p,0,0) = \int d^dl\,\frac{\log\left(\frac{(l+p)^2}{\mu^2}\right)}{[l^2]^2} =
- \frac{\pd}{\pd\omega} \, \int d^d l \,   \frac{\mu^{2\omega}}{[l^2]^2 \, [(l+p)^2]^\omega} \bigg|_{\omega=0} \, .
\eea
It is easy to see that the contributions in the last line in (\ref{simple}) cancel
after inserting the explicit value for the 2-point function in (\ref{B0general}).\\
The finite part of the expression is found to be, after removing some additional tadpoles,
\bea
&&
\langle T_{\mu\nu} \, O \, O  \rangle(p,q) =
(2\pi)^{3d} \, \delta^{(d)}(k+p+q)\, a\,
\nonumber \\
&\times&
\bigg\{
\frac{C(d/2-1)^3}{d\,(d-2)^2}\,
\bigg[
- 4\, (d-1)\, J_{\mu\nu}(p,-q) - 2 \, (d-1) \, \bigg( (q_\nu - p_\nu)\, J_\mu(p,-q)  + (q_\mu - p_\mu)\, J_\nu(p,-q) \bigg)
\nonumber \\
&&\hspace{25mm}
+\, \bigg(d\,(p_\mu q_\nu + p_\nu q_\mu ) - (d-2)\, (p_\mu p_\nu + q_\mu q_\nu) \bigg) \, J(p,-q) \bigg]
\nonumber \\
&-&
\delta_{\mu\nu}\,
\bigg[
\frac{C(d/2-1)^2}{d\,\pi^{d/2}\,2^{d}\,\Gamma(d/2)} \,
\bigg( \left(\gamma -\log 4 -\psi(d/2)\right)\, \big(I(p) + I(-q) \big) + \big(IL(p,0,-q) + IL(-q,0,p) \big) \bigg)
\nonumber \\
&& \hspace{-2mm}
+ \, \frac{C(d/2-1)}{3\,d\,2^{2d+1}\, \pi^d\,\Gamma(d/2)^2}\,
\bigg(
12\, \left(\gamma - \log 4 - \psi(d/2)\right)\, \big(IL(p,0,0) + IL(-q,0,0)\big)
\nonumber \\
&&\hspace{33mm}
+\, 3 \, \big(ILL(p,p,0,0) + 2\, ILL(p,-q,0) + ILL(-q,-q,0,0) \big)
\bigg)
\bigg]
\bigg\}
\eea
where now also double logarithmic integrals have appeared. Using the relations (\ref{B0general}) and (\ref{IntegralTrick}), the terms proportional to
$(\gamma - \log 4 - \psi(d/2))$, which are just a remain of the regularization procedure,
cancel out, leaving us with the simplified result
\bea
\label{toologs}
&&
\langle T_{\mu\nu} \, O \, O  \rangle(p,q) =
(2\pi)^{3d} \, \delta^{(d)}(k+p+q)\, a\,
\nonumber \\
&\times&
\bigg\{
\frac{C(d/2-1)^3}{d\,(d-2)^2}\,
\bigg[
- 4\, (d-1)\, J_{\mu\nu}(p,-q) - 2 \, (d-1) \, \bigg( (q_\nu - p_\nu)\, J_\mu(p,-q)  + (q_\mu - p_\mu)\, J_\nu(p,-q) \bigg)
\nonumber \\
&&\hspace{25mm}
+\, \bigg(d\,(p_\mu q_\nu + p_\nu q_\mu ) - (d-2)\, (p_\mu p_\nu + q_\mu q_\nu) \bigg) \, J(p,-q) \bigg]
\nonumber \\
&-&
\delta_{\mu\nu}\, \frac{C(d/2-1)}{d\,(4\pi)^d\,\Gamma(d/2)^2}
\bigg[
(4\pi)^{d/2}\,\Gamma(d/2) \,C(d/2-1)\, \bigg(\big(IL(p,0,-q) + IL(-q,0,p) \big) \bigg)
\nonumber \\
&& \hspace{30mm}
+ \, (d-4)\, \big(ILL(p,p,0) + 2\, ILL(p,-q,0,0) + ILL(-q,-q,0,0) \big)
\bigg]
\bigg\}.
\eea
It is slightly lengthy but quite straightforward to show that (\ref{toologs}) can be re-expressed in terms of ordinary Feynman integrals. This can be obtained by reducing all the tensor integrals
(logarithmic and non-logarithmic) to scalar forms.  After the reduction, one can check directly that specific combinations of logarithmic integrals can be expressed in terms of ordinary master integrals. In this case these relations hold since the integrands of the logarithmic expansion (linear combinations thereof) are equivalent to non-logarithmic ones, given the finiteness of the correlators. Obviously for a correlator which is not integrable such a correspondence does not exist and the logarithmic integrals cannot be avoided. This would be another signal, obviously, that the theory does not have a realization in terms of a local Lagrangian, since a Lagrangian field theory has a diagrammatic description only in terms of ordinary Feynman integrals.

We conclude this section with few more remarks concerning the treatment of correlators with more general
scaling dimensions $(2 \Delta)$.
For instance one could consider correlators of the  generic form
\beq
\langle O_i(x_i) O_j(x_j)O_k(x_k)\rangle =\frac{\lambda_{ijk}}{
((x_i-x_j)^2)^{\Delta_i + \Delta_j - \Delta_k} ((x_j - x_k)^2)^{\Delta_j
+ \Delta_k - \Delta_i}((x_k-x_i)^2)^{\Delta_k + \Delta_i - \Delta_j}}.
\eeq

In this case their expression in momentum space can be found by applying Mellin-Barnes methods. They can be reconducted to
integrals in momentum space of the form
\beq
J(\nu_1,\nu_2,\nu_3)=\int \frac{d^d l}{(l^2)^{\nu_1}((l-k)^2)^{\nu_2}((l+p)^2)^{\nu_3}}
\eeq
\beq
\nu_1=d/2-\Delta_i - \Delta_j + \Delta_k
\qquad \nu_2=d/2-\Delta_j - \Delta_k + \Delta_i
\qquad \nu_3=d/2-\Delta_k - \Delta_i + \Delta_j
\eeq
which can be expressed \cite{Davydychev:1992xr} in terms of generalized hypergeometric functions $F_4[a,b,c,d;x,y]$
of two variables $(x,y)$, the two ratios of the 3 external momenta. The computation of these integrals  with arbitrary exponents at the denominators is by now standard lore in perturbation theory, with recursion
relations which allow to relate shifts in the exponents in a systematic way. The problem is more involved for correlators which require an intermediate regularization in order to be transformed to momentum space. In this case one can show,  in general, that the pole structure (in $1/\omega$) of these can be worked out closely, but the finite $O(1)$ contributions involve derivatives of generalized hypergeometric functions respect to their indices $a,b,c,d$.
 The latter can be re-expressed in terms of poly-logarithmic functions, which are typical and common in ordinary perturbation theory, only in some cases.  The possibility to achieve this is essentially related to finding simple expansions of the hypergeometric functions around non integer (and not just rational) indicial points.
 For integrable correlators the analysis of Mellin-Barnes methods remains, however, a significant option,
 which will probably deserve a closer look.


\section{Conclusions}

In this work we have tried to close the gap between two analysis of several CFT correlators, such as the $TVV$ and $TTT$ vertices,
characterized by the presence of one, two and three gravitons on the external lines.
We have tried to map position space and momentum space approaches, showing their interrelation. We have used free field theory
realizations of the general solutions of these correlators in order to establish their expression in momentum space. These
expressions, obviously, remain valid for any CFT.
We have also drawn a parallel between the approach to renormalization typical of standard perturbation theory and the same
approach based on the solution of the anomalous Ward identities, as discussed in \cite{Osborn:1993cr,Erdmenger:1996yc}.
As a nontrivial test of the equivalence of both methods in 4 dimensions, we have verified that the counterterms predicted
by the general analysis in position space coincide with those obtained from momentum space in the Lagrangian predictions
obtained from one-loop  free field theory calculations.

In our approach, based on dimensional regularization, the anomaly is generated by tracing in 4 dimensions the renormalized
vertex, and in some cases, such as in the $TVV$ vertex, it can be thought as due to a single specific tensor structure. This is
characterized by the appearance of an anomaly pole. In the $TTT$ case, the explicit expression of this vertex that we have
presented is the starting point for further analysis. For instance it is a necessary intermediate step in demonstrating the correspondence between general CFT calculations in d-dimensional Euclidean position space, perturbative calculations by Feynman diagrams in momentum space,  and the anomaly effective action of \cite{Riegert:1984kt,MazurMottola:2001,Mottola:2006ew, Mottola:2010}.  
This will remove a possible objection to the anomaly effective action raised in \cite{Erdmenger:1996yc}) by the consistent inclusion of all the terms required by conformal invariance, including the non-anomalous ones for which the anomaly effective action is mute. The origin of an effective massless degree of freedom (an effective ``dilaton-like" field) coupled to gravity in the Standard Model will then be made fully explicit. As we have mentioned, this point has already been proven in the $TVV$ case \cite{Giannotti:2008cv,Armillis:2009pq, Armillis:2010qk}  and is expected on general grounds of anomalous Ward identities and the associated non-trivial cohomology of Weyl transformations \cite{MazurMottola:2001}.

We have also discussed a general algorithm that should prove useful to regulate and map correlators from position space to momentum
space, and we have illustrated how to perform such a mapping in a systematic way with a number of examples. The method can be applied in the analysis of more complex correlators, for which a manifest proof of finiteness may not be available. The power of the
approach has been shown by re-analysing finite conformal correlators investigated in the first part, offering a complete test
of its consistency.

\vspace{1cm}
\centerline{\bf Acknowledgements}
We thank H. Osborn for several exchanges while developing this project. We thank J. Erdmenger, A. Davydychev, B. Konopeltchenko, M. Bianchi and Y. Stanev for discussions.

\appendix

\section{The computation of TTT}
\label{ComputeTTT}
\subsection{Definitions and conventions}

The covariant derivatives of a contravariant vector $A^\mu$ and of a covariant one $B_\mu$ are respectively
\beqa
\nabla_{\nu} A^\mu \equiv \pd_\nu A^\mu + \Gamma^\mu_{\nu\r}A^\r\, ,\\
\nabla_{\nu} B_\mu \equiv \pd_\nu B_\mu - \Gamma^\r_{\nu\mu}B_\r\, ,
\eeqa
with the Christoffel symbols defined as
\beq\label{Christoffel}
\Gamma^{\a}_{\b\g}(z) = \frac{1}{2}g^{\a\k}(z)\left[-\pd_\k g_{\b\g}(z) + \pd_\b g_{\k\g}(z) + \pd_\g g_{\k\b}(z)\right]\, .
\eeq
Our definition of the Riemann tensor is
\bea \label{Tensors}
{R^\lambda}_{\mu\kappa\nu}
&=&
\pd_\nu \Gamma^\lambda_{\mu\kappa} - \pd_\kappa \Gamma^\lambda_{\mu\nu}
+ \Gamma^\lambda_{\nu\eta}\Gamma^\eta_{\mu\kappa} - \Gamma^\lambda_{\kappa\eta}\Gamma^\eta_{\mu\nu}.
\eea
The Ricci tensor is defined by the contraction $R_{\mu\nu} = {R^{\lambda}}_{\mu\lambda\nu}$ 
and the scalar curvature by $R = g^{\mu\nu}R_{\mu\nu}$.\\
The traceless part of the Riemann tensor in $d$ dimension is the Weyl tensor, 
\beq
C_{\alpha\beta\gamma\delta} = R_{\alpha\beta\gamma\delta} -
\frac{2}{d-2}( g_{\alpha\gamma} \, R_{\delta\beta} + g_{\alpha\delta} \, R_{\gamma\beta}
- g_{\beta\gamma} \, R_{\delta\alpha} - g_{\beta\delta} \, R_{\gamma\alpha} ) +
\frac{2}{(d-1)(d-2)} \, ( g_{\alpha\gamma} \, g_{\delta\beta} + g_{\alpha\delta} \, g_{\gamma\beta}) R\, .
\eeq
and its square, $F^d$, whose $d=4$ realization, called simply $F$, 
appears in the trace anomaly equation (\ref{TraceAnomaly}), is
\beq\label{Geometry1}
F^d \equiv
C^{\alpha\beta\gamma\delta}C_{\alpha\beta\gamma\delta}
=
R^{\alpha\beta\gamma\delta}R_{\alpha\beta\gamma\delta} -\frac{4}{d-2}R^{\alpha\beta}R_{\alpha\beta}+\frac{2}{(d-2)(d-1)}R^2
\eeq
The Euler density is instead
\beq\label{Euler}
G =
R^{\alpha\beta\gamma\delta}R_{\alpha\beta\gamma\delta} - 4\,R^{\alpha\beta}R_{\alpha\beta} + R^2\, .
\eeq
The functional variations with respect to the metric tensor are computed using the relations
\beqa\label{Tricks}
\delta \sqrt{-g} = -\frac{1}{2} \sqrt{-g}\, g_{\a\b}\,\delta g^{\a \b}\quad &&
\delta \sqrt{-g} = \frac{1}{2} \sqrt{-g}\, g^{\a\b}\,\delta g_{\a \b}  \nonumber \\
\delta g_{\mu\nu} = - g_{\mu\a} g_{\nu\b}\, \delta g^{\a\b} \quad&&
\delta g^{\mu\nu} = - g^{\mu\a} g^{\nu\b}\, \delta g_{\a\b}\,
\eeqa
The following structure has been repeatedly used throughout the calculations
\beqa\label{Tricks2}
s^{\a\b\g\delta} \, \delta(z,x) &\equiv& - \frac{\d g^{\a\b}(z)}{\d g_{\g\d}(x)} =
\frac{1}{2}\left[\delta^{\a\g}\delta^{\b\delta} + \delta^{\a\delta}\delta^{\b\g}\right]\delta(z,x)\, .
\eeqa

\section{Functional derivation of invariant integrals}
\label{FunctionalIntegral}

In this appendix we briefly show how to evaluate the functional variation of the invariant integral
$\mathcal{I}(a,b,c)$
\beq
\mathcal{I}(a,b,c)
\equiv
\int\,d^d x\,\sqrt{-g}\, K\,
\equiv
\int\,d^d x\,\sqrt{-g}\,
\big(a\,R^{\alpha\beta\gamma\delta}R_{abcd} + b\,R^{\alpha\beta}R_{\alpha\beta} + c\, R^2 \big)\, ,
\eeq
needed to compute the counterterms found in section \ref{Renormalization}. \\

Our index conventions for the Riemann and Ricci tensors are those in (\ref{Tensors}).
We have
\beqa
\delta (R^{\alpha\beta\gamma\delta}R_{\alpha\beta\gamma\delta})
&=&
\delta(g_{\alpha\sigma}g^{\beta\eta}g^{\gamma\zeta}g^{\delta\rho}{R^\alpha}_{\beta\gamma\delta}{R^\sigma}_{\eta\zeta\rho}) \nn\\
&=&
\delta (g_{\alpha\sigma}g^{\beta\eta}g^{\gamma\zeta}g^{\delta\rho}){R^\alpha}_{\beta\gamma\delta}{R^\sigma}_{\eta\zeta\rho}
      + g_{\alpha\sigma}g^{\beta\eta}g^{\gamma\zeta}g^{\delta\rho}\delta
       ({R^\alpha}_{\beta\gamma\delta}{R^\sigma}_{\eta\zeta\rho}) \nn\\
&=&
\delta (g_{\alpha\sigma}g^{\beta\eta}g^{\gamma\zeta}g^{\delta\rho}){R^\alpha}_{\beta\gamma\delta}{R^\sigma}_{\eta\zeta\rho}
      + \,2\,\delta ({R^\alpha}_{\beta\gamma\delta}){R_\alpha}^{\beta\gamma\delta}\, ,
\eeqa
Using (\ref{Tricks}) and (\ref{Tricks2}) and the product rule for derivatives one easily finds out that
the variation can be written at first as
\beqa
\delta \mathcal I(a,b,c)
&=& \int\,d^dx\,\sqrt{-g}\,\bigg\{ \bigg[ \frac{1}{2}g^{\mu\nu}K - 2a\, R^{\mu\alpha\beta\gamma}{R^\nu}_{\alpha\beta\gamma}
   - 2b\,R^{\mu\alpha}{R^\nu}_\alpha - 2c\,R R^{\mu\nu}\bigg]\delta g_{\mu\nu}\nn\\
&&\hspace{20mm}
+ \, 2a\, {R_\alpha}^{\beta\gamma\delta}\delta {R^\alpha}_{\beta\gamma\delta}
+ 2b\, R^{\alpha\beta}\delta R_{\alpha\beta} + 2c\, R g^{\alpha\beta}\delta R_{\alpha\beta}\bigg\}\, .
\eeqa
Exploiting  the Palatini identities,
\beq \label{Palatini}
\delta {R^\alpha}_{\beta\gamma\delta}
=
(\delta\Gamma^\alpha_{\beta\gamma})_{;\delta} - (\delta\Gamma^a_{\beta\delta})_{;\gamma} \quad \Rightarrow \quad
\delta R_{\beta\delta}
=
(\delta\Gamma^\lambda_{\beta\lambda})_{;\delta} - (\delta\Gamma^\lambda_{\beta\delta})_{;\lambda}\, ,
\eeq
and the Bianchi identities we get
\beqa\label{Bianchi}
R_{\alpha\beta\gamma\delta;\eta} + R_{\alpha\beta\eta\gamma;\delta} + R_{\alpha\beta\delta\eta;\gamma}
&=& 0
\quad \Rightarrow \quad
R_{\beta\delta;\eta} - R_{\beta\eta;\delta} + {R^\gamma}_{\beta\delta\eta;\gamma} =  0 \nn \\
\Rightarrow \quad
R_{;\delta}
&=&
2\,{R^\alpha}_{\delta;\alpha}
\quad \Leftrightarrow \quad
\big( R^{\alpha\beta} - \frac{1}{2}g^{\alpha\beta} R \big)_{;\beta} = 0 \, .
\eeqa
After an integration by parts and a reshuffling of indices we get
\beqa\label{deltaISecond}
\delta \mathcal I(a,b,c)
&=&
\int\,d^dx\,\sqrt{-g}\,\bigg\{\bigg[ \frac{1}{2}g^{\mu\nu}K - 2\big( a\, R^{\mu\alpha\beta\gamma}{R^\nu}_{\alpha\beta\gamma}
                                     + b\,R^{\mu\alpha}{R^\nu}_\alpha + c\,R R^{\mu\nu}\big)\bigg]\delta g_{\mu\nu}\nn\\
&&\hspace{20mm}
+ \, 4a\,g_{\gamma\beta}g^{\delta\eta}(\delta\Gamma^c_{\alpha\delta})_{;\eta}
- (4a+2b)\,(\delta \Gamma^\gamma_{\alpha\beta})_{;\gamma}
+ (4c+2b)\, (\delta \Gamma^\lambda_{\alpha\lambda})_{;\beta}\bigg\}.\nn\\
\eeqa
The variations of the Christoffel symbols and of their covariant derivatives
in terms of covariant derivatives of the metric tensors variations are
\beqa\label{deltaChristoffel}
\delta \Gamma^\alpha_{\beta\gamma}
&=&
\frac{1}{2}\,g^{\alpha\delta}\big[-(\delta g_{\beta\gamma})_{;\delta}
+ (\delta g_{\beta\delta})_{;\gamma} +(\delta g_{\gamma\delta})_{;\beta} \big]\, ,\nn\\
(\delta\Gamma^\alpha_{\beta\gamma})_{;\delta}
&=&
\frac{1}{2}\,g^{\alpha\eta}\big[-(\delta g_{\beta\gamma})_{;\eta;\delta} + (\delta g_{\beta\eta})_{;\gamma;\delta}
                                + (\delta g_{\gamma\eta})_{;\beta;\delta} \big]\, .
\eeqa
Now we use them to rewrite (\ref{deltaISecond}) as
\beqa\label{Bivio}
\delta \mathcal I(a,b,c)
&=&
\int\,d^dx\,\sqrt{-g}\bigg\{\bigg[ \frac{1}{2}g^{\mu\nu}K - 2\big( a\, R^{\mu\alpha\beta\gamma}{R^\nu}_{\alpha\beta\gamma}
                                  + b\,R^{\mu\alpha}{R^\nu}_\alpha + c\,R R^{\mu\nu}\big)\bigg]\delta g_{\mu\nu}\nn\\
&&\hspace{18,5mm}
+ \, \bigg[2a\,\big[-(\delta g_{\alpha\delta})_ {;\beta;\gamma} +(\delta g_{\alpha\beta})_ {;\gamma;\delta}
+ (\delta g_{\beta\delta})_ {;\alpha;\gamma} \big]\nn\\
&& \hspace{19mm}
- \, (2a + b)\,\big[- (\delta g_{\alpha\beta})_ {;\delta;\gamma} +(\delta g_{\alpha\delta})_ {;\beta;\gamma}
+ (\delta g_{\beta\delta})_ {;\alpha;\gamma} \big] + (2c + b)\, (\delta g_{\gamma\delta})_ {;\alpha;\beta}\nn\\
&& \hspace{19mm}
- \, 2c\, \big[- (\delta g_{\gamma\delta})_ {;\alpha;\beta} +(\delta g_{\alpha\delta})_ {;\gamma;\beta}
+ (\delta g_{\alpha\gamma})_ {;\delta;\beta} \big]\bigg]g^{\gamma\delta}\,R^{\alpha\beta} \bigg\}\, .
\eeqa
The presence of the factor $g^{cd}R^{ab}$ imposes two symmetry constraints on the terms in the last contribution in square brackets. By adding and subtracting  $-(4a+2b)\,(\delta g_{ac})_{;d;b}$ we obtain the expression
\beqa
\delta \mathcal I(a,b,c)
&=& \int\,d^dx\,\sqrt{-g}\,\bigg\{\bigg[\frac{1}{2}g^{\mu\nu}K
- 2\big( a\, R^{\mu\alpha\beta\gamma}{R^\nu}_{\alpha\beta\gamma}
+ b \, R^{\mu\alpha}{R^\nu}_\alpha + c\,R R^{\mu\nu}\big)\bigg]\delta g_{\mu\nu}\nn\\
&&\hspace{18,5mm}
+ \, \bigg[(4a+2b)\,\big[(\delta g_{\alpha\gamma})_ {;\beta;\delta}
- (\delta g_{\alpha\gamma})_ {;\delta;\beta}\big]+
(4a + b)(\delta g_{\alpha\beta})_ {;\gamma;\delta} + (4c + b)\, (\delta g_{\gamma\delta})_ {;\alpha;\beta}\nn\\
&& \hspace{18,5mm}
- \, (4a+2b+4c)\, (\delta g_{\alpha\gamma})_ {;\delta;\beta}\bigg]g^{\gamma\delta}\,R^{\alpha\beta} \bigg\}\, .
\label{HalfFirstFunctional}
\eeqa
The commutation of covariant derivatives allows us to write
\beqa
g^{\gamma\delta} \big[ (\delta g_{\alpha\gamma})_{;\beta;\delta}
- (\delta g_{\alpha\gamma})_{;\delta;\beta} \big]R^{\alpha\beta}
&=&
g^{\gamma\delta} \big[ -\delta g_{\alpha\sigma} {R^\sigma}_{\gamma\delta\beta}
- \delta g_{\gamma\sigma} {R^\sigma}_{\alpha\beta\delta} \big]R^{\alpha\beta}\nn\\
&=&
g^{\gamma\delta} \big[-s^{\mu\nu}_{\alpha\sigma}{R^\sigma}_{\gamma\beta\delta}
- s^{\mu\nu}_{c\sigma}{R^\sigma}_{\alpha\beta\delta}\big]R^{\alpha\beta} \,
\delta g_{\mu\nu}\nn\\
&=&
(- R^{\mu\alpha}{R^\nu}_\alpha + R^{\mu\alpha\nu\beta}R_{\alpha\beta}) \delta g_{\mu\nu}\, .
\eeqa
Inserting this back into (\ref{HalfFirstFunctional}) we get
\bea
\delta \mathcal I(a,b,c) =
\nonumber\\
&& \hspace{-20mm}
\int\,d^dx\,\sqrt{-g}\,\bigg\{\bigg[ \frac{1}{2}g^{\mu\nu}K
- 2a\, R^{\mu\alpha\beta\gamma}{R^\nu}_{\alpha\beta\gamma} + 4a\,R^{\mu\alpha}{R^\nu}_\alpha
-(4a+2b)\, R^{\mu\alpha\nu\beta}R_{\alpha\beta} - 2c \, R R^{\mu\nu}\bigg]\delta g_{\mu\nu}\nn\\
&&
+ \, \bigg[(4a + b)(\delta g_{\alpha\beta})_ {;\gamma;\delta}
+ (4c + b)\, (\delta g_{\gamma\delta})_ {;\alpha;\beta}
- (4a+2b+4c)\, (\delta g_{\alpha\gamma})_ {;\delta;\beta}\bigg]g^{\gamma\delta}\,R^{\alpha\beta}\bigg\}\, .
\nonumber\\
\eea
If the coefficients are $a = c = 1$ and $b=-4$, i.e. if the integrand is the Euler density,
the last three terms are zero. \\
All that is left to do is a double integration by parts for each one of the last three terms,
to factor out $\delta g_{\mu\nu}$.
This is easily performed and the final result can be written as
\beqa  \label{Magic}
\frac{\delta}{\delta g_{\mu\nu}} \mathcal I(a,b,c)
&=&
\frac{\delta}{\delta g_{\mu\nu}} \int\,d^d x\,\sqrt{-g\,}
\big( a\,R^{\alpha\beta\gamma\delta}R_{\alpha\beta\gamma\delta} + b\,R^{\alpha\beta}R_{\alpha\beta}
+ c\,R^2 \big)\nn\\
&=&
\sqrt{-g}\, \bigg\{\frac{1}{2}g^{\mu\nu}K
- 2a\, R^{\mu\alpha\beta\gamma}{R^\nu}_{\alpha\beta\gamma}
+ 4a\,R^{\mu\alpha}{R^\nu}_\alpha -(4a+2b)\, R^{\mu\alpha\nu\beta}R_{\alpha\beta} - 2c \, R R^{\mu\nu}\nn\\
&& \hspace{8mm}
+ \, (4a + b)\,\Box{R^{\mu\nu}} + (4c + b)\,g^{\mu\nu}{R^{\alpha\beta}}_{;\alpha;\beta}
- (4a+2b+4c){{R^{\nu\beta}}_{;\beta}}^{;\mu}\bigg\}\, .
\eeqa
%


\section{ List of functional derivatives}
\label{Functionals}

We list here the contributions to the trace anomalies for three point function
coming from the elementary quadratic objects. They are given by
\beqa\label{QuadraticFunctionals}
\big[\Box\,R\big]^{\alpha\beta\rho\sigma}(p,q)
&=& \big[g^{\mu\nu}(\pd_\mu\pd_\nu - \Gamma^\lambda_{\mu\nu}\pd_\lambda)R\big]^{\alpha\beta\rho\sigma}(p,q)\nn\\
&=& i^2 \,(p+q)^2\, \big[R\big]^{\alpha\beta\rho\sigma}(p,q)
- \big\{ i^2\, q^\alpha q^\beta - \delta^{\mu\nu}\big[\Gamma^\lambda_{\mu\nu}\big]^{\alpha\beta}(p)\,i\,
q_\lambda\big\}R^{\rho\sigma}(q)\nn\\
&-& \big\{ i^2\, p^\rho p^\sigma - \delta^{\mu\nu}\big[\Gamma^\lambda_{\mu\nu}\big]^{\rho\sigma}(q)\,i \,
p_\lambda\big\}R^{\alpha\beta}(p)\nn\\
&=&(p+q)^2\bigg\{-\frac{1}{2}\delta^{\alpha\beta}\big(p^\rho q^\sigma + p^\sigma q^\rho + 2\,p^\rho p^\sigma\big)
- \frac{1}{2}\delta^{\rho\sigma}\big(q^\alpha p^\beta + q^\beta q^\alpha + 2\,q^\alpha q^\beta \big)\nn\\
&+&
\frac{1}{2} p \cdot q \, \delta^{\alpha\beta}\delta^{\rho\sigma}
+\frac{1}{4}\big(p^\rho q^\beta \delta^{\alpha\sigma} + p^\rho q^\alpha \delta^{\beta\sigma}
+ p^\sigma q^\beta \delta^{\alpha\rho} + p^\sigma q^\alpha \delta^{\beta\rho}\big)\nn\\
&+&
\frac{1}{2}\bigg[\big(q^\rho p^\beta \delta^{\alpha\sigma} + q^\rho p^\alpha \delta^{\beta\sigma}
+ q^\sigma p^\beta \delta^{\alpha\rho} + q^\sigma p^\alpha \delta^{\beta\rho}\big)\nn\\
&+&
\delta^{\alpha\rho}\big(p^\beta p^\sigma + q^\beta q^\sigma \big)
 + \delta^{\alpha\sigma}\big(p^\beta p^\rho + q^\beta q^\rho \big)
 +\delta^{\beta\rho}\big(p^\alpha p^\sigma + q^\alpha q^\sigma \big)\nn\\
&+&
\delta^{\beta\sigma}\big(p^\alpha p^\rho + q^\alpha q^\rho \big)
-\big(\delta^{\alpha\sigma}\delta^{\beta\rho} + \delta^{\alpha\rho}\delta^{\beta\sigma}\big)
\big(p^2 +q^2 + \frac{3}{2}p \cdot q\big)\bigg]\bigg\}\nn\\
&+&
\frac{1}{2}\big(p^2 \delta^{\alpha\beta} - p^\alpha p^\beta\big)\big(p \cdot q \, \delta^{\rho\sigma}
- (p^\rho q^\sigma + p^\sigma q^\rho) - 2\,p^\rho p^\sigma \big)\nn\\
&+& \frac{1}{2}\big(q^2 \delta^{\rho\sigma} - q^\sigma q^\rho\big)\big(p \cdot q \, \delta^{\alpha\beta}
- (p^\alpha q^\beta + p^\beta q^\alpha) - 2\,q^\alpha q^\beta \big)\, .\nn
\eeqa
\beqa
\big[R_{\lambda\mu\kappa\nu}R^{\lambda\mu\kappa\nu}\big]^{\alpha\beta\rho\sigma}(p,q)
&=& 2\,\big[R_{\lambda\mu\kappa\nu}\big]^{\alpha\beta}(p)\big[R^{\lambda\mu\kappa\nu}\big]^{\rho\sigma}(q)\nn\\
&=& p \cdot q\, \big[p \cdot q \big(\delta^{\alpha\rho}\delta^{\beta\sigma} + \delta^{\alpha\sigma}\delta^{\beta\rho}\big)
- \big(\delta^{\alpha\rho}p^{\sigma}q^{\beta} + \delta^{\alpha\sigma}p^{\rho}q^{\beta}\nn\\
&+& \delta^{\beta\rho}p^{\sigma}q^{\alpha} + \delta^{\beta\sigma}p^{\rho}q^{\alpha}\big)\big]
+ 2 \, p^{\rho}p^{\sigma}q^{\alpha}q^{\beta}\, , \nn
\eeqa
\beqa
\big[R_{\mu\nu}R^{\mu\nu}\big]^{\alpha\beta\rho\sigma}(p,q)
&=& 2\,\big[R_{\mu\nu}\big]^{\alpha\beta}(p)\big[R^{\mu\nu}\big]^{\rho\sigma}(q)\nn\\
&=& \frac{1}{4} p \cdot q \big(\delta^{\alpha\rho}p^\beta q^\sigma + \delta^{\alpha\sigma}p^\beta q^\rho
+ \delta^{\beta\rho}p^\alpha q^\sigma + \delta^{\beta\sigma}p^\alpha q^\rho \big)\nn\\
&+&
\frac{1}{2}(p \cdot q)^2 \delta^{\alpha\beta}\delta^{\rho\sigma}
+ \frac{1}{4}p^2 q^2\big(\delta^{\alpha\rho}\delta^{\beta\sigma} + \delta^{\alpha\sigma}\delta^{\beta\rho}\big)\nn\\
&-&
\bigg[\frac{1}{4} p^2\big(q^\alpha q^\rho \delta^{\beta\sigma}+ q^\alpha q^\sigma \delta^{\beta\rho}
+ q^\beta q^\rho \delta^{\alpha\sigma}+ q^\beta q^\sigma \delta^{\alpha\rho}\big)\nn\\
&+&
\frac{1}{2}\delta^{\alpha\beta}\big( p\cdot q\,(p^\rho q^\sigma + p^\sigma q^\rho) - q^2 p^\rho p^\sigma \big)
+ (\alpha,\beta,p)\leftrightarrow(\rho,\sigma,q)\bigg]\, ,\nn
\eeqa
\beqa
\big[R^2\big]^{\alpha\beta\rho\sigma}(p,q)
&=& 2\,\delta^{\mu\nu}\big[R_{\mu\nu}\big]^{\alpha\beta}(p)\delta^{\tau\omega}\big[R_{\tau\omega}\big]^{\rho\sigma}(q)\nn\\
&=& 2\big(p^\alpha p^\beta q^\rho q^\sigma - p^2 q^\rho q^\sigma \delta^{\alpha\beta}
- q^2 p^\alpha p^\beta \delta^{\rho\sigma} + p^2 \, q^2 \delta^{\alpha\beta}\delta^{\rho\sigma} \big)\, ,
\eeqa
The dependence on the momenta is obviously determined by (\ref{3PFMom}).

\section{Vertices}
\label{VERTICES}
 
We have shown in Fig. \ref{listV} a list of all the vertices which
are needed for the momentum space computation of the various correlators in $d$ dimensions.
We list them below: notice that they are computed differentiating the first and second functional 
derivatives of the action, because this allows to keep multi-graviton correlators symmetric (see \ref{3PF}).
\bea
V_{T\phi\phi}^{\mu\nu}(p,q)
&=&
\frac{1}{2}\,p_\alpha \, q_\beta \, C^{\mu\nu\alpha\beta}
+\chi \bigg( \delta^{\mu\nu} \left( p + q \right)^2 - \left( p^\mu + q^\mu \right)\,\left( p^\nu + q^\nu \right) \bigg)
\, ,
\nonumber
\eea
\bea
V_{T \bar{\psi}\psi}^{\mu\nu}(p,q)
&=&
\frac{1}{8} \, A^{\mu\nu\alpha\lambda}\, \gamma_\alpha \,\left(p_\lambda - q_\lambda \right)\, ,
\nonumber
\eea
\bea
V_{TAA}^{\mu\nu\tau\omega}(p,q)
&=&
- \frac{1}{2}\,\bigg[
p \cdot q\, C^{\mu\nu\tau\omega} + D^{\mu\nu\tau\omega}(p,q) + \frac{1}{\xi}E^{\mu\nu\tau\omega}(p,q)
\bigg] = \left(\tilde{V}_{TAA} + \frac{1}{\xi}\,\bar{V}_{TAA}\right)^{\mu\nu\tau\omega}(p,q) ,
\nonumber
\eea
\bea
V_{T\bar{c}c}^{\mu\nu}(p,q)
&=&
V_{T\phi\phi}^{\mu\nu}(p,q)\bigg|_{\chi=0}\, ,
\nonumber
\eea
for the graviton $(T)$- to two scalars $(\phi)$, fermions, photons and ghost pairs. Quadrilinear interactions involving 2
gravitons are far more involved and are given by the expressions
\bea
V_{TT\phi\phi}^{\mu\nu\rho\sigma}(p,q,l)
&=&
  \frac{1}{2}\, p\cdot q s^{\mu\nu\rho\sigma} - \frac{1}{4}\, G^{\mu\nu\rho\sigma}(p,q)
+ \frac{1}{4}\, \delta^{\rho\sigma}\, p_\alpha\, q_\beta\,  C^{\mu\nu\alpha\beta}
\nonumber\\
&+&
\chi \, \bigg\{\bigg[  \bigg(\delta^{\mu\lambda}\,\delta^{\alpha\kappa}\,\delta^{\nu\beta}
+ \delta^{\mu\alpha}\,\delta^{\nu\kappa}\,\delta^{\beta\lambda}
- \delta^{\mu\kappa}\,\delta^{\nu\lambda}\,\delta^{\alpha\beta}
- \delta^{\mu\nu}\,\delta^{\alpha\lambda}\,\delta^{\beta\kappa}\bigg)
s^{\rho\sigma}_{\lambda\kappa}
\nonumber \\
&& \hspace{4mm}
+\, \frac{1}{2} \, \delta^{\rho\sigma} \, \bigg(\delta^{\mu\alpha}\,\delta^{\nu\beta} -
\delta^{\mu\nu}\,\delta^{\alpha\beta}\bigg)\bigg]
\left(p_\alpha \, q_\beta + p_\beta \, q_\alpha + p_\alpha \, p_\beta + q_\alpha \, q_\beta \right)
\nonumber \\
&-&
\bigg[\bigg(\delta^{\mu\alpha}\,\delta^{\nu\beta} - \delta^{\mu\nu}\,\delta^{\alpha\beta}\bigg)
\big[\Gamma^\lambda_{\alpha\beta}\big]^{\rho\sigma}(l)\, \left( -i \, p_\lambda - i \, q_\lambda\right)
+ \frac{1}{2} \, \bigg(\delta^{\mu\alpha}\,\delta^{\nu\beta} - \frac{1}{2}\,\delta^{\mu\nu}\,\delta^{\alpha\beta}\bigg)\,
  \big[R_{\alpha\beta}\big]^{\rho\sigma}(l) \,  \bigg]\bigg\}\, ,
\nonumber
\eea
\bea
\big[\Gamma^\lambda_{\alpha\beta}\big]^{\rho\sigma}(l)
&=&
\frac{1}{2} \delta^{\lambda\kappa} \,
i \, \left[ s_{\alpha\beta}^{\rho\sigma}\, l_\kappa
- s_{\alpha\kappa}^{\rho\sigma}\, l_\beta - s_{\beta\kappa}^{\rho\sigma}\, l_\alpha \right]\, ,
\nonumber \\
\big[R_{\alpha\beta}\big]^{\rho\sigma}(l)
&=&
- i \, l_\alpha\,[\Gamma^\lambda_{\lambda\beta}]^{\rho\sigma}(l)
+ i \, l_\lambda\,[\Gamma^\lambda_{\alpha\beta}]^{\rho\sigma}(l)\, ,
\nonumber
\eea
\bea
V_{TT\bar{\psi}\psi}^{\mu\nu\rho\sigma}(p,q)
&=&
\frac{1}{16} \bigg[ -4 \, s^{\mu\nu\rho\sigma} - 2 \, \delta^{\mu\nu}\,s^{\alpha\lambda\rho\sigma}
+ 2 \, \delta^{\alpha\mu}\,s^{\nu\lambda\rho\sigma} + 2\, \delta^{\alpha\nu}\,s^{\mu\lambda\rho\sigma}
+ \delta^{\mu\lambda}\,s^{\alpha\nu\rho\sigma} + \delta^{\nu\lambda}\,s^{\alpha\mu\rho\sigma}
\nonumber \\
&&\hspace{7mm}
+ \,\delta^{\rho\sigma} \, A^{\mu\nu\alpha\lambda}\,\gamma_{\alpha}\,(p_\lambda - q_\lambda)\bigg] \, ,
\nonumber
\eea
\bea
V_{TTAA}^{\mu\nu\rho\sigma\tau\omega}(p,q,l)
&=&
- \frac{1}{2} \, \bigg\{
\bigg[ B^{\alpha\mu\rho\sigma\beta\lambda\gamma\nu} + \frac{1}{4}\, B^{\mu\nu\rho\sigma\alpha\lambda\gamma\beta} \bigg]\,
{F_{\alpha\beta\gamma\lambda}}^{\tau\omega} (p,q)
+ \frac{1}{\xi}\bigg( H^{\mu\nu\rho\sigma\tau\omega}(p,q,l) + I^{\mu\nu\rho\sigma\tau\omega}(p,q,l) \bigg) \bigg\}
\nonumber\\
&-&
\frac{1}{4}\, \delta^{\rho\sigma}\, \bigg[ p \cdot q \,  C^{\mu\nu\tau\omega}
+ D^{\mu\nu\tau\omega}(p,q) + \frac{1}{\xi}E^{\mu\nu\tau\omega}(p,q) \bigg]
= \big(\tilde{V}_{TTAA}(p,q) + \bar{V}_{TTAA}(p,q,l)\big)^{\mu\nu\rho\sigma\tau\omega}\, ,
\nonumber
\eea
\bea
V_{TT\bar{c}c}^{\mu\nu\rho\sigma}(p,q,l)
&=&
V_{TT\phi\phi}^{\mu\nu\rho\sigma}(p,q,l)\bigg|_{\chi=0}\, ,
\label{MomVertices}
\eea
describing interactions similar to those shown above in the trilinear case, but now with the insertion of one extra graviton.
We have simplified the notation by introducing, for convenience, the tensor components
\bea
s^{\mu\nu\rho\sigma}
&=&
\frac{1}{2} \, (\delta^{\mu\rho} \delta^{\nu\sigma} + \delta^{\mu\sigma} \delta^{\nu\rho})\, ,
\nonumber
\eea
\bea
A^{\mu\nu\alpha\lambda} = 2\,\delta^{\mu\nu}\,\delta^{\alpha\lambda}
- \delta^{\alpha\mu}\,\delta^{\lambda\nu} - \delta^{\alpha\nu}\,\delta^{\lambda\mu}
\nonumber
\eea
\bea
B^{\alpha\mu\rho\sigma\beta\lambda\gamma\nu}
&=&
  s^{\alpha\mu\rho\sigma}    \, \delta^{\beta\lambda} \, \delta^{\gamma\nu}
+ s^{\beta\lambda\rho\sigma} \, \delta^{\alpha\mu}    \, \delta^{\gamma\nu}
+ s^{\gamma\nu\rho\sigma}    \, \delta^{\alpha\mu}    \, \delta^{\beta\lambda}
\nonumber
\eea
\bea
C^{\mu\nu\rho\sigma}
&=&
\delta^{\mu\rho} \delta^{\nu\sigma} + \delta^{\mu\sigma} \delta^{\nu\rho} - \delta^{\mu\nu} \delta^{\rho\sigma}\, ,
\nonumber
\eea
\bea
D^{\mu\nu\rho\sigma} (p,q)
&=&
\delta^{\mu\nu} p^\sigma q^\rho + \delta^{\rho\sigma} \big(p^\mu q^\nu + p^\nu q^\mu\big)
- \delta^{\mu\sigma} p^\nu q^\rho - \delta^{\mu\rho} p^\sigma q^\nu
- \delta^{\nu\sigma}p^\mu q^\rho - \delta^{\nu\rho} p^\sigma q^\mu
\nonumber
\eea
\bea
E^{\mu\nu\rho\sigma} (p, q)
&=&
\delta^{\mu\nu}\, \big[p^\rho p^\sigma + q^\rho q^\sigma + p^\rho q^\sigma \big]
- \big[\delta^{\nu\sigma} p^\mu p^\r + \delta^{\nu\rho}
q^\mu q^\sigma + \delta^{\mu\sigma} p^\nu p^\rho + \delta^{\mu\rho} q^\nu q^\sigma \big]\, ,
\nonumber
\eea
\bea
F^{\mu\nu\rho\sigma\tau\omega} (p, q)
&=&
-  \delta^{\tau\rho}    \delta^{\omega\mu} p^{\sigma} q^{\nu} + \delta^{\tau\rho} \delta^{\omega\nu} p^{\sigma} q^{\mu}
+  \delta^{\tau\sigma}  \delta^{\omega\mu} p^{\rho} q^{\nu} - \delta^{\tau\sigma} \delta^{\omega\nu} p^{\rho} q^{\mu}
+ (\tau,p) \leftrightarrow (\omega,q)
\nonumber
\eea
\bea
G^{\mu\nu\rho\sigma}(p,q)
&=&
  \delta^{\mu\sigma} \big[p^{\rho} q^{\nu}  +  q^{\rho}   p^{\nu}\big]
+ \delta^{\nu\sigma}\big[p^{\rho}  q^{\mu}  +  q^{\rho}   p^{\mu}\big]+
  \delta^{\mu\rho}\big[p^{\sigma}  q^{\nu}  +  q^{\sigma} p^{\nu}\big]
+ \delta^{\nu\rho}\big[p^{\sigma}  q^{\mu}  +  q^{\sigma} p^{\mu}\big]
\nonumber\\
&-&
\delta^{\mu\nu}\big[p^{\rho} q^{\sigma} + q^{\rho} p^{\sigma} \big]
\nonumber
\eea
\bea
H^{\mu\nu\rho\sigma\tau\omega}(p,q,l)
&=&
\bigg[
\bigg(s^{\mu\omega\rho\sigma}\, \delta^{\nu\lambda} + s^{\nu\lambda\rho\sigma}\, \delta^{\mu\omega} \bigg)\, p_\lambda\, p^\tau
+  \delta^{\mu\omega}\,\bigg(s^{\lambda\tau\rho\sigma}\, l^\nu + s^{\lambda\tau\rho\sigma}\, p^\nu\bigg)\, p_\lambda
\nonumber \\
&+&
 \frac{1}{2} \delta^{\mu\omega}\,
\left(p + l\right)^\nu\, \bigg(- l^{\tau}\, \delta^{\rho\sigma} + 2\, l_\lambda\, s^{\tau\lambda\rho\sigma}\bigg)
+ (\mu \leftrightarrow \nu)\bigg] + (\tau,p) \leftrightarrow (\omega,q)
\nonumber
\eea
\bea
I^{\mu\nu\rho\sigma\tau\omega}(p,q,l)
&=&
\delta^{\mu\nu}\,\bigg\{
\frac{1}{2}\, \delta^{\rho\sigma}\, l^\tau\, \left(p + q + l\right)^\omega
-  s^{\lambda\tau\rho\sigma}   \, \bigg[q^\omega\, p_\lambda + l_\lambda\, \left( p + q + l \right)^\omega \bigg]
-  s^{\lambda\omega\rho\sigma} \, \bigg[p^\tau  \, p_\lambda + q_\lambda\, \left( q + l     \right)^\tau   \bigg]
\bigg\}
\nonumber\\
&-&
s^{\mu \nu \rho \sigma}\, \bigg(p^\omega\, p^\tau + q^\omega\, p^\tau\bigg) + (\tau,p) \leftrightarrow (\omega,q).
\label{TensorStructures}
\eea
We have performed all our computations in the Feynman gauge ($\xi=1$)
The Euclidean propagators of the fields in this case are
\bea
\langle\phi \, \phi \rangle (p)
&=&
\frac{1}{p^2}
\nonumber\\
\langle \bar{\psi}\, \psi \rangle (p)
&=&
\frac{1}{\slash{p}} = \frac{\slash{p}}{p^2} \, .
\nonumber \\
\langle A^\mu \, A^\nu \rangle (p)
&=&
- \frac{\delta^{\mu\nu}}{p^2} \, ,
\nonumber\\
\langle \bar{c}\, c \rangle (p)
&=&
\frac{1}{p^2} \, .
\eea

\section{Comments on the inverse mapping}
\label{InverseTTT}

In this appendix we offer some calculational details in the derivation of the expression of the $TTT$
correlator in position space. The remarks apply as well to any other correlator.\\
For example  Eq. (\ref{ScalarTriangle}) refers to the contribution coming from the triangle diagram shown in Fig.
\ref{Fig.diagramsTTT}. We assign the loop momentum $l$ to flow from the upper external point
($x_3$) to the lower one ($x_2$) on the right, the other two flows being determined by momentum conservation.
We denote the third external point as $x_1$.
For the scalar case, for instance, the complete one-loop triangle diagram is
\beq\label{TriangleLoop}
\int\, \frac{d^dl}{(2\pi)^d} \,
\frac{V^{\mu\nu}_{T\phi\phi}(l-q,-l-p)V^{\rho\sigma}_{T\phi\phi}(l,-l+q)V^{\alpha\beta}_{T\phi\phi}(l+p,-l)}{l^2\,(l-q)^2\,(l+p)^2}
\eeq
The vertices are defined in Eq. (\ref{MomVertices}).
The first argument in each vertex denotes the momentum of the incoming particle, the second argument
is the momentum of the outgoing one.
A typical term appearing in the loop integral is then
 \beq\label{TypicalInverse3}
I\equiv\int\, \frac{d^dl}{(2\pi)^d} \,
\frac{l^{\mu}\,l^{\nu}\,(l+p)^{\rho}\,(l+p)^{\sigma}(l-q)^{\alpha}\,(l-q)^{\beta}}{l^2\,(l-q)^2\,(l+p)^2}\, .
\eeq
From (\ref{fund}) the propagators in configuration space are
\beq
\frac{1}{l^2\,(l-q)^2\,(l+p)^2} =
C(1)^3 \, \int\,d^d x_{12}\,d^d x_{23}\,d^d x_{31}\,
\frac{e^{i\,[l\cdot x_{23}+(l-q)\cdot x_{12}+(l+p)\cdot x_{31}]}}{(x^2_{12})^{d/2-1}\,(x^2_{23})^{d/2-1}\,(x^2_{31})^{d/2-1}}
\, ,
\eeq
where $C\alpha)$ has been defined in (\ref{fund}).
It is straightforward to see that (\ref{TypicalInverse3}) is given by
\bea\label{TypicalInverse3Coord}
\int\, \frac{d^dl}{(2\pi)^d} \,
\frac{l^{\mu}\,l^{\nu}\,(l+p)^{\rho}\,(l+p)^{\sigma}(l-q)^{\alpha}\,(l-q)^{\beta}}{l^2\,(l-q)^2\,(l+p)^2}
&=&
\nonumber\\
&& \hspace {-70mm}
C(1)^3\,\int\, \frac{d^dl}{(2\pi)^d}\,d^d x_{12}\,d^d x_{23}\,d^d x_{31}\,
\frac{(-i)^6\,\partial^\mu_{23}\,\partial^\nu_{23}\,\partial^\rho_{31}\,\partial^\sigma_{31}\,
\partial^\alpha_{12}\,\partial^\beta_{12} \,e^{i\,[l\cdot x_{23}+(l-q)\cdot x_{12}+(l+p)\cdot x_{31}]}}
{(x^2_{12})^{d/2-1}\,(x^2_{23})^{d/2-1}\,(x^2_{31})^{d/2-1}} \nn \\
\, .
\eea
We can now integrate by parts moving the derivatives onto the propagators, getting
\bea
I&=&
C(1)^3\,\int\, \frac{d^dl}{(2\pi)^d}\,d^d x_{12}\,d^d x_{23}\,d^d x_{31}\,
e^{i\,[l\cdot x_{23}+(l-q)\cdot x_{12}+(l+p)\cdot x_{31}]}
\nonumber\\
&\times&
i^6\,\partial^\mu_{23}\,\partial^\nu_{23}\,\partial^\rho_{31}\,\partial^\sigma_{31}\,
\partial^\alpha_{12}\,\partial^\beta_{12}\,\frac{1}{(x^2_{12})^{d/2-1}\,(x^2_{23})^{d/2-1}\,(x^2_{31})^{d/2-1}} \, .
\eea
The second line is immediately identified with the coordinate space Green's function.\\
This can be done for each term of (\ref{TriangleLoop}), justifying the rule quoted
in section \ref{InverseMappingTTT}, that we have used for all the inverse mappings of the paper.
According to this the correlators in coordinate space can be obtained replacing the momenta in the vertices with ``$i$" times
the respective derivative which then act directly on the propagators after a partial integration.

The same arguments could be applied to the bubbles. Nevertheless, we have seen in \ref{InverseMappingTTT}
that derivatives of delta functions appear in the scalar case.
These are generated by the dependence of the $V^{\mu\nu\rho\sigma}_{TT\phi\phi}(p,q,l)$ from the momentum $l$
of the graviton bringing the pair of indices $\rho\sigma$ (see Eq. (\ref{MomVertices})).
They are due to coupling of the scalar with derivatives of the metric through the Ricci scalar $R$ in
the improvement term (see Eq. (\ref{scalarAction})) and state that the graviton feels the metric gradient.
We discuss this below, showing how to inverse-map the third bubble in Fig. (\ref{Fig.diagramsTTT}), getting
(\ref{ScalarKBubble}). \\
This bubble can be seen as the ($x_2\rightarrow x_3$) limit of the triangle
and its diagrammatic momentum-space expression at one-loop is
\beq\label{KBubbleLoop}
\int\, \frac{d^dl}{(2\pi)^d} \,
\frac{V^{\mu\nu}_{T\phi\phi}(l-q,-l-p)V^{\alpha\beta\rho\sigma}_{TT\phi\phi}(l+p,-l+q,-q)}{(l-q)^2\,(l+p)^2}\, .
\eeq
As the two propagators are expressed by
\beq
\frac{1}{(l+q)^2\,(l+p)^2} =
C(1)^2 \, \int\,d^d x_{12}\,d^d x_{31}\,
\frac{e^{i\,[(l-q)\cdot x_{12}+(l+p)\cdot x_{31}]}}{(x^2_{12})^{d/2-1}\,(x^2_{31})^{d/2-1}} \, ,
\eeq
the dependence of the second vertex on $p$ cannot be ascribed to neither of them.\\
Two typical terms encountered in (\ref{KBubbleLoop}) are
\bea\label{TypicalInverse2}
\int\, \frac{d^dl}{(2\pi)^d} \,
\frac{(l+p)^{\rho}\,(l+p)^{\sigma}(l-q)^{\alpha}\,(l-q)^{\beta}}{(l-q)^2\,(l+p)^2} \, ,
\nonumber\\
\int\, \frac{d^dl}{(2\pi)^d} \,
\frac{(l+p)^{\rho}\,(l+p)^{\sigma}(l-q)^{\alpha}\,p^{\beta}}{(l-q)^2\,(l+p)^2} \, .
\eea
The first one is treated at once restricting the procedure used for the three point function to the case of two
propagators.\\
For the second one, the following relation is immediately checked:
\bea\label{TypicalInverse2Coord}
\int\, \frac{d^dl}{(2\pi)^d} \,
\frac{(l+p)^{\rho}\,(l+p)^{\sigma}(l-q)^{\alpha}\,p^{\beta}}{(l-q)^2\,(l+p)^2}
&=&
\nonumber\\
&& \hspace {-70mm}
C(1)^2\,\int\, \frac{d^dl}{(2\pi)^d}\,d^d x_{12}\,d^d x_{23}\,d^d x_{31}\,
\delta^{(d)}(x_{23})\,\frac{(-i)^4\,\partial^\rho_{31}\,\partial^\sigma_{31}\,\partial^\alpha_{12}
\,(\partial_{31}-\partial_{23})^\beta\,
e^{i\,[l\cdot x_{23}+(l-q)\cdot x_{12}+(l+p)\cdot x_{31}]}}
{(x^2_{12})^{d/2-1}\,(x^2_{31})^{d/2-1}} \, . \nn \\
\eea
Notice that an integration by parts brings in a derivative on the delta functions giving
\bea
C(1)^2\,\int\, \frac{d^dl}{(2\pi)^d}\,d^d x_{12}\,d^d x_{23}\,d^d x_{31}\,
e^{i\,[l\cdot x_{23}+(l-q)\cdot x_{12}+(l+p)\cdot x_{31}]}\,
(i)^4\,\partial^\rho_{31}\,\partial^\sigma_{31}\,\partial^\alpha_{12}\,(\partial_{31}-\partial_{23})^\beta\,
\frac{\delta^{d}(x_{23})}{(x^2_{12})^{d/2-1}\,(x^2_{31})^{d/2-1}} \, . \nn \\
\eea
This approach has been followed in all the derivations of the expressions given in (\ref{InverseMappingTTT}).\\
The integration on $l$ brings about a $\delta^{(d)}(x_{12}+x_{23}+x_{31})$, so that it is natural to chose
the parameterization
\beq
x_{12} = x_1 - x_2\, , \quad x_{23} = x_2 - x_3\, , \quad x_{31} = x_3 - x_1\, .
\eeq
A more inviolved example is the 4-particle vertex. For instance the 
$V_{TT\phi\phi}(i\, \partial_{31},- i\, \partial_{12},i \, (\partial_{12}-\partial_{23}))$
is obtained from $V_{TT\phi\phi}(p,q,l)$ with the functional replacements 
\beq
p\to \hat{p}=i\, \partial_{31},\qquad  q \to \hat{q}=- i\, \partial_{12}\qquad l\to \hat{l}=i \, (\partial_{12}-\partial_{23})
\eeq
giving 
\bea
V_{TT\phi\phi}^{\mu\nu\rho\sigma}(i\, \partial_{31},- i\, \partial_{12},i \, (\partial_{12}-\partial_{23})) =
&&
\nonumber\\
&& \hspace{-60mm} \frac{1}{2}\, i\, \partial_{31}\cdot (-i)\,\partial_{12} s^{\mu\nu\rho\sigma} 
  - \frac{1}{4}\, G^{\mu\nu\rho\sigma}(i\,\partial_{31},-i\,\partial_{12})
+ \frac{1}{4}\, \delta^{\rho\sigma}\, i\,\partial_{31\,\alpha} \, (-i)\,\partial_{12\,\beta}\, C^{\mu\nu\alpha\beta}
\nonumber\\
&&\hspace{-60mm}
+ \, \chi \, \bigg\{\bigg[  \bigg(\delta^{\mu\lambda}\,\delta^{\alpha\kappa}\,\delta^{\nu\beta}
+ \delta^{\mu\alpha}\,\delta^{\nu\kappa}\,\delta^{\beta\lambda}
- \delta^{\mu\kappa}\,\delta^{\nu\lambda}\,\delta^{\alpha\beta}
- \delta^{\mu\nu}\,\delta^{\alpha\lambda}\,\delta^{\beta\kappa}\bigg)
s^{\rho\sigma}_{\lambda\kappa}
\nonumber \\
&& \hspace{-60mm}
+\, \frac{1}{2} \, \delta^{\rho\sigma} \, \bigg(\delta^{\mu\alpha}\,\delta^{\nu\beta} -
\delta^{\mu\nu}\,\delta^{\alpha\beta}\bigg)\bigg]
\left(i\,\partial_{31\,\alpha} \, (-i)\partial_{12\,\beta} + i\,\partial_{31\,\beta} \,(-i)\partial_{12\,\alpha} 
+ i\,\partial_{31\,\alpha} \, i\,\partial_{31\,\beta} + (-i)\partial_{12\,\alpha} \,(-i)\partial_{12\,\beta} \right)
\nonumber \\
&&\hspace{-60mm}
- \, \bigg[\bigg(\delta^{\mu\alpha}\,\delta^{\nu\beta} - \delta^{\mu\nu}\,\delta^{\alpha\beta}\bigg)
\bigg(\big[\Gamma^\lambda_{\alpha\beta}\big]^{\rho\sigma}\left(i\,(\partial_{12}-\partial_{23})\right)\bigg)\, 
(-i) \, \left( i \,  \partial_{31\,\lambda} + \, (-i)\, \partial_{12\,\lambda}\right)
\nonumber	\\
&& \hspace{-60mm} 
+ \, \frac{1}{2} \, \bigg(\delta^{\mu\alpha}\,\delta^{\nu\beta} - \frac{1}{2}\,\delta^{\mu\nu}\,\delta^{\alpha\beta}\bigg)\,
 \bigg( \big[R_{\alpha\beta}\big]^{\rho\sigma}\left(i\,(\partial_{12}-\partial_{23})\right)\bigg) \,  \bigg]\bigg\} \, .
\eea

\section{Regularizations and distributional identities}
\label{Distributional}

We add few more comments and examples which illustrate the regularization that we have applied in the
computation of the various correlators.

The computation of the logarithmic integrals requires some care due to the distributional nature of some of these formulas.
As an example we consider the integrals
\bea
H_1 = \int d^d l\, e^{i l\cdot x}\,  \frac{\mu^{2 \omega}}{[l^2]^{1 + \omega}} \qquad
H_2 = \int d^d l\, e^{i l\cdot x}\,  \frac{\mu^{2 \omega}}{[l^2]^{ \omega}} \qquad
H_3 = \int d^d l\, e^{i l\cdot x}\,  \log\left(\frac{l^2}{\mu^2}\right)
\eea
We can relate them in the form
\beq
H_3 = -\frac{\partial}{\partial \omega}H_2 \bigg|_{\omega=0}
    = \Box \left(\frac{\partial}{\partial \omega}H_1 \bigg|_{\omega=0}\right)
\eeq
In the two cases we get, using (\ref{fund})
\beq
-\frac{\partial}{\partial \omega}H_2 \bigg|_{\omega=0} = -  \frac{(4\,\pi)^{d/2} \, \Gamma(d/2)}{(x^2)^{d/2}}
\eeq
and
\beq
\frac{\partial}{\partial \omega}H_1 \bigg|_{\omega=0}
= \frac{2^{d-2} \pi^{d/2} \Gamma(d/2-1)}{[x^2]^{d/2-1}}
  \bigg( \log(x^2 \mu^2) + \gamma - \log 4  - \psi \left( \frac{d-2}{2}\right) \bigg)
\label{intermed}
\eeq
By redefining the regularization scale $\mu$ with Eq. (\ref{massscale}) we clearly obtain from (\ref{intermed})
\beq
\int d^d l \frac{\log( l^2/\mu^2) e^{i l\cdot x}}{l^2} =
2^{d-2}\pi^{d/2} \Gamma(d/2-1) \frac{\log x^2 \bar{\mu}^2}{[x^2]^{d/2-1}}
\eeq
and
\beq \label{First}
H_3 = \Box \left(\frac{\pd}{\pd\omega} H_1\bigg|_{\omega=0} \right) =
2^{d - 2} \pi^{d/2} \Gamma(d/2-1) \Box \left( \frac{\log x^2 \bar{\mu}^2}{[x^2]^{d/2-1}}\right)
\eeq
The use of $H_2$ instead gives
\beq
H_3 = -\frac{\partial}{\partial \omega}H_2 \bigg|_{\omega=0}
    = - \frac{2^d \pi^{d/2} \Gamma(d/2)}{[x^2]^{d/2}}
\eeq
Notice that this second relation coincides with (\ref{First}) away from the point $x=0$,
but differs from it right on the singularity, since
\beq
\Box \frac{\log x^2\mu^2}{[x^2]^{d/2-1}}
= - 2 \, (d-2) \bigg( \frac{\pi^{d/2}}{\Gamma(d/2)} \, \log( x^2\mu^2) \,  \delta^d (x)
  + \frac{1}{[x^2]^{d/2}} \bigg)
\eeq
For this reason we take (\ref{First}) as the regularized expression of $H_3$, in agreement
with the standard approach of differential regularization.

\subsection{Evaluation of the single log integrals}

The direct method discussed in the second part of the paper, though very general and applicable to any correlator, introduces in 
momentum space some logarithmic integrals which are more difficult to handle. They take the role of the ordinary master integrals of 
perturbation theory. The scalar integrals needed for the tensor reduction of the logarithmic contributions in the text are defined
in (\ref{StandInt}).
After a shift of the momentum in the argument of the logarithm,  a standard tensor reduction gives
\bea
IL_\mu(0,p_1,p_2)
&=&
{CL}_1(p_1,p_2)\, p_{1\,\mu} + {CL}_2(p_1,p_2)\, p_{2\,\mu} \, ,
\nonumber\\
{CL}_1(p_1,p_2)
&=&
\frac{({p_1}^2 - p_1\cdot p_2){p_2}^2\, IL(0,p_1,p_2) + ({p_2}^2 - p_1\cdot p_2)\, {IL^\mu}_\mu(0,p_1,p_2)}
{2\, (p_1\cdot p_2)^2 - {p_1}^2\, {p_2}^2}\, ,
\nonumber \\
{CL}_2(p_1,p_2)
&=&
\frac{({p_2}^2 - p_1\cdot p_2){p_1}^2\, IL(0,p_1,p_2) + ({p_1}^2 - p_1\cdot p_2)\, {IL^\mu}_\mu(0,p_1,p_2)}
{2\, (p_1\cdot p_2)^2 - {p_1}^2\, {p_2}^2}\, .
\eea
To complete the computation of the $VVV$ correlator we need the explicit form of the logarithmic integrals
in terms of ordinary logarithmic and polylogarithmic functions.
We define
\beq
\mathcal I \equiv \int d^d l \frac{\log \left(l^2 / \mu^2\right)}{(l+p_1)^2 (l-p_2)^2} =
- \frac{\partial}{\partial \lambda}
\int d^dl\,  \frac{\mu^{2 \lambda}}{(l^2)^\lambda\,(l+p_1)^2\,(l-p_2)^2}_{\lambda = 0} \, .
\eeq
The logarithmic integral is identified from the term of O($\lambda$) in the series expansion of the previous expression.
Because the coefficient in front of the parametric integral starts at this order, we just need to know the zeroth
order expansion of the integrand, which we separate into two terms. The first one is integrable
\bea
I_1 = \int_0^1 d t  \frac{t^{-\epsilon}  (y t)^{1-\epsilon-\lambda}}{A(t)^{1-\epsilon}} =
\int_0^1 d t  \frac{t^{-\epsilon}  (y t)^{1-\epsilon}}{A(t)^{1-\epsilon}}  + O(\lambda) \equiv  I_1^{(0)}  + O(\lambda)\, ,
\eea
while the last term has a singularity in $t=0$ which must be factored out and re-expressed in terms of a pole in $\lambda$
\bea
I_2 &=& - \int_0^1 d t  \frac{t^{-\epsilon}  (x / t)^{1-\epsilon-\lambda}}{A(t)^{1-\epsilon}} =
- \frac{ x^{1-\epsilon-\lambda}}{\lambda} \int_0^1 d t \frac{1}{A(t)^{1-\epsilon}} \frac{d }{dt} t^\lambda \nn \\
&=&
- \frac{ x^{1-\epsilon-\lambda}}{\lambda} \bigg[ 1 - (\epsilon -1) \int_0^1 d t \frac{t^\lambda}{A(t)^{1-\epsilon}}
\left(\frac{1}{t-t_1} + \frac{1}{t-t_2}\right)\bigg] \nn \\
&=&
\frac{x^{1-\epsilon} }{\lambda} \bigg\{ -1 +  (\epsilon -1) \int_0^1 d t \frac{1}{A(t)^{1-\epsilon}}
\left(\frac{1}{t-t_1} + \frac{1}{t-t_2}\right)\bigg] \bigg\} \nn \\
&+&
x^{1-\epsilon} \bigg[ \log x + (\epsilon -1) \int_0^1 d t \frac{\log \left(t/x\right)}{A(t)^{1-\epsilon}}
\left(\frac{1}{t-t_1} + \frac{1}{t-t_2}\right)\bigg] + O(\lambda)  \equiv \frac{1}{\lambda} I_2^{(-1)} +  I_2^{(0)}
+ O(\lambda)\, , \nonumber \\
\eea
where $t_1$ and $t_2$ are the two roots of $A(t) = y t^2 + (1-x-y)t + x$.
We are now able to write down the full $\lambda$-expansion of $J(1,1,\lambda)$ and to extract the logarithmic
integral $\mathcal I$
\bea
\mathcal I = -  \frac{\pi^{2-\epsilon} i^{1+2\epsilon}}{(p_3^2)^{\epsilon}} \frac{ \Gamma(1-\epsilon) \Gamma(2-\epsilon)
\Gamma(\epsilon )}{ \Gamma(2-2\epsilon)} \frac{1}{\epsilon-1} \bigg\{ I_1^{(0)} + I_2^{(0)} \bigg\} \,.
\eea
The previous expression can be expanded in $d=4-2\epsilon$ dimensions in which it manifests a $1/\epsilon$
pole of ultraviolet origin %
\bea
\mathcal I =
\frac{\pi^{2-\epsilon} i^{1+2\epsilon}}{(p_3^2)^{\epsilon}} \left( -\frac{1}{\epsilon}
+ \gamma \right)\bigg[ A(x,y) + \epsilon \, B(x,y) \bigg]  + O(\epsilon) \,,
\eea
where $A(x,y)$ and $B(x,y)$ are defined from the $\epsilon$-expansion of the two integrals $I_1^{(0)}$ and $I_2^{(0)}$ as
\bea
A(x,y) &=&  x \log x + \int_0^1  \frac{dt}{A(t)} \bigg[ y t - x \log \left( t/x \right)
\left(\frac{1}{t-t_1} + \frac{1}{t-t_2}\right)\bigg] \,, \\
B(x,y) &=& - x \log^2 x + \int_0^1  \frac{dt}{A(t)}  \bigg[ y t \left( \log\left(t-t_1 \right)
+ \log\left(t-t_2 \right) - 2 \log t \right) \nn \\
&-& x \log\left(t/x \right) \left(\frac{1}{t-t_1} + \frac{1}{t-t_2}\right)
\left( \log\left(t-t_1 \right) + \log\left(t-t_2 \right) -  \log \left( x/y\right) -1\right) \bigg].
\eea




\end{document}